\documentclass[aps, prd, reprint]{revtex4-2}
\usepackage[pdftex]{graphicx}
\usepackage{amsmath,amssymb}
\usepackage{mathtools}
\usepackage{slashed}
\usepackage{xcolor}
\usepackage{hyperref}
 \usepackage{caption}
\usepackage{subcaption} 
 \newcommand{\llangle}{\Big\langle \!\! \Big\langle}
\newcommand{\rrangle}{\Big\rangle \!\! \Big\rangle}

\begin{document}
\title{Quasi-Classical Evaluation of Gluon Saturation Induced Helicity Effects}
\date{\today}
\author{Ming Li}
\email{li.13449@osu.edu}
\affiliation{Department of Physics, The Ohio State University, Columbus, 43210, Ohio USA}

\begin{abstract}
At sub-eikonal order in the high-energy limit, helicity dependences are generally contained in the transverse components of the gluon field. In the gluon saturation regime, novel helicity effects arise from the nonlinear interaction between the eikonal-order longitudinal gluon field and the sub-eikonal-order transverse gluon field. We derive this saturation induced helicity-dependent field both by solving the classical Yang-Mills equations and through direct diagrammatic calculations. Our analysis shows that the saturation induced helicity effect is intrinsically a two-particle (or multi-particle) correlation effect, rather than a single-particle distribution effect. Furthermore, we evaluate this effect in the context of double-spin asymmetry for incoherent diffractive dijet production in longitudinally polarized electron-nucleus collisions, using a simplified helicity-extended McLerran-Venugopalan model. We find that the saturation induced helicity effect further suppresses the back-to-back peak in the dijet azimuthal angle correlation, with a magnitude comparable to that of the direct helicity effect. This gluon saturation induced helicity effect may offer a novel avenue to probe gluon saturation in polarized collisions.
\end{abstract}
\maketitle

\section{Introduction}
Discovering and studying the properties of gluon saturation is one of the major scientific goals of the upcoming Electron-Ion Collider (EIC) to be constructed at Brookhaven National Laboratory \cite{AbdulKhalek:2021gbh}. Gluon saturation concerns the small $x$ (longitudinal momentum fraction) regime of the nuclear wavefunction and originates from nonlinear gluon interactions. As the collisional energies in high energy nuclear collisions are increased, more and more small $x$ gluons are generated from gluon splittings due to increased phase space. When the gluon density becomes sufficiently high, inverse processes of gluon merging eventually balance out the gluon spliting processes, resulting in saturated gluon state \cite{Iancu:2003xm}. The typical transverse momentum of the saturated gluons is characterized by a dynamically generated momentum scale, $Q_s$, the gluon saturation scale. The $Q_s$ depends on both the nuclear size (atomic mass number $A$) and the collision energy (center-of-mass collision energy $s$) \cite{Kovchegov:2012mbw}. In the gluon saturation regime in which the gluon field is of order $A^{\mu} \sim \mathcal{O}(1/g)$, scatterings involving multiple gluon exchanges have to be resumed. One compelling signature of gluon saturation is the suppression of di-hadron (or dijet) azimuthal angle correlations in high energy electron-proton/nucleus collisions \cite{AbdulKhalek:2021gbh, Zheng:2014vka}. The amount of this suppression is determined by the gluon saturation scale, which varies with both the atomic mass number and the collision energy.

Most experimental observables proposed to investigate gluon saturation are based on unpolarized nuclear collisions \cite{AbdulKhalek:2021gbh}.  It is because, at the eikonal order in the high energy limit, the gluon field characterizing the saturated gluons is insensitive to the spin states of the proton or nucleus. However, in this paper, we aim to demonstrate that novel helicity effects induced by gluon saturation at sub-eikonal order can emerge in polarized nuclear collisions. Investigating gluon saturation in polarized nuclear collisions will thus complement studies in unpolarized nuclear collisions, providing new insights into the nature of gluon saturation. Furthermore, future EIC experiments will have the capability to perform polarized electron-ion collisions, in addition to polarized electron-proton collisions, potentially providing valuable data to test these novel gluon saturation effects.   To study gluon saturation in polarized nuclear collisions, it is necessary to go beyond the eikonal approximation. In recent years, there have been many efforts in developing the effective theories to describe small $x$ physics at the sub-eikonal order  \cite{Chirilli:2018kkw, Kovchegov:2021iyc, Altinoluk:2021lvu, Li:2023tlw, Agostini:2023cvc}.

For a highly boosted proton or nucleus moving in the positive-$z$ direction, the $A^+(x)$ component of the gluon field $A^{\mu}(x)$ dominates at the eikonal order in the Lorenz gauge. At sub-eikonal order, both $A^+(x)$ and $A^i(x)$ components contribute. In the quasi-classical approximation, these components of gluon field are effectively described by quasi-classical fields, which can be obtained by solving the classical Yang-Mills equations, treating the large $x$ partonic modes as external color currents \cite{Li:2024fdb, Cougoulic:2020tbc}. At the eikonal order, the functional form of $A^+_{\mathrm{eik}}(x)$ in terms of color currents is the same in both the dilute regime $ \mathcal{O}(g)$ and the dense regime $ \mathcal{O}(1/g)$. However, at the sub-eikonal order, the expressions for both $A^+_{\mathrm{sub}}(x)$ and $ A^i_{\mathrm{sub}}(x)$ differ depending on whether the dilute or dense regime is considered. In general, explicit helicity dependence is found in $A^i_{\mathrm{sub}}(x)$ but not in $A^+_{\mathrm{sub}}(x)$. However, in the dense regime, an additional contribution to $A^{+ a}_{\mathrm{sub}}(x) \sim gf^{abc} \partial^i A^{+, b}_{\mathrm{eik}}(x) A^{i, c}_{\mathrm{sub}}(x)$ renders $A^+_{\mathrm{sub}}(x)$ sensitive to helicity.  This additional sub-eikonal order helicity dependence arises from the nonlinear interaction between the eikonal order field $A^+_{\mathrm{eik}}(x)$, which is of order $\mathcal{O}(1/g)$, and the sub-eikonal order field $A^i_{\mathrm{sub}}(x)$. We refer to this as the gluon saturation induced helicity-dependent field. The goal of this paper is to evaluate the gluon saturation induced helicity effect encoded in $A^+_{\mathrm{sub}}(x)$ using a simplified helicity-extended McLerran-Venugopalan (MV) model \cite{Cougoulic:2020tbc, McLerran:1993ni, McLerran:1993ka} and make a comparison with direct helicity effect from $A^i_{\mathrm{sub}}(x)$. 

We begin by examining the single gluon helicity distribution and demonstrate that the gluon saturation induced helicity effects do not contribute. Specifically, we calculate the transverse momentum-dependent \textit{dipole} gluon helicity distribution \cite{Cougoulic:2022gbk} at both leading and next-to-leading order within the quasi-classical approximation, showing that these terms vanish. Additionally, we present a straightforward argument demonstrating that the \textit{Weizsacker-Williams} (WW) gluon helicity distribution \cite{Kovchegov:2017lsr} is unaffected by helicity-dependent fields induced by gluon saturation. These results suggest that the gluon saturation induced helicity effects are inherently two-particle (or multi-particle) correlations and do not influence single-particle distributions.

We analyze two-particle correlated helicity distributions, focusing on the observable of double-spin asymmetry for incoherent diffractive quark-antiquark dijet production in longitudinally polarized electron-proton or electron-nucleus collisions. By examining the incoherent component of diffractive dijet production, we can isolate two-particle correlations by subtracting single-particle distributions. In unpolarized collisions, incoherent diffractive dijet production is related to unpolarized Wilson line dipole-dipole correlator and has been shown to probe gluon Bose correlations \cite{Kar:2023jkn} and to provide potential test of gluon saturation \cite{Mantysaari:2019hkq, Rodriguez-Aguilar:2023ihz, Rodriguez-Aguilar:2024efj}. In polarized collisions, this observable is related to correlations between polarized and unpolarized Wilson line dipoles. We focus on the azimuthal angle correlation between the momenta of the two jets. 

For the dijet azimuthal angle correlation, the leading contribution is of order $Q_s^4$, arising solely from the direct helicity effect due to $A^i_{\mathrm{sub}}(x)$. This correlation is positive, peaking at a back-to-back configuration with $\Delta \phi = \pi$. At next-to-leading order, $Q_s^6$, both the direct helicity effect and the helicity effect induced by gluon saturation contribute negatively, with comparable magnitudes. Consequently, the gluon saturation-induced helicity effect is expected to further suppress the back-to-back peak in addition to the suppression observed when only direct helicity effect is considered. More quantitative studies will require computing higher order contributions beyond the order $Q_s^6$, likely necessitating an all-order resummation. The result at order $Q_s^6$ already demonstrate that the gluon saturation induced helicity effect causes further suppression of the back-to-back peak, just like the direct helicity effect, highlighting the need for more detailed investigations. 

The paper is organized as follows: In sec.~\ref{sec:deriving_induced}, we review the derivation of the gluon saturation induced helicity-dependent field using the method of solving classical Yang-Mills equations.  We then provide a second derivation through explicit diagrammatic calculations.  The quasi-classical approximation is described in sec.~\ref{sec:quasi_classical_app}, emphasizing the three-field averaging as the fundamental building block. The gluon saturation induced helicity effect is evaluated in single-particle distribution in sec.~\ref{sec:single-particle_dis} and in the two-particle correlations in sec.~\ref{sec:two_particle_dis}. Analysis in the context of longitudinal double-spin asymmetry for incoherent diffractive dijet production is presented in sec.~\ref{sec:ALL_dijet}. Both analytic expressions and numerical calculations are included. Conclusion and outlook are provided in sec.~\ref{sec:summary_outlook}.

\section{Gluon Saturation Induced Helicity-Dependent Field}\label{sec:deriving_induced}
In this section, we review the derivation of the gluon saturation induced helicity-dependent field from solving classical Yang-Mills equations first presented in \cite{Li:2024fdb}. We then give an alternative derivation by explicit diagrammatic calculations. 

\subsection{From solving classical Yang-Mills equations}
For a pure gluonic system, the quasi-classical gluon fields, $\mathcal{A}^{\mu}$, which describe the small $x$ gluons, are governed by the classical equations of motion:
\begin{equation}\label{eq:EOM}
\mathcal{D}_{\nu} \mathcal{F}^{\nu\mu}_b   = J_b^{\mu}
\end{equation}
with the current
\begin{equation}
J^{\mu}  =  ig[A_{\nu}, \overline{F}^{\mu\nu}] + ig\mathcal{D}^{\nu}[A^{\mu}, A_{\nu}]
\end{equation}
defined by the large-$x$ gluon field $A^{\mu}$. 
The covariant derivative is given in terms of  the small-$x$ gluon field as $\mathcal{D}_{\nu} = \partial_{\nu} + ig[\mathcal{A}_{\nu}, \,\,\,]$. The field strength tensor, in the background of quasi-classical gluon field, is then expressed as
\begin{equation}
\overline{F}^{\mu\nu} = \mathcal{D}^{\mu} A^{\nu} - \mathcal{D}^{\nu} A^{\mu} + ig[A^{\mu}, A^{\nu}].
\end{equation}
The equations of motion given in eq.~\eqref{eq:EOM} are manifestly gauge-invariant, and the color current transforms covariantly under gauge transformations. Color current conservation $\mathcal{D}_{\mu} J^{\mu} =0$ follows straightforwardly. 

We work in the Lorenz gauge $\partial_{\mu} \mathcal{A}^{\mu}=0$. Eq.~\eqref{eq:EOM} becomes
\begin{equation}
\partial_{\nu} \partial^{\nu} \mathcal{A}^{\mu} + ig \left[\mathcal{A}^{\nu}, \partial_{\nu}\mathcal{A}^{\mu}\right] + ig\left[\mathcal{A}_{\nu}, \mathcal{F}^{\nu\mu}\right] = J^{\mu}
\end{equation}
The four components of the equations are 
\begin{align}\label{eq:EOM_+} 
&\partial^2 \mathcal{A}^{+} + ig[\mathcal{A}^+, \partial_+ \mathcal{A}^+] + ig[\mathcal{A}^-, \partial_- \mathcal{A}^+] + ig[\mathcal{A}^i, \partial_i \mathcal{A}^+]  \notag \\
&\qquad+ ig[\mathcal{A}_i, \mathcal{F}^{i+}] + ig[\mathcal{A}_-, \mathcal{F}^{-+}]= J^+, 
\end{align}
\begin{align}\label{eq:EOM_i}
&\partial^2 \mathcal{A}^{i} + ig[\mathcal{A}^+, \partial_+ \mathcal{A}^i] + ig[\mathcal{A}^-, \partial_- \mathcal{A}^i] + ig[\mathcal{A}^j, \partial_j \mathcal{A}^i] \notag \\
&\qquad + ig[\mathcal{A}_j, \mathcal{F}^{ji}] + ig[\mathcal{A}_-, \mathcal{F}^{-i}] + ig[\mathcal{A}_+, \mathcal{F}^{+i}]= J^i,
\end{align}
\begin{align}\label{eq:EOM_-}
&\partial^2 \mathcal{A}^{-} + ig[\mathcal{A}^+, \partial_+ \mathcal{A}^-] + ig[\mathcal{A}^-, \partial_- \mathcal{A}^-] + ig[\mathcal{A}^i, \partial_i \mathcal{A}^-] \notag\\
&\qquad + ig[\mathcal{A}_i, \mathcal{F}^{i-}] + ig[\mathcal{A}_+, \mathcal{F}^{+-}] = J^-.
\end{align}
Here $\partial^2 = 2\partial_+\partial_--\partial^2_{\perp}$.

The equations in eq.~\eqref{eq:EOM_+}, ~\eqref{eq:EOM_i}, and ~\eqref{eq:EOM_-} will be solved up to sub-eikonal order. The approach for determining the eikonality of gluon fields is based on their transformation properties under Lorentz boosts. 
Specifically, under a Lorentz boost, the gluon field transforms as 
\begin{equation}
\begin{split}
&\mathcal{A}^+\longrightarrow \frac{1}{\xi} \mathcal{A}^+ (\xi x^+, \frac{1}{\xi} x^-, \mathbf{x}), \\
&\mathcal{A}^- \longrightarrow \xi A^-(\xi x^+, \frac{1}{\xi} x^-, \mathbf{x}),\\
&\mathcal{A}^i \longrightarrow A^i(\xi x^+, \frac{1}{\xi} x^-, \mathbf{x}),\\
\end{split}
\end{equation}
Here $\xi = e^{-\omega}$, where $\omega$ characterizes the amount of the Lorentz boost. In the high energy limit, $\xi\rightarrow 0$. 
A similarly transformation applies for the color current $J^{\mu}$ and the field strength tensor $\mathcal{F}^{\mu\nu}$. 

The gluon fields are expanded in a Taylor series in powers of $\xi$. (note that one redefines $\tilde{x}^- = \frac{1}{\xi} x^-$ so the arguments of the gluon fields are $(0^+, \tilde{x}^-, \mathbf{x})$ after expansions ).
At the eikonal order, it is evident that the only nonvanishing component of the gauge field is $\mathcal{A}^+$, which is independent of the light-cone time $x^+$.  At sub-eikonal order, both $\mathcal{A}^+$  and  $\mathcal{A}^i$ are nonvanishing components of the field.

At the eikonal order, from eq.~\eqref{eq:EOM_+}, one obtains
\begin{equation}
\mathcal{A}^+_{\mathrm{eik}} = \frac{-1}{\partial^2_{\perp}}J^+_{\mathrm{eik}}.
\end{equation}
To be more explicit,
\begin{equation}\label{eq:eik_A^+_J^+}
\mathcal{A}^+_{\mathrm{eik}}(0^+, x^-, \mathbf{x}) = \int d^2\mathbf{z}\, \phi(\mathbf{x}-\mathbf{z}) J^+_{\mathrm{eik}}(0^+, x^-, \mathbf{z})
\end{equation}
with
\begin{equation}
\phi(\mathbf{x}-\mathbf{z}) =-\frac{1}{\partial^2_{\perp}}(\mathbf{x}, \mathbf{z}) = \frac{1}{2\pi} \ln\frac{1}{|\mathbf{x}-\mathbf{z}|\Lambda}.
\end{equation} 
Here $\Lambda$ is the infrared cut-off. Eq.~\eqref{eq:eik_A^+_J^+} is the well-known eikonal order solution.\\

 At the sub-eikonal order, we look for solutions $\mathcal{A}^+_{\mathrm{sub}}, \mathcal{A}^i_{\mathrm{sub}}$  in the background of eikonal order field $\mathcal{A}^+_{\mathrm{eik}}$ and assuming 
$ \mathcal{A}^-=0$. 
The Lorenz gauge condition reduces to 
\begin{equation}\label{eq:cov_gauge_sub-eikonal}
\partial_{+}\mathcal{A}^+_{\mathrm{sub}} = -\partial_i \mathcal{A}^i_{\mathrm{sub}}.
\end{equation}
It is apparent that at sub-eikonal order $\mathcal{A}^+_{\mathrm{sub}}$ contains terms that are  linearly dependent on the light-cone time $x^+$ while $\mathcal{A}^i_{\mathrm{sub}}$ is a static field.  At the sub-eikonal order, eq.~\eqref{eq:EOM_+}  reduces to 
\begin{align}
&2\partial_+\partial_- \mathcal{A}^+_{\mathrm{sub}} - \partial^2_{\perp} \mathcal{A}^+_{\mathrm{sub}} + 2ig\left[\mathcal{A}^+_{\mathrm{eik}}, \partial_+\mathcal{A}^+_{\mathrm{sub}}\right]  \notag \\
&\qquad +2 ig\left[\mathcal{A}^i_{\mathrm{sub}}, \partial_i \mathcal{A}^+_{\mathrm{eik}}\right] = J^+_{\mathrm{sub}},
\end{align}
and eq.~\eqref{eq:EOM_i} reduces to
\begin{equation}
-\partial^2_{\perp} \mathcal{A}^i_{\mathrm{sub}}  = J^i_{\mathrm{sub}}. 
\end{equation}
With the help of eq.~\eqref{eq:cov_gauge_sub-eikonal}, one obtains the solutions 
\begin{align}
&\mathcal{A}^+_{\mathrm{sub}} = - \frac{1}{\partial_{\perp}^2}\left(2\partial_i \mathcal{D}_-\mathcal{A}^i_{\mathrm{sub}} + J^+_{\mathrm{sub}}\right), \label{eq:A^+_exp_sub}\\
&\mathcal{A}^i_{\mathrm{sub}} = -\frac{1}{\partial^2_{\perp}}J^i_{\mathrm{sub}}.\label{eq:A^i_exp_sub}
\end{align}
The covariant derivative is defined with respect to the eikonal order field $\mathcal{D}_- = \partial_- + ig[\mathcal{A}^+_{\mathrm{eik}}, \,\,\,]$.
The $J^+_{\mathrm{sub}}$ contain terms that are linearly dependent on the light-cone time $x^+$ as can be seen from the current conservation equation $\partial_{+}J^+_{\mathrm{sub}}= - \partial_i J^i_{\mathrm{sub}}$ at the sub-eikonal order.  

At the sub-eikonal order, the transverse color current can be decomposed into 
\begin{equation}\label{eq:Ji_decompose}
J^{i, c}_{\mathrm{sub}}(x^-, \mathbf{x}) = -\epsilon^{il} \partial^l_{\mathbf{x}} \, j_c(x^-, \mathbf{x}) + j_c^i(x^-, \mathbf{x}).
\end{equation}
The helicity dependence is contained in $j(x^-, \mathbf{x})$ but not in $j^i(x^-, \mathbf{x})$. In the following, we only focus on the helicity dependent part of the transverse gluon field so that one approximates
\begin{equation}\label{eq:A^j_exp}
\mathcal{A}^{j, c}_{\mathrm{sub}}(x^-, \mathbf{x})  \simeq - \epsilon^{jl}\partial^l_{\mathbf{x}} \beta^c(x^-, \mathbf{x})
\end{equation}
with 
\begin{equation}
\beta^c(x^-, \mathbf{x}) = \int_{\mathbf{z}} \phi(\mathbf{x}-\mathbf{z}) j^c(x^-,\mathbf{z}).
\end{equation}
The $\mathcal{A}^+_{\mathrm{sub}}(x^-, \mathbf{x})$  in eq.~\eqref{eq:A^+_exp_sub} contains the gluon saturation induced helicity-dependent term, 
\begin{equation}\label{eq:induced_eik+_subi}
\mathcal{A}^{+, c}_{\mathrm{ind}}(x^-, \mathbf{x}) \simeq 2g f^{cde} \int_{\mathbf{z}} \phi(\mathbf{x}-\mathbf{z}) \, \partial^i \mathcal{A}^{+, d}_{\mathrm{eik}} (x^-, \mathbf{z}) \, \mathcal{A}^{i, e}_{\mathrm{sub}}(x^-, \mathbf{z}).
\end{equation}
Note that $\partial_i \mathcal{A}^i_{\mathrm{sub}}$ is independent of helicity per eq.~\eqref{eq:Ji_decompose}. 
Relabeling eq.~\eqref{eq:eik_A^+_J^+} as
\begin{equation}\label{eq:A^+_in_alpha}
\mathcal{A}^{+, d}_{\mathrm{eik}} (x^-, \mathbf{z})\equiv \alpha^d(x^-, \mathbf{z}) =  \int_{\mathbf{z}}\phi(\mathbf{x}-\mathbf{z}) \rho^d(x^-, \mathbf{z}),
\end{equation}
one gets the expression for gluon saturation induced helicity-dependent field
\begin{equation}\label{eq:final_induced_field_exp}
\mathcal{A}^{+, c}_{\mathrm{ind}}(x^-, \mathbf{x}) = -2gf^{cde}\epsilon^{il} \int_{\mathbf{z}}\phi(\mathbf{x}-\mathbf{z}) \partial^i \alpha^d(x^-, \mathbf{z}) \partial^l \beta^e(x^-, \mathbf{z}).
\end{equation}
This field originates from the nonlinear interaction between the eikonal order longitudinal gluon field and sub-eikonal order transverse gluon field. In the dilute regime, where $\alpha^d(x^-, \mathbf{z}) \sim g$, the contribution is higher order in the strong coupling constant and can therefore be neglected. However, in the dense regime, where $\alpha^d(x^-, \mathbf{z}) \sim 1/g$, it is of the same order as the direct helicity-dependent field expressed in eq.~\eqref{eq:A^j_exp} and must be considered when addressing helicity-related effects. Our goal is to evaluate the relative importance of $A^+_{\mathrm{ind}}$ compared to $A^j_{\mathrm{sub}}$ in helicity dependent observables.  A key feature of the induced helicity-dependent field is that two different color sources are needed to generate $A^+_{\mathrm{ind}}(x^-, \mathbf{x})$. These two color sources have the same longitudinal coordinate $x^-$, but differ in their transverse coordinates and their color indices. This property will become evident through the diagrammatic derivations presented in the following subsection.

\subsection{Diagrammatic Calculations}
In this section, we present an alternative derivation of the gluon saturation induced helicity-dependent field as given in eq.~\eqref{eq:final_induced_field_exp}. This new derivation provides a deeper  understanding of how the induced field emerges and highlights its key features. In this derivation, we consider large $x$ color sources from both quarks and gluons.
\subsubsection{Quark Source}
\begin{figure*}
    \includegraphics[width=0.75\textwidth]{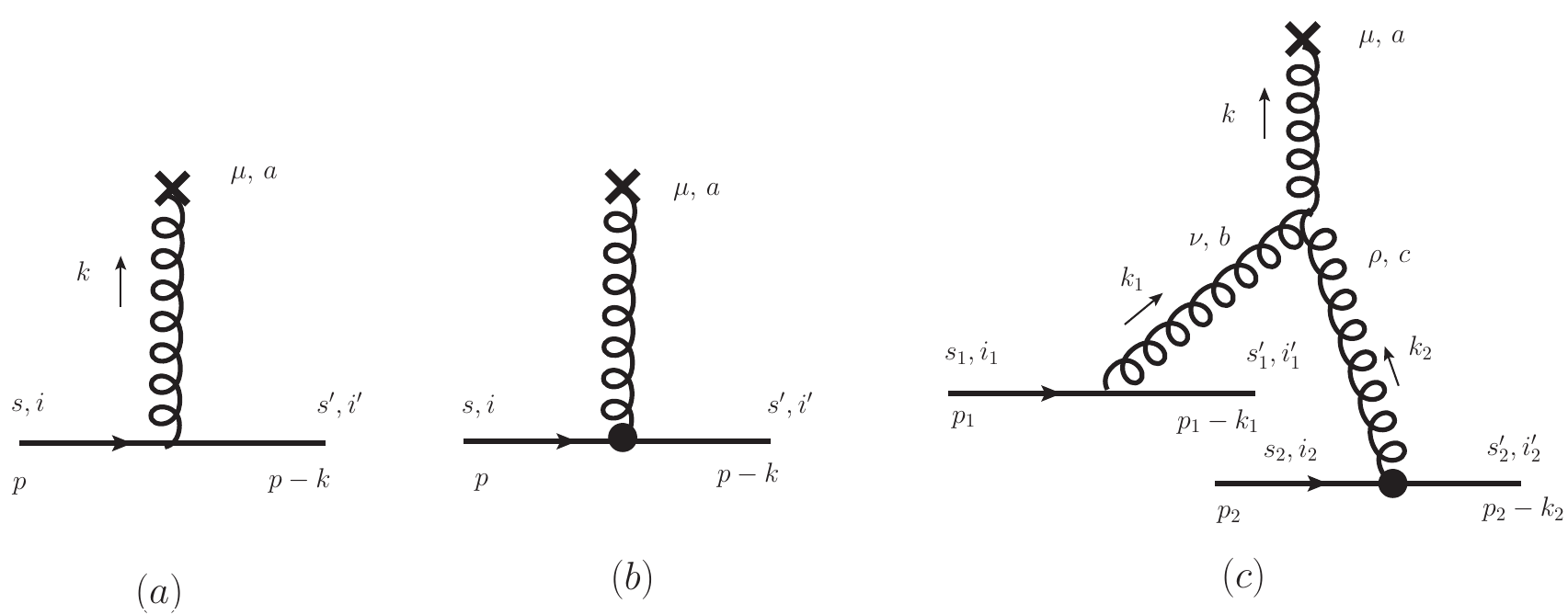}
    \caption{(a) Eikonal order gluon field $A^+$; (b) sub-eikonal order helicity-dependent gluon field $A^i$; (c) Gluon saturation induced helicity-dependent gluon field $A^+$ at sub-eikonal order.}
\label{fig:quark_induced}		
\end{figure*}
We begin by considering the situation that the gluon field is generated by quark sources. The relevant diagrams are shown in Fig.~\ref{fig:quark_induced}. Diagrams (a) and (b) in Fig.~\ref{fig:quark_induced} can be analyzed collectively, as they represent the gluon fields, up to sub-eikonal order,  which are generated by a single quark source. To be consistent with the result obtained from solving classical Yang-Mills equations, we also work in the Lorenz gauge. The quarks are assumed to be massless. The incoming quark is on-shell $p^2=0$. We further require the outgoing quark to remain on-shell $(p-k)^2=0$. The amplitude can be computed by
\begin{equation}\label{eq:A(k)_one}
\begin{split}
\mathcal{A}^{\mu}_a(k)
 =& \bar{u}(p-k, s') (-ig \gamma_{\nu} t^a) u(p, s) \frac{-i}{k^2+i\epsilon}\\
 &\quad \times  \left[g^{\mu\nu} - \frac{k^{\mu} k^{\nu}}{k^2}\right] (2\pi) \delta((p-k)^2)\\
  =&-g t^a_{i'i} \, \bar{u}(p-k, s') \gamma^{\mu} u(p, s)  \frac{1}{k^2+i\epsilon} (2\pi) \delta((p-k)^2)
\end{split}
\end{equation}
We work in the frame that the incoming quark travels along the $+z$ direction. The incoming and outgoing momenta are
\begin{equation}
\begin{split}
& p^{\mu} = (p^+, 0^-, \mathbf{0}), \\
&p^{\mu} -k^{\mu} = (p^+-k^+, -k^-, -\mathbf{k}).\\
\end{split}
\end{equation}
We also assume that the produced gluon field has small longitudinal momentum fraction  $k^+\ll p^+$.  The spinors are taken as eigenstates of helicity \cite{Thomson:2013zua}. It is a straightforward calculation to obtain ($s=\pm \frac{1}{2}$)
\begin{subequations}
\begin{align}
&\bar{u}(p-k, s') \gamma^{+} u(p, s)  = 2\sqrt{(p^+-k^+) p^+} \delta_{ss'}\\
&\bar{u}(p-k, s') \gamma^{i} u(p, s)  = \delta_{ss'} \sqrt{\frac{p^+}{p^+-k^+}} \Big[-\mathbf{k}^i + i (2s) \epsilon^{ij} \mathbf{k}^j\Big] \, \label{eq:ubar_i_u}\\
&\bar{u}(p-k, s') \gamma^{-} u(p, s)  =0.
\end{align}
\end{subequations}
The ``$-$" component is nonzero only when the transverse components of both the incoming and outing momenta have nonvanishing values. Even in that case, the expression is of sub-sub-eikonal order in powers of $k^+/p^+$. We can therefore safely neglect the ``$-$" component because we are only interested in gluon fields up to sub-eikonal order. 

To get the gluon field in coordinate space, we make Fourier transformation of eq.~\eqref{eq:A(k)_one}\begin{equation}
\mathcal{A}^{\mu}_a(x) = \int \frac{d^4 k}{(2\pi)^4} e^{-i k^+ (x^--b^-)-ik^-(x^+-b^+) + i\mathbf{k}\cdot(\mathbf{x}-\mathbf{b})} \mathcal{A}^{\mu}_a(k)
\end{equation} 
Here $b^{\mu} = (b^+, b^-, \mathbf{b})$ denotes the position of the quark source. 
The eikonal order gluon field, corresponding to diagram $(a)$ in Fig.~\ref{fig:quark_induced}, is computed to be
\begin{equation}
\mathcal{A}^{+ a}_{\mathrm{eik}}(x) 
=g t^{a}_{i' i} \delta_{ss'}  \delta(x^--b^-) \frac{1}{2\pi} \ln \frac{1}{|\mathbf{x}-\mathbf{b}|\Lambda}.
\end{equation}
This eikonal order expression has been derived in \cite{Kovchegov:2012mbw}. We used the approximation
\begin{equation}
 (p-k)^2  \simeq -2p^+k^- -\mathbf{k}^2 \simeq -2p^+k^- 
 \end{equation}
 so that $\delta((p-k)^2) \simeq \frac{1}{2p^+ }\delta(k^-)$. 
The sub-eikonal order helicity dependent gluon field, corresponding to diagram $(b)$ in Fig.~\ref{fig:quark_induced} can be similarly computed using eq.~\eqref{eq:ubar_i_u}. Its expression is
\begin{equation}
\mathcal{A}^{i, a}_{\mathrm{sub}}(x) \notag
 =-g t^{a}_{i' i} (2s\,\delta_{ss'})  \delta(x^--b^-) \frac{1}{2p^+} \frac{1}{2\pi} \frac{\epsilon^{ij} (\mathbf{x}-\mathbf{b})^j}{|\mathbf{x}-\mathbf{b}|^2}.
\end{equation}
Both the eikonal order gluon field, $\mathcal{A}_{\mathrm{eik}}^+$, and the sub-eikonal order helicity dependent gluon field, $\mathcal{A}_{\mathrm{sub}}^i$, depend on $\delta(x^--b^-)$, indicating that the longitudinal coordinate of the gluon field is the same as that of the quark source. 
The sub-eikonal order gluon field $\mathcal{A}_{\mathrm{sub}}^i(x)$ includes a factor $1/p^+$, which is characteristic of sub-eikonal order fields. 

We now compute the gluon saturation induced helicity-dependent field. This corresponds to diagram $(c)$ in Fig.~\ref{fig:quark_induced}. In this case, there are two quark sources that in general locate at different positions. Both outgoing quarks are required to be on-shell.  The diagram is computed by
\begin{widetext}
\begin{align}\label{eq:Amu_quark_general}
\mathcal{A}^{\mu}_a(k, k_1, k_2)
 =& \bar{u}(p_1-k_1, s'_1) (-ig\gamma^{\nu} t^b) u(p_1, s_1) \frac{-i}{k_1^2+i\epsilon}  \bar{u}(p_2-k_2, s'_2) (-ig\gamma^{\rho} t^c) u(p_2, s_2) \frac{-i}{k_2^2+i\epsilon} \notag\\
&\times (-gf^{abc})\Big[g_{\mu'\nu} (-k-k_1)_{\rho} + g_{\nu\rho} (k_1-k_2)_{\mu'} + g_{\rho\mu'}(k_2+k)_{\nu} \Big]  \frac{-i}{k^2+i\epsilon}\left[g^{\mu\mu'} - \frac{k^{\mu}k^{\mu'}}{k^2}\right] \notag \\
&\times (2\pi)^4 \delta^{(4)}(k_1+k_2-k) (2\pi) \delta((p_1-k_1)^2) (2\pi) \delta((p_2-k_2)^2) \notag\\
\simeq & ig^3 f^{abc} t^b_{i'_1 i_1} t^c_{i'_2 i_2}\, \Big[ \bar{u}(p_1-k_1, s'_1)\gamma^{\nu} u(p_1, s_1) \Big]\Big[\bar{u}(p_2-k_2, s'_2) \gamma^{\rho} u(p_2, s_2)\Big]\frac{1}{k_1^2+i\epsilon} \frac{1}{k_2^2+i\epsilon} \notag\\
&\times \Big[ g_{\mu'\nu} (-2k_{\rho})  + g_{\rho\mu'}(2k_{\nu})\Big]\frac{g^{\mu\mu'}}{k^2+i\epsilon}  (2\pi)^4 \delta^{(4)}(k_1+k_2-k) (2\pi) \delta((p_1-k_1)^2) (2\pi) \delta((p_2-k_2)^2).
\end{align}
We have used the spinor identities 
\begin{equation}
\bar{u}(p_1-k_1, s'_1)\slashed{k}_1 u(p_1, s_1) =0, \quad \bar{u}(p_2-k_2, s'_2) \slashed{k}_2 u(p_2, s_2)=0.
\end{equation}
Note that the potential contribution from the  $k^{\mu}k^{\mu'}/k^2$ term in the gluon propagator, although nonvanishing,  are sub-sub-eikonal. We therefore ignored the $k^{\mu}k^{\mu'}/k^2$ part in the gluon propagator.  The same argument also applies to the $g_{\nu\rho}(k_1-k_2)_{\mu'} $ term when considering the $g^{\mu\nu'}$ in the gluon propagator. 

When $\mu=i$ in eq.~\eqref{eq:Amu_quark_general}, one has the factor 
\begin{equation}
 \Big[ \bar{u}(p_1-k_1, s'_1)\gamma^{\nu} u(p_1, s_1) \Big]\Big[\bar{u}(p_2-k_2, s'_2) \gamma^{\rho} u(p_2, s_2)\Big]  \Big[ g_{i \nu} (-2k_{\rho})  + g_{\rho i}(2k_{\nu})\Big]
\end{equation}
whose helicity dependent contributions start from sub-sub-eikonal order and we discard these terms.

When $\mu=+$ in eq.~\eqref{eq:Amu_quark_general}, using eq.~\eqref{eq:ubar_i_u}, the field in momentum space becomes
\begin{equation}
\begin{split}
\mathcal{A}^{+, a}_{\mathrm{sub}}(k, k_1, k_2) 
=&-g^3 f^{abc} t^b_{i'_1 i_1} t^c_{i'_2 i_2}\, \delta_{s'_1 s_1} \delta_{s'_2s_2}\Big[2p_1^+ (2s_2) + 2p_2^+ (2s_1)\Big] 2\epsilon^{ij}\mathbf{k}_1^i \mathbf{k}_2^j\frac{1}{k_1^2+i\epsilon} \frac{1}{k_2^2+i\epsilon} \frac{1}{k^2+i\epsilon}\\
&\times (2\pi)^4 \delta^{(4)}(k_1+k_2-k) (2\pi) \delta((p_1-k_1)^2) (2\pi) \delta((p_2-k_2)^2).\\
\end{split}
\end{equation}
Making Fourier transformation to coordinate space, one gets 
\begin{equation}
\begin{split}
\mathcal{A}^{+, a}_{\mathrm{sub}}(x) = &\int \frac{d^4k }{(2\pi)^4} e^{-ik^+(x^--b^-) - ik^-(x^+-b^+) + i\mathbf{k}\cdot(\mathbf{x}-\mathbf{b})} \int \frac{d^4 k_1}{(2\pi)^4} e^{- ik_1^+(b^--b_1^-) -ik_1^-(b^+-b_1^+) + i\mathbf{k}_1\cdot(\mathbf{b}-\mathbf{b}_1)}\\
&\quad \times \int \frac{d^4 k_2}{(2\pi)^4} e^{-ik_2^+(b^--b_2^-) - ik_2^-(b^+-b_2^+) + i\mathbf{k}_2\cdot(\mathbf{b}-\mathbf{b}_2)}\mathcal{A}^{+,a}_{\mathrm{sub}}(k, k_1,k_2) \\
=&2gf^{abc} \int d^2\mathbf{b} \frac{1}{2\pi} \ln \frac{1}{|\mathbf{x}-\mathbf{b}|\Lambda}  \Bigg[ \delta(x^--b_1^-) gt^b_{i'_1i_1} \delta_{s'_1s_1}  \frac{-1}{2\pi} \frac{(\mathbf{b}-\mathbf{b}_1)^i}{|\mathbf{b}-\mathbf{b}_1|^2}\times \delta(x^--b_2^-) gt^c_{i'_2i_2} \delta_{s'_2 s_2} \frac{2s_2}{2p_2^+} \frac{1}{2\pi} \frac{\epsilon^{ij} (\mathbf{b}-\mathbf{b}_2)^j}{|\mathbf{b}-\mathbf{b}_2|^2} \\
&+ \delta(x^--b_1^-)  gt^b_{i'_1i_1} \delta_{s'_1s_1} \frac{2s_1}{2p_1^+} \frac{1}{2\pi} \frac{\epsilon^{ij} (\mathbf{b}-\mathbf{b}_1)^i}{|\mathbf{b}-\mathbf{b}_1|^2}\times  \delta(x^--b_2^-) gt^c_{i'_2i_2} \delta_{s'_2 s_2}\frac{-1}{2\pi} \frac{(\mathbf{b}-\mathbf{b}_2)^j}{|\mathbf{b}-\mathbf{b}_2|^2}\Bigg]\\
=&2gf^{abc} \int d^2\mathbf{b} \frac{1}{2\pi} \ln \frac{1}{|\mathbf{x}-\mathbf{b}|\Lambda} \Big[ \partial^i_{\mathbf{b}} \mathcal{ A}^{+, b}_{\mathrm{eik}}(x^-, \mathbf{b}; b_1) A^{i, c}_{\mathrm{sub}}(x^-, \mathbf{b}; b_2)  - \mathcal{A}^{j, b}_{\mathrm{sub}}(x^-, \mathbf{b}; b_1) \partial^j_{\mathbf{b}} \mathcal{A}^{+, c}_{\mathrm{eik}}(x^-, \mathbf{b}; b_2)\Big]\\
=&2gf^{abc} \int d^2\mathbf{b} \frac{1}{2\pi} \ln \frac{1}{|\mathbf{x}-\mathbf{b}|\Lambda} \partial^i_{\mathbf{b}} \Big[\mathcal{A}^{+,b}_{\mathrm{eik}}(x^-, \mathbf{b}; b_1) + \mathcal{A}^{+, b}_{\mathrm{eik}}(x^-, \mathbf{b}; b_2) \Big]\Big[\mathcal{A}^{i, c}_{\mathrm{sub}}(x^-, \mathbf{b}; b_2) + \mathcal{A}^{i, c}_{\mathrm{sub}}(x^-, \mathbf{b}; b_1)\Big]\\
=&2gf^{abc} \int d^2\mathbf{b}\,  \phi(\mathbf{x}-\mathbf{b}) \, \partial^i \mathcal{A}^{+, b}_{\mathrm{eik}}(x^-, \mathbf{b}) \, \mathcal{A}^{i, c}_{\mathrm{sub}}(x^-, \mathbf{b}).
\end{split}
\end{equation}
\end{widetext}
In arriving at the last equality, we combined the gluon fields generated by two quark color sources locating at positions $b_1, b_2$ into a single gluon field, suppressing explicit labels for the color sources. This result is exactly the same as the expression given in eq.~\eqref{eq:induced_eik+_subi} that was derived from solving classical Yang-Mills equations. 
Notably, for the fields $\mathcal{A}_{\mathrm{sub}}^i$ and $\mathcal{A}_{\mathrm{eik}}^+$, if  they originate from the same quark source, they do not contribute to gluon saturation induced helicity-dependent field. The two quark sources must have different transverse coordinates and carry different colors to produce the induced field. Additionally, it is noteworthy that the two quark sources and the induced gluon field share the same longitudinal coordinate.

\subsection{Gluon Source}
\begin{figure*}
    \includegraphics[width=0.75\textwidth]{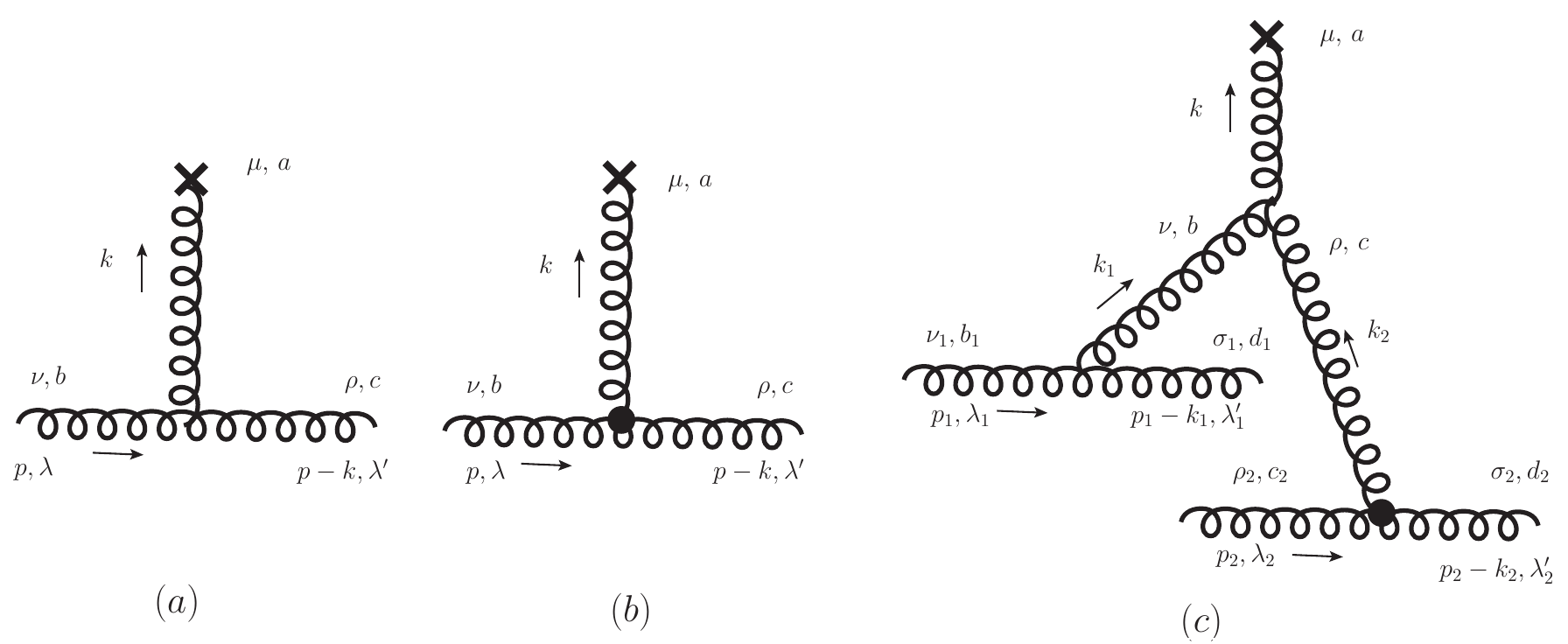}
    \caption{(a) Eikonal order gluon field $A^+$; (b) sub-eikonal order helicity-dependent gluon field $A^i$; (c) Gluon saturation induced helicity-dependent gluon field $A^+$ at sub-eikonal order.}
\label{fig:gluon_induced}		
\end{figure*}
We consider the case that the quasi-classical gluon fields are generated by gluonic sources. The incoming and outgoing gluons are treated as physical transversely polarized gluons. According to BRST symmetry \cite{Peskin:1995ev}, asymptotic incoming gluon state must be transversely polarized. Both the incoming and the outgoing gluons are required to be on-shell.
 
 Diagram $(a)$ and diagram $(b)$ in Fig.~\ref{fig:gluon_induced} can be analyzed collectively, the amplitude is computed by 
\begin{align}\label{eq:single_gluon_sourced_A(k)}
&\mathcal{A}^{\mu}_a(k; p) \notag\\
 = &-gf^{abc}\Big[g_{\mu'\nu}(-k-p)_{\rho} + g_{\nu\rho}(2p-k)_{\mu'} + g_{\rho\mu'}(2k-p)_{\nu}\Big] \notag\\
 &\times \frac{-i}{k^2+i\epsilon}\Big[ g^{\mu\mu'} - \frac{k^{\mu} k^{\mu'}}{k^2}\Big]  \varepsilon^{\nu} (p, \lambda) \varepsilon^{\rho \ast}(p-k, \lambda') \notag\\
 &\times (2\pi) \delta((p-k)^2) \notag \\
=&-gf^{abc}\Big[g^{\mu\nu}(-2k)^{\rho} + g^{\nu\rho}(2p-k)^{\mu} + g^{\rho\mu}(2k)^{\nu}\Big] \notag\\
&\times \frac{-i}{k^2+i\epsilon}\varepsilon_{\nu} (p, \lambda) \varepsilon^{\ast}_{\rho}(p-k, \lambda') (2\pi) \delta((p-k)^2)
\end{align}
We used the identities
\begin{equation}
p^{\nu} \varepsilon_{\nu}(p, \lambda) =0, \quad (p-k)^{\rho} \varepsilon_{\rho}(p-k, \lambda') =0. 
\end{equation}
The term in the gluon propagator involving $k_{\mu}k_{\mu'}/k^2$ vanishes.  
For further analysis, one needs an explicit expression for the gluon polarization vector that is eigenstate of the helicity operator \cite{Auvil:1966eao, Dreiner:2008tw,Leader:2011vwq}.  
\begin{equation}
\begin{split}
&\varepsilon^{\mu}(p, \lambda)\\
 = &\Big(\varepsilon^{t}(p, \lambda), \varepsilon^x(p,\lambda), \varepsilon^y(p, \lambda), \varepsilon^z(p, \lambda)\Big)\\
=&\frac{1}{\sqrt{2}} \Big(\, 0,\,  -\lambda \cos\theta + i\sin \phi  e^{i\lambda\phi} (1-\cos\theta), \\
&\, -i\cos\theta - i\cos\phi e^{i\lambda\phi}(1-\cos\theta), \, \lambda \sin \theta e^{i\lambda\phi}\, \Big).\\
\end{split}
\end{equation}
Here 
\begin{equation}
p^{\mu} = (p^t, p^x, p^y, p^z) = (p, p\sin\theta\cos\phi, p \sin\theta\sin\phi, p\cos\theta).
\end{equation}
In the situation that $p^z \gg p^x \sim p^y$, one can approximate
\begin{equation}
\varepsilon^{\mu}(p, \lambda)  = \frac{1}{\sqrt{2}}\Big(0, -\lambda, -i,  \frac{\lambda p^x + ip^y}{p^z}\Big) + \mathcal{O}(1/(p^z)^2)
\end{equation} 
The correction starts at  sub-eikonal order. In eq.~\eqref{eq:single_gluon_sourced_A(k)}, without loss of generality, we assume that the incoming gluon travels along $+z$ direction, $p^{\mu} = (p^+, 0^-, \mathbf{0})$ and it follows that $p-k = (p^+-k^+, -k^-, -\mathbf{k})$. We are only interested in quasi-classical gluon field at sub-eikonal order. Therefore, one can approximate the gluon polarization vector as
\begin{equation}
\varepsilon_{\nu} (p, \lambda) =(0^+, 0^-, \vec{\epsilon}_{\lambda}), \qquad \epsilon_{\lambda} = -\frac{1}{\sqrt{2}}(\lambda, i). 
\end{equation}
In eq.~\eqref{eq:single_gluon_sourced_A(k)}, when $\mu = +$, one gets the eikonal order gluon field in momentum space
\begin{equation}
\mathcal{A}^{+,a}_{\mathrm{eik}}(k; p) =  -g T^a_{cb} \delta_{\lambda\lambda'} \frac{1}{-\mathbf{k}^2 + i\epsilon} \delta(k^-)
\end{equation}
The corresponding expression in coordinate space is
\begin{equation}
\mathcal{A}^{+,a}_{\mathrm{eik}}(x) 
=g T^a_{cb} \delta_{\lambda\lambda'} \delta(x^--b^-) \frac{1}{2\pi} \ln \frac{1}{|\mathbf{x}-\mathbf{b}|\Lambda}.
\end{equation}

\begin{figure}
    \includegraphics[width=0.45\textwidth]{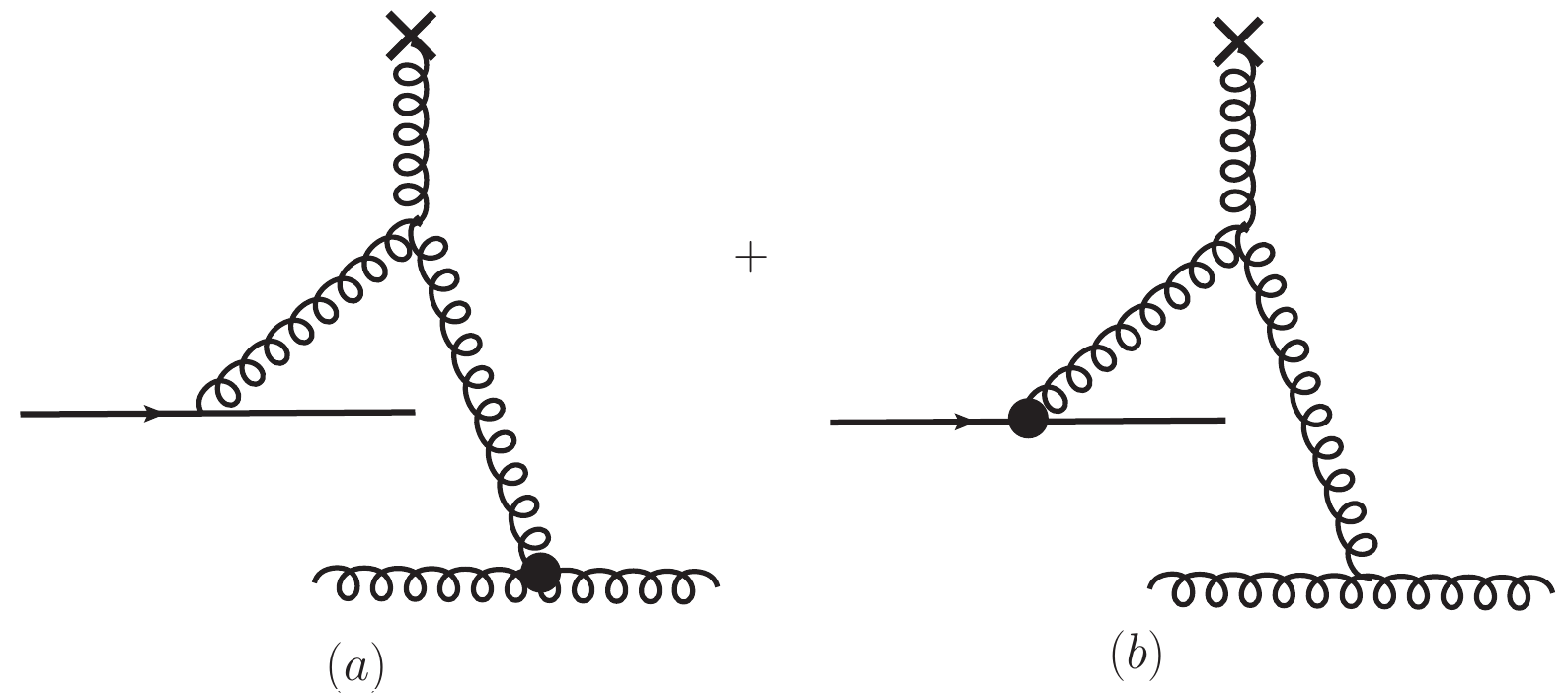}
    \caption{The two color sources can come from quarks and gluons separately.}
\label{fig:mixed_induced}		
\end{figure}

In eq.~\eqref{eq:single_gluon_sourced_A(k)}, when $\mu = i$, we only focus on sub-eikonal order fields that are dependent on helicity,
\begin{equation}\label{eq:gluon_induced_Ai(k)}
\mathcal{A}^{i,a}_{\mathrm{sub}}(k; p) 
=-gT^a_{cb} \lambda \delta_{\lambda\lambda'} \frac{i\epsilon^{ij}\mathbf{k}^j}{-\mathbf{k}^2+i\epsilon} \frac{1}{p^+} (2\pi) \delta(k^-).
\end{equation}
We  used the identity
$ \epsilon^i_{\lambda}\epsilon^{j\ast}_{\lambda'} - \epsilon^{j}_{\lambda} \epsilon^{i\ast}_{\lambda'} = -i\lambda \epsilon^{ij}\delta_{\lambda\lambda'}$. We have discarded terms that are of sub-eikonal order but insensitive to helicity. 
Making Fourier transformation, one gets the transverse gluon field  in coordinate space
\begin{equation}
\begin{split}
\mathcal{A}^{i,a}_{\mathrm{sub}}(x)
=&-\lambda \delta_{\lambda\lambda'} \, gT^a_{cb} \,  \delta(x^--b^-) \frac{1}{p^+} \frac{1}{2\pi} \frac{\epsilon^{ij} (\mathbf{x}-\mathbf{b})^j}{|\mathbf{x}-\mathbf{b}|^2}.\\
\end{split} 
\end{equation}
The diagram $(c)$ in Fig.~\ref{fig:gluon_induced} can be computed by 
\begin{widetext}
\begin{equation}\label{eq:gluon_diag_c_exp}
\begin{split}
\mathcal{A}^{\mu}_a(k, k_1, k_2) 
= &-gf^{abc}\Big[ g_{\mu'\nu} (-k-k_1)_{\rho} + g_{\nu\rho}(k_1-k_2)_{\mu'} + g_{\rho\mu'}(k_2+k)_{\nu}\Big] \frac{-i}{k^2+i\epsilon}\Big[ g^{\mu\mu'} - \frac{k^{\mu}k^{\mu'}}{k^2}\Big]\\
&\quad\times  \mathcal{A}^{\nu}_b(k_1; p_1)\,  \mathcal{A}^{\rho}_c(k_2; p_2) (2\pi)^4 \delta^{(4)}(k_1+k_2-k)\\
\simeq & igf^{abc}\Big[ g^{\mu\nu} (-k-k_1)^{\rho} + g^{\nu\rho}(k_1-k_2)^{\mu} + g^{\rho\mu}(k_2+k)^{\nu}\Big] \frac{1}{k^2+i\epsilon}\\
&\qquad \times \mathcal{A}_{\nu, b}(k_1; p_1)\,  \mathcal{A}_{\rho, c}(k_2; p_2) (2\pi)^4 \delta^{(4)}(k_1+k_2-k).\\
\end{split}
\end{equation}
For the term involving $k^{\mu}k^{\mu'}/k^2$ in the gluon propagator,  its contribution turns out to be starting from sub-sub-eikonal order and has been discarded. 

In eq.~\eqref{eq:gluon_diag_c_exp}, when $\mu=+$, one gets (only focusing on helicity-dependent terms at sub-eikonal order)
\begin{equation}
\begin{split}
\mathcal{A}^{+, a}_{\mathrm{sub}}(k, k_1, ,_2) =&- 2gf^{abc} \Big[ (-i\mathbf{k}_1^i) \mathcal{A}^{+, b}_{\mathrm{eik}}(k_1; p_1) \mathcal{A}^{i, c}_{\mathrm{sub}}(k_2; p_2) - (-i\mathbf{k}_2^i) \mathcal{A}^{+, c}_{\mathrm{eik}}(k_2; p_2) \mathcal{A}^{i, b}_{\mathrm{sub}}(k_1; p_1)\Big] \frac{1}{k^2+i\epsilon} (2\pi)^4 \delta^{(4)}(k_1+k_2-k).\\
\end{split}
\end{equation}
We have imposed the transverse momentum conservation $\mathbf{k}=\mathbf{k}_1+\mathbf{k}_2$.
Making Fourier transformations to coordinate space, one obtains
\begin{equation}
\begin{split}
\mathcal{A}^{+, a}_{\mathrm{sub}}(x) = &2gf^{ade} \int d^2\mathbf{b}\,  \phi(\mathbf{x}-\mathbf{b})  \Big[ \partial^i_{\mathbf{b}} \mathcal{A}^{+, d}_{\mathrm{eik}}(x^-, \mathbf{b}; b_1) \mathcal{A}^{i,e}_{\mathrm{sub}}(x^-, \mathbf{b}; b_2) - \partial^i_{\mathbf{b}} \mathcal{A}^{+, e}_{\mathrm{eik}}(x^-, \mathbf{b}; b_2) \mathcal{A}^{i, d}_{\mathrm{sub}}(x^-, \mathbf{b}; b_1)\Big]\\
=&2gf^{ade} \int d^2\mathbf{b}\, \phi(\mathbf{x}-\mathbf{b}) \partial^i_{\mathbf{b}}\Big[ \mathcal{A}^{+, d}_{\mathrm{eik}}(x^-, \mathbf{b}; b_1)+  \mathcal{A}^{+, d}_{\mathrm{eik}}(x^-, \mathbf{b}; b_2)\Big] \Big[ \mathcal{A}^{i,e}_{\mathrm{sub}}(x^-, \mathbf{b}; b_1) +  \mathcal{A}^{i,e}_{\mathrm{sub}}(x^-, \mathbf{b}; b_2)  \Big]\\
=&2gf^{ade} \int d^2\mathbf{b} \, \phi(\mathbf{x}-\mathbf{b}) \, \partial^i_{\mathbf{b}} \mathcal{A}^{+, d}_{\mathrm{eik}}(x^-, \mathbf{b})\, \mathcal{A}^{i,e}_{\mathrm{sub}}(x^-, \mathbf{b}).
\end{split}
\end{equation}
\end{widetext}
The result is identical to the expression in eq.~\eqref{eq:induced_eik+_subi} obtained from solving classical Yang-Mills equations.
Naturally, the gluon saturation induced helicity-dependent field can also be generated from two color sources: one originating from a quark and the other from a gluon, as depicted in Fig.~\ref{fig:mixed_induced}. Since the analysis follows a similar process, we will not repeat the calculations here.

\section{Quasi-Classical Approximation}\label{sec:quasi_classical_app}
\begin{figure*}
    \includegraphics[width=0.75\textwidth]{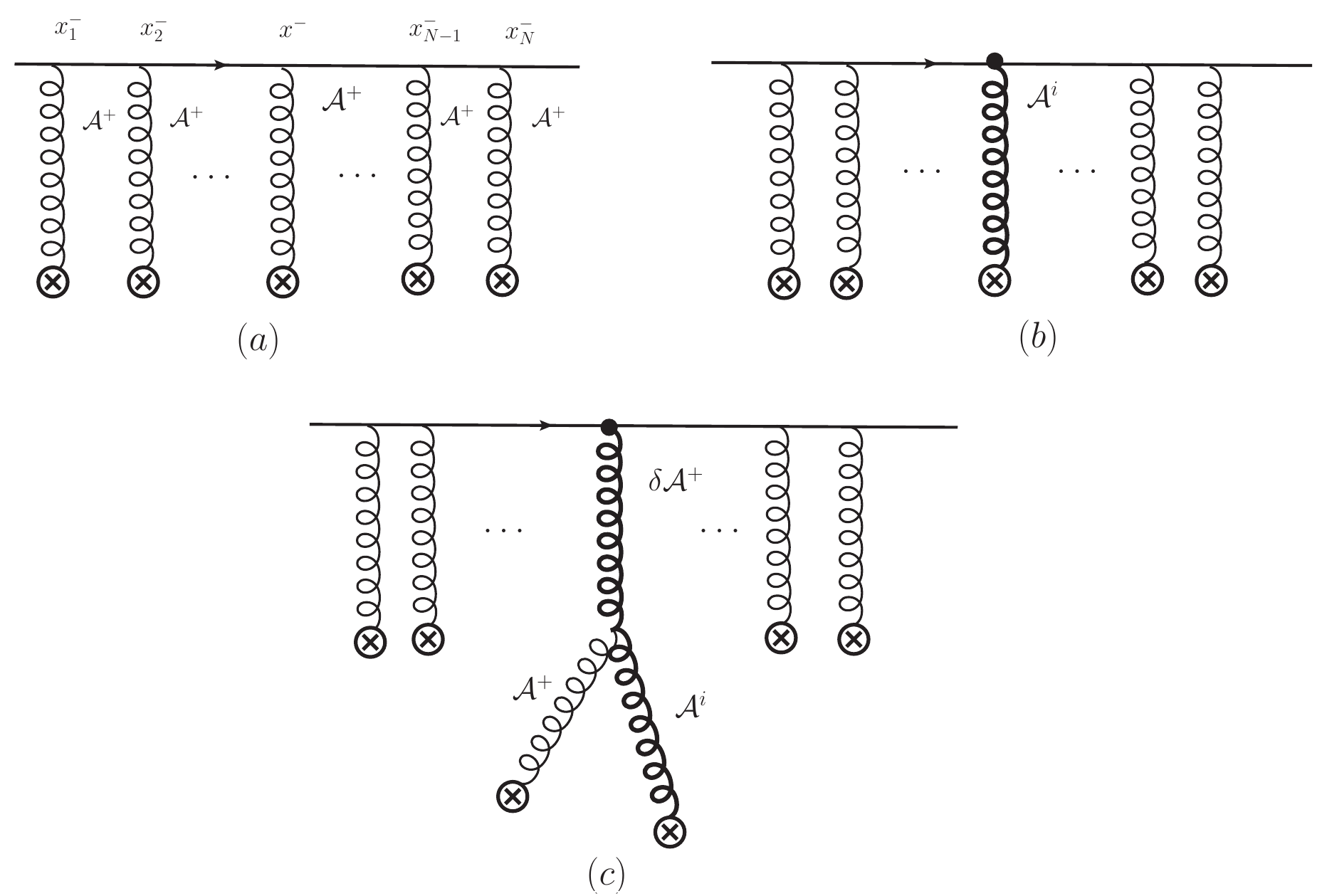}
    \caption{(a) Multiple eikonal gluon exchanges can be resumed into the eikonal Wilson line. (b) sub-eikonal order polarized Wilson line with one insertion of sub-eikonal order helicity-dependent gluon field $\mathcal{A}^i$.  (c) sub-eikonal order polarized Wilson line with one insertion of the sub-eikonal order gluon saturation induced helicity-dependent gluon field $\delta \mathcal{A}^+$. }
\label{fig:multigluon_exchanges}		
\end{figure*}
The gluon saturation induced helicity-dependent field contributes to spin related observables through the transverse chromo-electrically polarized Wilson line \cite{Chirilli:2018kkw, Altinoluk:2021lvu, Cougoulic:2022gbk, Li:2023tlw}
\begin{align}\label{eq:VjG[2]_def}
V^{j, G[2]}_{\mathbf{x}} 
= &\frac{igP^+}{s} \int_{-\infty}^{+\infty} dx^- V_{\mathbf{x}}[+\infty, x^-] \Big(x^- \partial^j A^+(x^-, \mathbf{x})\notag \\
& + A^j(x^-, \mathbf{x})\Big) V_{\mathbf{x}}[x^-, -\infty].
\end{align}
Here $P^+$ is the longitudinal momentum of the nuclear target and $s$ the center-of-mass collision energy. 

As schematically shown in diagram $(a)$ in Fig.~\ref{fig:multigluon_exchanges}, multiple $t$-channel gluon exchanges at the eikonal order can be resumed into the eikonal Wilson line 
\begin{equation}
V_{\mathbf{x}}[+\infty, -\infty] = \mathcal{P}\mathrm{exp}\left\{-ig\int_{-\infty}^{+\infty} dx^- \mathcal{A}^+(x^-, \mathbf{x})\right\},
\end{equation}
In the context of multiple gluon exchanges at sub-eikonal order, helicity dependent interaction is added by one insertion of the sub-eikonal transverse gluon field $\mathcal{A}^i$ as shown in diagram $(b)$ in Fig.~\ref{fig:multigluon_exchanges}. Helicity dependence can also arise from inserting the gluon saturation induced helicity-dependent field $\delta \mathcal{A}^+$ as shown in diagram $(c)$. 

The transverse chromo-electrically polarized Wilson line in eq.~\eqref{eq:VjG[2]_def}
 not only contributes to polarized Wilson line correlators related to single gluon helicity distribution but also contributes to two-particle (or multi-particle) correlated helicity distribution.   
In the following, we will evaluate the gluon saturation induced helicity effects in several physical quantities involving the chromo-electrically polarized Wilson line in the quasi-classical approximation. 
We perform quasi-classical averaging using Gaussian-like models. 
For the eikonal order gluon field, the MV model \cite{McLerran:1993ni,McLerran:1993ka} says
\begin{equation} \label{eq:MV}
\Big\langle \alpha_a(x^-, \mathbf{x}) \alpha_b(y^-, \mathbf{y}) \Big\rangle = \delta^{ab} \delta(x^--y^-) L(\mathbf{x}-\mathbf{y}) \mu_0^2,
\end{equation}
Here $\mu_0^2$ is related to the gluon saturation scale $Q_s^2$. We also use the following simplified ansatz for averaging over longitudinally polarized proton/nucleus, inspired by the helicity-extended MV model \cite{Cougoulic:2020tbc}, 
\begin{equation}\label{eq:hMV}
\llangle \beta_a(x^-, \mathbf{x}) \alpha_b(y^-, \mathbf{y}) \rrangle = \delta^{ab} \delta(x^--y^-) L(\mathbf{x}-\mathbf{y}) \frac{\mu_0^2}{P^+}.
\end{equation}
In both expressions, one has the function
\begin{equation}\label{eq:L(x)_exp}
L(\mathbf{x}-\mathbf{y}) =  \int \frac{d^2\mathbf{q}}{(2\pi)^2} e^{i\mathbf{p}\cdot(\mathbf{x}-\mathbf{y})} \frac{1}{\mathbf{q}^4}.
\end{equation}
We use single bracket $\langle \ldots \rangle$ to indicate quasi-classical ensemble averaging over unpolarized states and double bracket $\langle \langle \ldots \rangle\rangle$ for averaging over longitudinally polarized states.

In evaluating the gluon saturation induced helicity effects within the quasi-classical approximation, one encounters ensemble averaging of three $A^+$ fields over longitudinally polarized proton/nucleus states. 
\begin{equation}\label{eq:three_field_ave}
\begin{split}
&\llangle A^+_c(x^-, \mathbf{x}) A^+_b(y^-, \mathbf{y}) A^+_a(u^-, \mathbf{u}) \rrangle \\
=&\llangle \delta\mathcal{A}^+_c(x^-, \mathbf{x}) \mathcal{A}^+_b(y^-, \mathbf{y}) \mathcal{A}^+_a(u^-, \mathbf{u}) \rrangle \\
&+ \llangle \delta\mathcal{A}^+_b(y^-, \mathbf{y}) \mathcal{A}^+_c(x^-, \mathbf{x}) \mathcal{A}^+_a(u^-, \mathbf{u}) \rrangle  \\
&+ \llangle \delta \mathcal{A}^+_a(u^-, \mathbf{u})\mathcal{A}^+_c(x^-, \mathbf{x}) \mathcal{A}^+_b(y^-, \mathbf{y})  \rrangle. \\
\end{split}
\end{equation}
Any one of the three fields can be the sub-eikonal order gluon saturation induced field $\delta \mathcal{A}^+$ while the other two are eikonal order quasi-classical gluon fields $\mathcal{A}^+$.  There are three possibilities. Eq.~\eqref{eq:final_induced_field_exp} and eq.~\eqref{eq:A^+_in_alpha} are then plugged into eq.~\eqref{eq:three_field_ave}. Taking the first term in eq.~\eqref{eq:three_field_ave} as an example, the elementary three-field averaging in the quasi-classical approximation can be computed by
 \begin{widetext}
\begin{align}\label{eq:3A+_averaging}
&\llangle \delta \mathcal{A}^+_c(x^-, \mathbf{x}) \mathcal{A}^+_b(y^-, \mathbf{y}) \mathcal{A}^+_a(u^-, \mathbf{u}) \rrangle \notag\\
=&-2gf^{cde} \epsilon^{ml} \int_{\mathbf{z}} \phi(\mathbf{x}-\mathbf{z}) \llangle \partial^m_{\mathbf{z}}\alpha^d(x^-, \mathbf{z}) \partial^l_{\mathbf{z}}\beta^e(x^-, \mathbf{z}) \alpha^b(y^-, \mathbf{y}) \alpha^a(u^-, \mathbf{u})\rrangle  \notag\\
=&-2gf^{cde} \epsilon^{ml} \int_{\mathbf{z}} \phi(\mathbf{x}-\mathbf{z})\Big[ \Big\langle \partial^m_{\mathbf{z}}\alpha^d(x^-, \mathbf{z}) \alpha^b(y^-, \mathbf{y})\Big\rangle \llangle  \partial^l_{\mathbf{z}}\beta^e(x^-, \mathbf{z})  \alpha^a(u^-, \mathbf{u}) \rrangle \notag \\
&\qquad + \Big\langle \partial^m_{\mathbf{z}}\alpha^d(x^-, \mathbf{z})\alpha^a(u^-, \mathbf{u}) \Big\rangle \llangle  \partial^l_{\mathbf{z}}\beta^e(x^-, \mathbf{z})  \alpha^b(y^-, \mathbf{y}) \rrangle \Big] \notag\\
=& -2gf^{cba}\frac{2\mu_0^4}{P^+} \left[ \epsilon^{ml} \int_{\mathbf{z}}\phi(\mathbf{x}-\mathbf{z}) \partial^m_{\mathbf{z}}L(\mathbf{z}-\mathbf{y}) \partial^l_{\mathbf{z}}L(\mathbf{z}-\mathbf{u})\right]\delta(x^--y^-) \delta(x^--u^-). 
\end{align}
\end{widetext}
From this expression, it is easy to see that when the two eikonal order fields have the same transverse coordinates $\mathbf{u}=\mathbf{y}$, the averaging vanishes. 
\begin{equation}\label{eq:3A^+_equalx}
\llangle \delta \mathcal{A}^+_c(x^-, \mathbf{x}) \mathcal{A}^+_b(y^-, \mathbf{y}) \mathcal{A}^+_a(u^-, \mathbf{u}) \rrangle = 0.
\end{equation}
From the expression in momentum space 
\begin{equation}
\begin{split}
& \epsilon^{ml} \int_{\mathbf{z}}\phi(\mathbf{x}-\mathbf{z}) \partial^m_{\mathbf{z}}L(\mathbf{z}-\mathbf{y}) \partial^l_{\mathbf{z}}L(\mathbf{z}-\mathbf{u})\\
=&\int \frac{d^2\mathbf{p}}{(2\pi)^2} \frac{d^2\mathbf{q}}{(2\pi)^2} e^{i\mathbf{p}\cdot(\mathbf{x}-\mathbf{y})} e^{i\mathbf{q}\cdot(\mathbf{x}-\mathbf{u})} \frac{1}{(\mathbf{p}+\mathbf{q})^2} \frac{-\mathbf{p} \times \mathbf{q}}{\mathbf{p}^4\mathbf{q}^4},
\end{split}
 \end{equation}
it is also apparent that if $\mathbf{x}=\mathbf{y}$ or $\mathbf{x}=\mathbf{u}$ the above expression vanishes due to the angular integrations. One therefore has
\begin{equation}
\begin{split}
&\llangle \delta \mathcal{A}^+_c(x^-, \mathbf{x}) \mathcal{A}^+_b(y^-, \mathbf{x}) \mathcal{A}^+_a(u^-, \mathbf{u}) \rrangle \\
= &\llangle \delta \mathcal{A}^+_c(x^-, \mathbf{x}) \mathcal{A}^+_b(y^-, \mathbf{y}) \mathcal{A}^+_a(u^-, \mathbf{x})  \rrangle =0.\\
\end{split}
\end{equation}
We can conclude that eq.~\eqref{eq:3A+_averaging} vanishes when any two of the three transverse coordinates are the same.  It is only non-zero when all three transverse coordinates are different. 

Using eq.~\eqref{eq:3A+_averaging}, one obtains the expression for eq.~\eqref{eq:three_field_ave} in the quasi-classical approximation 
\begin{equation}\label{eq:three_field_ave_res}
\begin{split}
&\llangle A^+_c(x^-, \mathbf{x}) A^+_b(y^-, \mathbf{y}) A^+_a(u^-, \mathbf{u}) \rrangle\\
 =&-\frac{4g\mu_0^4}{P^+}f^{cba}\, \Gamma(\mathbf{x}, \mathbf{y}, \mathbf{u})\, \delta(x^--y^-) \delta(x^--u^-).\\
\end{split}
\end{equation}
We have introduced the auxiliary function of three transverse coordinates
\begin{widetext}
\begin{align}
\Gamma(\mathbf{x}, \mathbf{y}, \mathbf{u})
 = &\epsilon^{ml}\Bigg[\int_{\mathbf{z}}\phi(\mathbf{x}-\mathbf{z})\partial_{\mathbf{z}}^mL(\mathbf{z}-\mathbf{y}) \partial^l_{\mathbf{z}}L(\mathbf{z}-\mathbf{u}) + \int_{\mathbf{z}}\phi(\mathbf{y}-\mathbf{z}) \partial^m_{\mathbf{z}}L(\mathbf{z}-\mathbf{u}) \partial^l_{\mathbf{z}}L(\mathbf{z}-\mathbf{x})  \notag \\
 &\qquad +\int_{\mathbf{z}}\phi(\mathbf{u}-\mathbf{z})\partial^m_{\mathbf{z}}L(\mathbf{z}-\mathbf{x}) \partial^l_{\mathbf{z}}L(\mathbf{z}-\mathbf{y})\Bigg] \notag\\
=&\int_{\mathbf{p}, \mathbf{q}} \frac{-(\mathbf{p}\times \mathbf{q})}{(\mathbf{p}+\mathbf{q})^2 \mathbf{p}^4 \mathbf{q}^4} \Big[e^{i\mathbf{p}\cdot(\mathbf{x}-\mathbf{y})} e^{i\mathbf{q}\cdot(\mathbf{x}-\mathbf{u})}+e^{i\mathbf{p}\cdot(\mathbf{y}-\mathbf{u})} e^{i\mathbf{q}\cdot(\mathbf{y}-\mathbf{x})}+ e^{i\mathbf{p}\cdot(\mathbf{u}-\mathbf{x})} e^{i\mathbf{q}\cdot(\mathbf{u}-\mathbf{y})}\Big].
\end{align}
\end{widetext}
Note that the ordering of the arguments in $\Gamma(\mathbf{x}, \mathbf{y}, \mathbf{u})$ matters. One has the property
\begin{equation}
\Gamma(\mathbf{x}, \mathbf{y}, \mathbf{u}) = - \Gamma(\mathbf{x}, \mathbf{u}, \mathbf{y}) = -\Gamma(\mathbf{y}, \mathbf{x}, \mathbf{u}) = -\Gamma(\mathbf{u}, \mathbf{y}, \mathbf{x}),
\end{equation}
that is, exchanging the ordering of any two coordinates generate an additional minus sign.  As a result, the function vanishes if any two of the three coordinates are the same. 
\begin{equation}\label{eq:Gamma_func_vanish}
\Gamma(\mathbf{x}, \mathbf{u}, \mathbf{u}) = \Gamma(\mathbf{x}, \mathbf{x}, \mathbf{u})= \Gamma(\mathbf{x}, \mathbf{y}, \mathbf{x})=0
\end{equation}
Eq.~\eqref{eq:three_field_ave_res} serves as a basic building block for evaluating the gluon saturation induced helicity effects within the quasi-classical approximation. Additionally, the property expressed in eq.~\eqref{eq:Gamma_func_vanish} that the three-field averaging vanishes when any two coordinates coincide will also be very useful in simplifying the calculations.

\section{Single-Particle Helicity Distributions}\label{sec:single-particle_dis}
In this section, we evaluate the gluon saturation induced helicity effects in two types of single-gluon helicity distributions: dipole gluon helicity TMD and WW gluon helicity TMD. We demonstrate that gluon saturation induced helicity effects do not contribute to either of these distributions.

\subsection{Dipole Gluon Helicity TMD}\label{sec:dipole_TMD}
To see whether gluon saturation induced helicity effect contributes to the dipole gluon helicity TMD or not, one starts from its definition \cite{Cougoulic:2022gbk}
\begin{widetext}
\begin{equation}\label{eq:def_dipole_TMD}
\begin{split}
&\Delta G_{\mathrm{dip}}(x, \mathbf{k}^2) = \frac{1}{2g^2\pi^3} i\epsilon^{ij} \mathbf{k}^j \int d^2\mathbf{x} d^2\mathbf{y} e^{-i\mathbf{k}\cdot(\mathbf{x}-\mathbf{y})} \llangle \mathrm{tr}\left[V_{\mathbf{x}}^{i, G[2]} V_{\mathbf{y}}^{\dagger}\right]\rrangle + c.c.
\end{split}
\end{equation}
Using the expression in eq.~\eqref{eq:VjG[2]_def}, there are two different contributions to the chromo-electrically polarized Wilson line correlator
\begin{subequations}\label{eq:ViG[2]xy_full}
\begin{align}
 \llangle \mathrm{tr}\left[V_{\mathbf{x}}^{i, G[2]} V_{\mathbf{y}}^{\dagger}\right]\rrangle
 =&\frac{igP^+}{s} \int_{-\infty}^{+\infty} dx^- \llangle \mathrm{tr}\Big[V_{\mathbf{x}}[+\infty, x^-]  A^i(x^-, \mathbf{x}) V_{\mathbf{x}}[x^-, -\infty]V^{\dagger}_{\mathbf{y}}\Big]\rrangle  \label{eq:ViG[2]xy_Ai}\\
 &+ \frac{igP^+}{s} \int_{-\infty}^{+\infty} x^-dx^- \llangle \mathrm{tr}\Big[V_{\mathbf{x}}[+\infty, x^-]  \partial^i A^+(x^-,\mathbf{x})  V_{\mathbf{x}}[x^-, -\infty]V_{\mathbf{y}}^{\dagger}\rrangle. \label{eq:ViG[2]xy_diA+}
\end{align}
 \end{subequations}
 \end{widetext}
The term in eq.~\eqref{eq:ViG[2]xy_Ai} contains direct helicity effect from the transverse field $A^i(x^-, \mathbf{x})$.  For the term in eq.~\eqref{eq:ViG[2]xy_diA+}, if all the $A^+(x^-,\mathbf{x})$ fields involved were eikonal-order gluon fields, this term would have vanished as it is independent of helicity. However, if one of the $A^+(x^-, \mathbf{x})$ fields is considered as the gluon saturation induced helicity-dependent field,  eq.~\eqref{eq:ViG[2]xy_diA+} could have non-vanishing contribution.
In general, both terms in eq.~\eqref{eq:ViG[2]xy_Ai} and eq.~\eqref{eq:ViG[2]xy_diA+} contribute to the dipole gluon helicity TMD. The small-$x$ helicity evolution equation derived in \cite{Cougoulic:2022gbk} takes into account contributions from both terms.

We aim to evaluate the contributions from the term in eq.~\eqref{eq:ViG[2]xy_diA+}. Performing an analytic calculation that resums all orders in multiple gluon exchanges is challenging. Instead, we will compute the leading-order contribution involving three-gluon field exchange and the next-to-leading order involving five-gluon field exchange.
To achieve this, 
it is necessary to expand both $V^{i,G[2]}_{\mathbf{x}}$  and $V^{\dagger}_{\mathbf{y}}$ up to fourth order in $A^+$ from the (partial) Wilson lines. 
The Gaussian averaging within the MV model guarantees that, for unpolarized Wilson lines,  only terms containing even powers of the eikonal order gluon field $A^+$ contribute to the averaging. On the other hand, when analyzing the gluon saturation induced helicity effects using the helicity-extended MV model,  one should focus on terms containing odd powers of $A^+$. 

Note that the linear order term from expanding $V^{i,G[2]}_{\mathbf{x}}$ does not contribute because one can always integrate by parts regarding the derivative in $\partial^i_{\mathbf{x}}A^+(x^-, \mathbf{x})$ so that $\epsilon^{ij}\mathbf{k}^j \mathbf{k}^i =0$ in eq.~\eqref{eq:def_dipole_TMD}. 

To obtain terms that contain $(A^+)^3$ from $\mathrm{tr}[V^{i,G[2]}_{\mathbf{x}}V^{\dagger}_{\mathbf{y}}]$, there are two possible combinations: $
\{1, 2\}$, $\{2, 1\}$. We use a pair of numbers to indicate the number of $A^+$ fields from $V_{\mathbf{x}}^{i,G[2]}$ and $V^{\dagger}_{\mathbf{y}}$ respectively.  The combination $\{1, 2\}$ vanishes because of eq.~\eqref{eq:3A^+_equalx}. For the combination $\{2, 1\}$, all the terms containing three $A^+$ fields vanish after averaging because the integral over longitudinal coordinates gives zero.
\begin{align}\label{eq:x^-_integral_vanish}
& \int_{-\infty}^{+\infty} dy^- \int_{-\infty}^{+\infty} x^- dx^- \int_{x^-}^{+\infty} dx_1^- \delta(x_1^--x^-) \delta(x^--y^-) \notag\\
=&\frac{1}{2} \int_{-\infty}^{+\infty} x^- dx^-=0.
\end{align}
As a result, three-gluon field exchanges do not contribute to the dipole gluon helicity TMD. 
\begin{figure*}
    \includegraphics[width=0.8\textwidth]{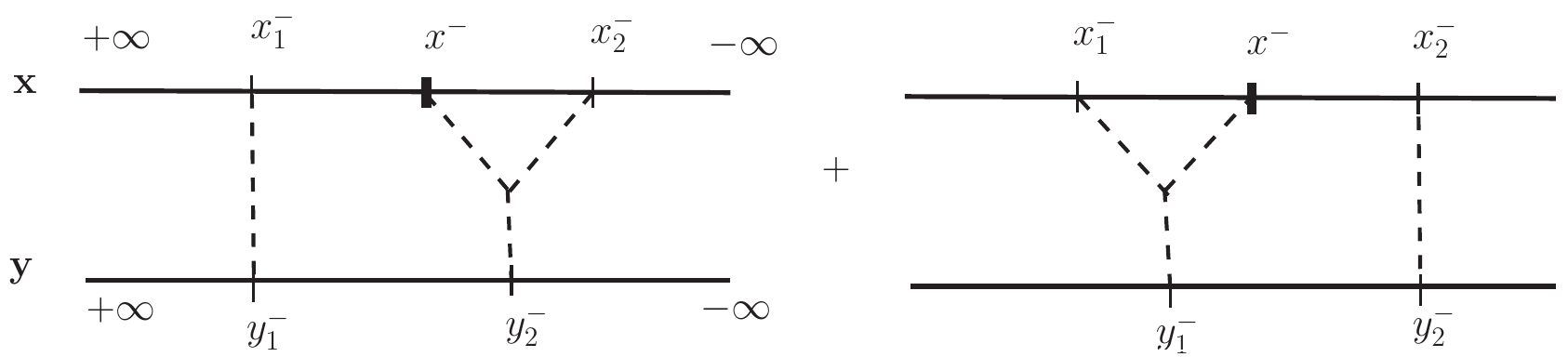}
    \caption{ Contraction diagrams characterized by longitudinal coordinates corresponding to terms in eq.~\eqref{eq:32_decomposition}. }
\label{fig:contraction_32_one}		
\end{figure*}

\begin{figure*}
\centering
\begin{subfigure}{0.49\textwidth}
  \centering
  \includegraphics[width=0.9\textwidth]{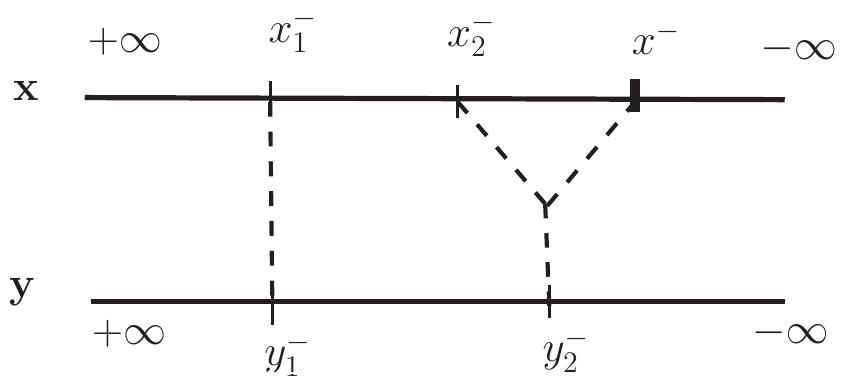}
  \caption{}
  \label{fig:contraction_32_two}
\end{subfigure}
\begin{subfigure}{0.49\textwidth}
  \centering
  \includegraphics[width=0.9\textwidth]{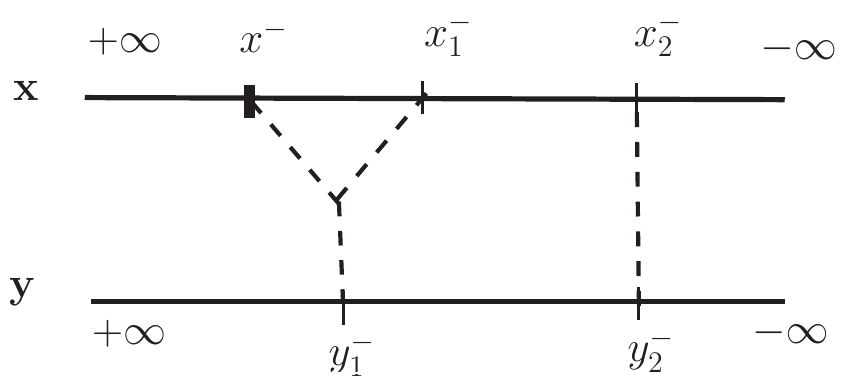}
  \caption{}
  \label{fig:contraction_32_three}
\end{subfigure}
\caption{Contraction diagram characterized by longitudinal coordinates with (a) corresponding to eq.~\eqref{eq:exp_diagram6} and (b) corresponding to eq.~\eqref{eq:exp_diagram7} .}
\label{fig:contraction_32_twothree}
\end{figure*}

For terms containing five $A^+$ fields, the possible combinations are $\{1, 4\}$, $\{2, 3\}$, $\{3, 2\}$ and $\{4, 1\}$. The combinations $\{1, 4\}$, $\{2, 3\}$ yield vanishing contributions. For the combination $\{3, 2\}$, there are three terms. None of the three terms vanishes but they combine to cancel each other. 
\begin{widetext}
\begin{align}
\{3, 2\}=& \frac{ig^5P^+}{s} \int_{-\infty}^{+\infty} x^- dx^- \int_{x^-}^{+\infty} dx_1^- \int_{-\infty}^{x^-} dx_2^- \int_{-\infty}^{+\infty} dy_1^- \int_{-\infty}^{y_1^-}dy_2^- \, \mathrm{tr}\Big[ t^{c_1} t^c t^{c_2} t^{b_2}t^{b_1}\Big]\notag\\
&\qquad \times \llangle A^+_{c_1}(x_1^-, \mathbf{x}) \partial^i_{\mathbf{x}} A^+_c(x^-, \mathbf{x}) A^+_{c_2}(x_2^-, \mathbf{x}) A^+_{b_2}(y_2^-, \mathbf{y}) A^+_{b_1}(y_1^-, \mathbf{y}) \rrangle \label{eq:32_first} \\
&+ \frac{ig^5P^+}{s} \int_{-\infty}^{+\infty} x^- dx^- \int_{x^-}^{+\infty} dx_1^- \int^{x_1^-}_{x^-} dx_2^- \int_{-\infty}^{+\infty} dy_1^- \int_{-\infty}^{y_1^-}dy_2^- \, \mathrm{tr}\Big[ t^{c_1} t^{c_2}t^c t^{b_2}t^{b_1}\Big]\notag\\
&\qquad \times \llangle A^+_{c_1}(x_1^-, \mathbf{x})A^+_{c_2}(x_2^-, \mathbf{x}) \partial^i_{\mathbf{x}} A^+_c(x^-, \mathbf{x})  A^+_{b_2}(y_2^-, \mathbf{y}) A^+_{b_1}(y_1^-, \mathbf{y}) \rrangle  \label{eq:32_second}\\
&+ \frac{ig^5P^+}{s} \int_{-\infty}^{+\infty} x^- dx^- \int^{x^-}_{-\infty} dx_1^- \int_{-\infty}^{x_1^-} dx_2^- \int_{-\infty}^{+\infty} dy_1^- \int_{-\infty}^{y_1^-}dy_2^- \, \mathrm{tr}\Big[  t^c t^{c_1}t^{c_2} t^{b_2}t^{b_1}\Big]\notag\\
&\qquad \times \llangle  \partial^i_{\mathbf{x}} A^+_c(x^-, \mathbf{x}) A^+_{c_1}(x_1^-, \mathbf{x})A^+_{c_2}(x_2^-, \mathbf{x}) A^+_{b_2}(y_2^-, \mathbf{y}) A^+_{b_1}(y_1^-, \mathbf{y}) \rrangle. \label{eq:32_third}
\end{align}
We give detailed calculations for the term in eq.~\eqref{eq:32_first} to demonstrate how the averaging is performed in the quasi-classical approximation. For the ensemble averaging in eq.~\eqref{eq:32_first},
\begin{align}\label{eq:32_decomposition}
&\llangle A^+_{c_1}(x_1^-, \mathbf{x}) \partial^i_{\mathbf{x}} A^+_c(x^-, \mathbf{x}) A^+_{c_2}(x_2^-, \mathbf{x}) A^+_{b_2}(y_2^-, \mathbf{y}) A^+_{b_1}(y_1^-, \mathbf{y}) \rrangle \notag \\
=&\Big\langle A^+_{c_1}(x_1^-, \mathbf{x}) A^+_{b_1}(y_1^-, \mathbf{y})\Big\rangle \llangle  \partial^i_{\mathbf{x}} A^+_c(x^-, \mathbf{x}) A^+_{c_2}(x_2^-, \mathbf{x}) A^+_{b_2}(y_2^-, \mathbf{y})  \rrangle \notag\\
&+\llangle A^+_{c_1}(x_1^-, \mathbf{x}) \partial^i_{\mathbf{x}} A^+_c(x^-, \mathbf{x})  A^+_{b_1}(y_1^-, \mathbf{y}) \rrangle \Big\langle A^+_{c_2}(x_2^-, \mathbf{x}) A^+_{b_2}(y_2^-, \mathbf{y}) \Big\rangle \notag \\
=&-\frac{4g\mu_0^6}{P^+}\delta^{c_1b_1} f^{cb_2c_2}  \delta(x_1^- -y_1^-) \delta(x^--x_2^-) \delta(x^--y_2^-) L(\mathbf{x}-\mathbf{y}) \Big[\partial^i_{\mathbf{x}}\Gamma(\mathbf{x}, \mathbf{y}, \mathbf{u})\Big]_{\mathbf{u}\rightarrow\mathbf{x}} \notag \\
&-\frac{4g\mu_0^6}{P^+} \delta^{c_2b_2} f^{cb_1c_1} \delta(x_2^--y_2^-) \delta(x^--x_1^-) \delta(x^--y_1^-)L(\mathbf{x}-\mathbf{y}) \Big[\partial^i_{\mathbf{x}} \Gamma(\mathbf{x}, \mathbf{y}, \mathbf{u})\Big]_{\mathbf{u}\rightarrow \mathbf{x}}
\end{align}
\end{widetext}
The five-field averaging is expressed as the product of three-field averaging and two-field averaging. It is not difficult to see that the only non-vanishing combinations are shown in Fig.~\ref{fig:contraction_32_one}, corresponding to the two terms in the first equality of eq.~\eqref{eq:32_decomposition}. Only contractions that avoid line crossings contribute. In arriving at the second equality, we have used the expressions in eq.~\eqref{eq:MV} and eq.~\eqref {eq:three_field_ave_res}. 
Substituting eq.~\eqref{eq:32_decomposition} into eq.~\eqref{eq:32_first},  one needs to carry out the integrals involving longitudinal coordinates. Since all the $A^+$ fields are confined within the shockwave, characterized by the longitudinal extent $[-L^-/2, L^-/2]$, the integration is restricted to this range.  
 The longitudinal coordinate integral of the first term in eq.~\eqref{eq:32_decomposition} is then determined as follows 
\begin{align}
& \int_{-\infty}^{+\infty} x^- dx^- \int_{x^-}^{+\infty} dx_1^- \int_{-\infty}^{x^-} dx_2^- \int_{-\infty}^{+\infty} dy_1^- \int_{-\infty}^{y_1^-}dy_2^- \notag\\
&\quad  \times \delta(x^--x_2^-)\delta(x^--y_2^-) \delta(x_1^--y_1^-)\notag\\
=&\frac{1}{2}\int_{-L^-/2}^{+L^-/2} x^- dx^- \int_{x^-}^{+L^-/2} dx_1^- \notag\\
=&-\frac{1}{2} \frac{(L^-)^3}{12}.
\end{align}
For the second term in eq.~\eqref{eq:32_decomposition}, it is
\begin{align}
& \int_{-\infty}^{+\infty} x^- dx^- \int_{x^-}^{+\infty} dx_1^- \int_{-\infty}^{x^-} dx_2^- \int_{-\infty}^{+\infty} dy_1^- \int_{-\infty}^{y_1^-}dy_2^- \notag\\
&\times \delta(x_1^--x^-)\delta(x_1^--y_1^-) \delta(x_2^--y_2^-)\notag\\
=&\frac{1}{2} \int_{-L^-/2}^{+L^-/2} x^- dx^-  \int_{-L^-/2}^{x^-} dx_2^-\notag\\
= &\frac{1}{2} \frac{(L^-)^3}{12}
\end{align}
The corresponding two color factors are evaluated by
 \begin{align}
 &f^{cb_2c_2} \delta^{c_1b_1}  \mathrm{tr}\Big[ t^{c_1} t^c t^{c_2} t^{b_2}t^{b_1}\Big]= - \frac{i}{2} N_c^2 C_F^2,  \\
 & f^{cb_1c_1} \delta^{c_2b_2} \mathrm{tr}\Big[ t^{c_1} t^c t^{c_2} t^{b_2}t^{b_1}\Big]= \frac{i}{2}N_c^2 C_F^2.
 \end{align}
The final result  for the term in eq.~\eqref{eq:32_first} is 
\begin{equation}\label{eq:32_cancel_1}
2\times \frac{1}{12s} g^6(L^-\mu_0^2)^3 N_c^2 C_F^2\,  L(\mathbf{x}-\mathbf{y}) \Big[\partial^i_{\mathbf{x}} \Gamma(\mathbf{x}, \mathbf{y}, \mathbf{u})\Big]_{\mathbf{u}\rightarrow \mathbf{x}}.
\end{equation} 
For the term in eq.~\eqref{eq:32_second},  the five-field averaging can be computed by
 \begin{equation}\label{eq:exp_diagram6}
 \begin{split}
& \llangle A^+_{c_1}(x_1^-, \mathbf{x})A^+_{c_2}(x_2^-, \mathbf{x}) \partial^i_{\mathbf{x}} A^+_c(x^-, \mathbf{x})  A^+_{b_2}(y_2^-, \mathbf{y}) A^+_{b_1}(y_1^-, \mathbf{y}) \rrangle \\
=& \Big\langle A^+_{c_1}(x_1^-, \mathbf{x})A^+_{b_1}(y_1^-, \mathbf{y})\Big\rangle \llangle A^+_{c_2}(x_2^-, \mathbf{x}) \partial^i_{\mathbf{x}} A^+_c(x^-, \mathbf{x})  A^+_{b_2}(y_2^-, \mathbf{y})  \rrangle \\
=&-\frac{4g\mu_0^6}{P^+} \delta^{c_1b_1} f^{cb_2c_2} \delta(x_1^--y_1^-)\delta(x^--x_2^-)\delta(x^--y_2^-)\\
&\times  L(\mathbf{x}-\mathbf{y}) \Big[ \partial^i_{\mathbf{x}} \Gamma(\mathbf{x}, \mathbf{y}, \mathbf{u})\Big]_{\mathbf{u}\rightarrow \mathbf{x}}
\end{split}
\end{equation} 
The corresponding non-vanishing contraction is shown in Fig.~\ref{fig:contraction_32_two}. Working out the integral over longitudinal coordinates and the color factors, one obtains the final result for eq.~\eqref{eq:32_second}
\begin{equation}\label{eq:32_cancel_2}
-\frac{1}{12s} g^6(L^-\mu_0^2)^3 N_c^2 C_F^2\,  L(\mathbf{x}-\mathbf{y}) \Big[\partial^i_{\mathbf{x}} \Gamma(\mathbf{x}, \mathbf{y}, \mathbf{u})\Big]_{\mathbf{u}\rightarrow \mathbf{x}}.
\end{equation} 
For the term in eq.~\eqref{eq:32_third}
\begin{align}\label{eq:exp_diagram7}
&\llangle  \partial^i_{\mathbf{x}} A^+_c(x^-, \mathbf{x}) A^+_{c_1}(x_1^-, \mathbf{x})A^+_{c_2}(x_2^-, \mathbf{x}) A^+_{b_2}(y_2^-, \mathbf{y}) A^+_{b_1}(y_1^-, \mathbf{y}) \rrangle \notag\\
=&\llangle  \partial^i_{\mathbf{x}} A^+_c(x^-, \mathbf{x}) A^+_{c_1}(x_1^-, \mathbf{x})A^+_{b_1}(y_1^-, \mathbf{y}) \rrangle \\
&\times \Big\langle A^+_{c_2}(x_2^-, \mathbf{x}) A^+_{b_2}(y_2^-, \mathbf{y}) \Big\rangle  \notag\\
=&-\frac{4g\mu_0^6}{P^+} \delta^{c_2b_2} f^{cb_1 c_1} \delta(x_2^--y_2^-) \delta(x^--x_1^-) \delta(x^--y_1^-)  \notag\\
&\quad \times  L(\mathbf{x}-\mathbf{y}) \Big[\partial^i_{\mathbf{x}} \Gamma(\mathbf{x}, \mathbf{y}, \mathbf{u})\Big]_{\mathbf{u}\rightarrow \mathbf{x}}
\end{align}
The nonvanishing contraction is shown in Fig.~\ref{fig:contraction_32_three}. Carrying out the integral over longitudinal coordinates and working out the color factor, one obtains the final result for
 eq.~\eqref{eq:32_third}, 
\begin{equation}\label{eq:32_cancel_3}
-\frac{1}{12s} g^6(L^-\mu_0^2)^3 N_c^2 C_F^2\,  L(\mathbf{x}-\mathbf{y}) \Big[\partial^i_{\mathbf{x}} \Gamma(\mathbf{x}, \mathbf{y}, \mathbf{u})\Big]_{\mathbf{u}\rightarrow \mathbf{x}}.
\end{equation} 
Diagrammatically, it is not difficult to see that the contraction diagram in Fig.~\ref{fig:contraction_32_two} and the first diagram in Fig.~\ref{fig:contraction_32_one} cancel each other.  Similarly,  the contraction diagram in Fig.~\ref{fig:contraction_32_three} and the second diagram in Fig.~\ref{fig:contraction_32_one} also cancel each other.  The three terms in eq.~\eqref{eq:32_cancel_1}, ~\eqref{eq:32_cancel_2} and ~\eqref{eq:32_cancel_3} cancel so that the combination $\{3,2\}$ vanishes.

\begin{figure*}
    \includegraphics[width=.8\textwidth]{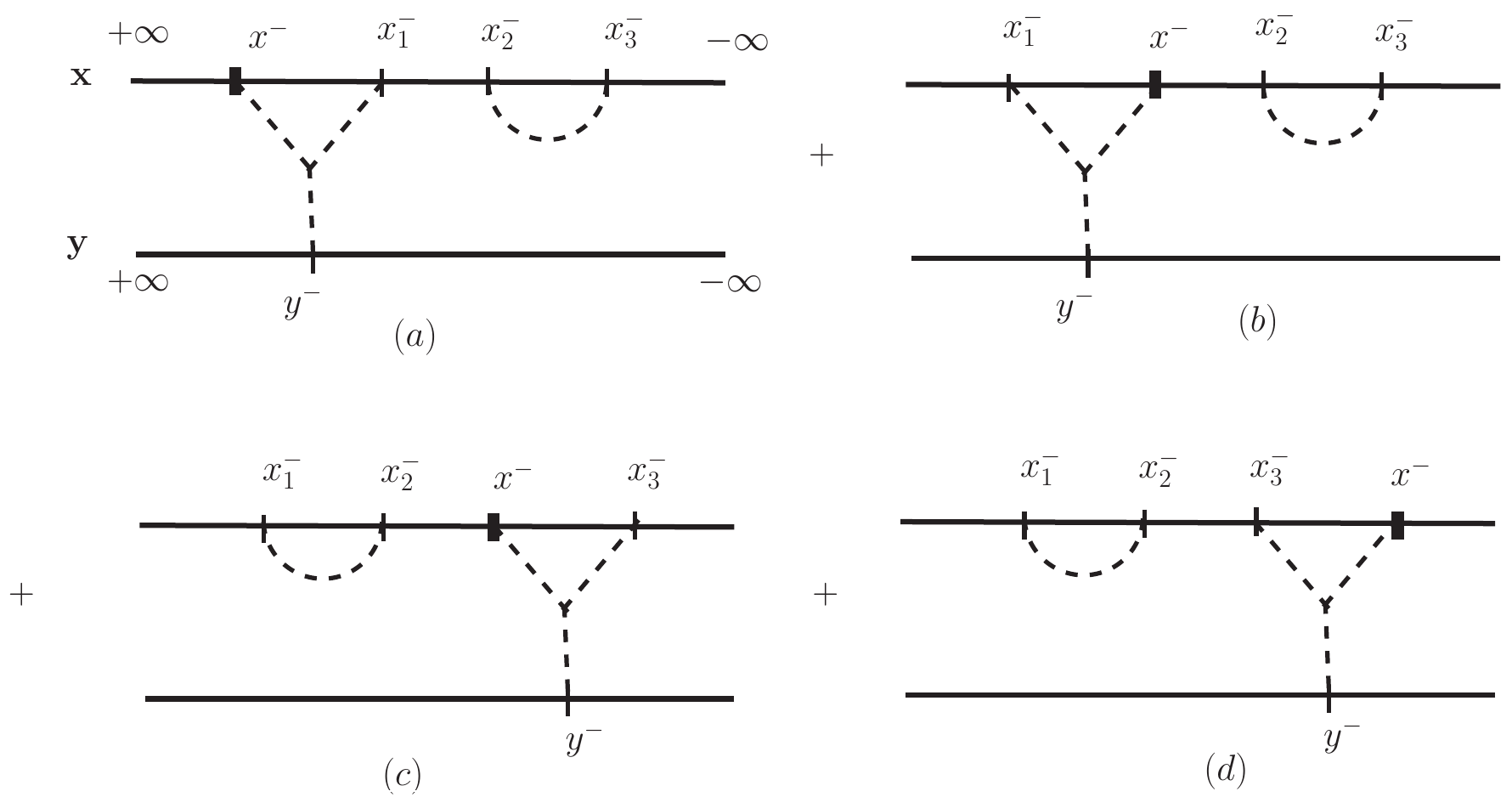}
    \caption{Non-vanishing contraction diagrams for the four terms in $\{4, 1\}$. Relative positions of the longitudinal coordinates are explicitly shown.}
\label{fig:contraction_41}		
\end{figure*}

Having shown that the combination $\{3, 2\}$ does not contribute, we continue to analyze the combination $\{4, 1\}$.  We have 
\begin{widetext}
\begin{align}
\{4, 1\} = & \frac{-ig^5P^+}{s}  \int_{-\infty}^{+\infty} x^- dx^-\int_{-\infty}^{x^-} dx_1^- \int_{-\infty}^{x_1^-} dx_2^-\int_{-\infty}^{x_2^-} dx_3^- \int_{-\infty}^{+\infty} dy^-  \mathrm{tr} \Big[t^{c}t^{c_1}t^{c_2}t^{c_3} t^b\Big] \notag\\
&\qquad \times \llangle  \partial_{\mathbf{x}}^i A^+_c(x^-, \mathbf{x})A^+_{c_1}(x_1^-, \mathbf{x}) A^+_{c_2}(x_2^-, \mathbf{x})A^+_{c_3}(x_3^-, \mathbf{x}) A^+_b(y^-, \mathbf{y})\rrangle  \label{eq:41_first}\\
&+\frac{-ig^5P^+}{s}  \int_{-\infty}^{+\infty} x^- dx^-\int_{x^-}^{+\infty} dx_1^- \int_{-\infty}^{x^-} dx_2^- \int_{-\infty}^{x_2^-} dx_3^- \int_{-\infty}^{+\infty} dy^- \mathrm{tr}\Big[ t^{c_1} t^c t^{c_2} t^{c_3} t^b\Big]\notag\\
&\qquad \times \llangle  A^+_{c_1}(x_1^-, \mathbf{x}) \partial_{\mathbf{x}}^i A^+_c(x^-, \mathbf{x})A^+_{c_2}(x_2^-, \mathbf{x}) A^+_{c_3}(x_3^-, \mathbf{x}) A^+_b(y^-, \mathbf{y})\rrangle \label{eq:41_second}\\
&+\frac{-ig^5P^+}{s} \int_{-\infty}^{+\infty} x^- dx^-\int_{x^-}^{+\infty} dx_1^- \int_{x^-}^{x_1^-} dx_2^- \int_{-\infty}^{x^-} dx_3^- \int_{-\infty}^{+\infty} dy^- \mathrm{tr}\Big[  t^{c_1} t^{c_2}t^c  t^{c_3} t^b\Big]\notag\\
&\qquad \times \llangle  A^+_{c_1}(x_1^-, \mathbf{x})A^+_{c_2}(x_2^-, \mathbf{x}) \partial_{\mathbf{x}}^i A^+_c(x^-, \mathbf{x}) A^+_{c_3}(x_3^-, \mathbf{x}) A^+_{b}(y^-, \mathbf{y})\rrangle \label{eq:41_third}\\
&+\frac{-ig^5P^+}{s}\int_{-\infty}^{+\infty} x^- dx^-\int_{x^-}^{+\infty} dx_1^- \int_{x^-}^{x_1^-} dx_2^- \int_{x^-}^{x_2^-} dx_3^- \int_{-\infty}^{+\infty} \mathrm{tr}\Big[   t^{c_1} t^{c_2}t^{c_3}  t^{c}t^b\Big]\notag\\
&\qquad \times \llangle  A^+_{c_1}(x_1^-, \mathbf{x})A^+_{c_2}(x_2^-, \mathbf{x})A^+_{c_3}(x_3^-, \mathbf{x}) \partial_{\mathbf{x}}^i A^+_c(x^-, \mathbf{x}) A^+_b(y^-, \mathbf{y})\rrangle. \label{eq:41_fourth}
\end{align}
\end{widetext}
In evaluating the five-field averaging, the non-vanishing contractions are shown in Fig.~\ref{fig:contraction_41}. 
The cancellation of these diagrams is clear. 
In fact, one can straightforwardly carry out full calculations.  The term in eq.~\eqref{eq:41_first} has the final expression
\begin{equation}\label{eq:41_cancel_first}
\frac{1}{24s} g^6 (L^- \mu_0^2)^3 N_c^2 C_F^2 L(\mathbf{0}) \Big[\partial^i_{\mathbf{x}}\Gamma(\mathbf{x}, \mathbf{y}, \mathbf{u})\Big]_{\mathbf{\mu}\rightarrow\mathbf{x}}.
\end{equation}
The term in eq.~\eqref{eq:41_third} becomes
\begin{equation}\label{eq:41_cancel_third}
-\frac{1}{24s} g^6 (L^- \mu_0^2)^3 N_c^2 C_F^2 L(\mathbf{0}) \Big[\partial^i_{\mathbf{x}}\Gamma(\mathbf{x}, \mathbf{y}, \mathbf{u})\Big]_{\mathbf{\mu}\rightarrow\mathbf{x}}.
\end{equation}
The term in eq.~\eqref{eq:41_second} is
\begin{equation}\label{eq:41_cancel_second}
-\frac{1}{24s} g^6 (L^-\mu_0^2)^3 N_c^2C_F^2L(\mathbf{0})\Big[ \partial^i_{\mathbf{x}}\Gamma(\mathbf{x}, \mathbf{y}, \mathbf{u})\Big]_{\mathbf{u}\rightarrow \mathbf{x}}   .
\end{equation}
The term in eq.~\eqref{eq:41_fourth} is 
\begin{equation}\label{eq:41_cancel_fourth}
\frac{1}{24s} g^6 (L^-\mu_0^2)^3 N_c^2C_F^2L(\mathbf{0})\Big[ \partial^i_{\mathbf{x}}\Gamma(\mathbf{x}, \mathbf{y}, \mathbf{u})\Big]_{\mathbf{u}\rightarrow \mathbf{x}}.
\end{equation}
It is apparent that eq.~\eqref{eq:41_cancel_third} cancels eq.~\eqref{eq:41_cancel_first}
and eq.~\eqref{eq:41_cancel_fourth} cancels eq.~\eqref{eq:41_cancel_second}.

We have explicitly checked that, up to the order of $\mu_0^6$ within quasi-classical approximation, the gluon saturation induced helicity-dependent field does not contribute to the dipole gluon helicity distribution. These explicit calculations, although lacking an all order proof, strongly suggest that the gluon saturation induce helicity effects play no role in the dipole gluon helicity distribution. From a technical perspective, this vanishing result can be attributed to two factors during the ensemble averaging of the two-point correlator involving polarized Wilson lines. 
The first is the cancellation due to the longitudinal coordinate integral, as shown in eq.~\eqref{eq:x^-_integral_vanish}, and the second is the insufficient number (only two) of distinct transverse coordinates, as indicated in eq.~\eqref{eq:Gamma_func_vanish}.

\subsection{WW gluon helicity distribution}
To evaluate gluon saturation induced helicity effects in the WW gluon helicity distribution within the quasi-classical approximation, one starts from the definition \cite{Kovchegov:2025, Kovchegov:2017lsr} 
\begin{widetext}
\begin{equation}\label{eq:DeltaGWW}
\Delta G_{WW}(x, \mathbf{k}^2) = -\frac{1}{2g^2\pi^3} \epsilon^{ij} \int d^2\mathbf{x} d^2\mathbf{y} e^{-i\mathbf{k}\cdot(\mathbf{x}-\mathbf{y})}
\llangle \mathrm{tr}\left[V_{\mathbf{y}} \partial^iV^{\dagger}_{\mathbf{y}} V^{j, G[2]}_{\mathbf{x}} V_{\mathbf{x}}^{\dagger}\right]\rrangle + c.c.
\end{equation}
The explicit expressions of the two factors in the Wilson line structure are 
\begin{equation}
V_{\mathbf{y}} \partial^i V_{\mathbf{y}}^{\dagger} = (ig)\int_{-\infty}^{+\infty} dy^-  U_{\mathbf{y}}^{\dagger ab}[+\infty, y^-] \partial^i_{\mathbf{y}} A^+_a(y^-, \mathbf{y}) t^b.
\end{equation}
\begin{equation}
V^{j, G[2]}_{\mathbf{x}} V_{\mathbf{x}}^{\dagger} = \frac{igP^+}{s} \int_{-\infty}^{+\infty} dx^- U_{\mathbf{x}}^{\dagger cd}[+\infty, x^-] \Big(x^- \partial^j_{\mathbf{x}} A_c^+(x^-, \mathbf{x}) + A^j_c(x^-, \mathbf{x})\Big) t^d.
\end{equation}
It should be noted that these expressions are operatorial expressions. The gluon fields are not necessarily quasi-classical gluon fields. One needs to evaluate
\begin{equation}\label{eq:WW_starting_exp}
\begin{split}
&\epsilon^{ij}\llangle \mathrm{tr}\left[V_{\mathbf{y}} \partial^iV^{\dagger}_{\mathbf{y}} V^{j, G[2]}_{\mathbf{x}} V_{\mathbf{x}}^{\dagger}\right]\rrangle\\
=& -\frac{g^2P^+}{2s}\epsilon^{ij} \int_{-\infty}^{+\infty} dx^- dy^- \llangle U_{\mathbf{x}}^{bc}[+\infty, x^-] U^{ba}_{\mathbf{y}}[+\infty, y^-]\Big(x^- \partial^j_{\mathbf{x}} A_c^+(x^-, \mathbf{x}) + A^j_c(x^-, \mathbf{x})\Big) \partial^i_{\mathbf{y}} A^+_a(y^-, \mathbf{y})\rrangle
\end{split}
\end{equation}
In eq.~\eqref{eq:WW_starting_exp}, the potential gluon saturation induced helicity effects are contained in
\begin{equation}\label{eq:WW_helicity_dep}
\begin{split}
 -\frac{g^2P^+}{2s} \epsilon^{ij}\int_{-\infty}^{+\infty} x^-dx^-\int_{-\infty}^{+\infty} dy^- \llangle U_{\mathbf{x}}^{bc}[+\infty, x^-] U^{ba}_{\mathbf{y}}[+\infty, y^-]\partial^j_{\mathbf{x}} \mathcal{A}_c^+(x^-, \mathbf{x}) \partial^i_{\mathbf{y}}\mathcal{ A}^+_a(y^-, \mathbf{y})\rrangle
\end{split}
\end{equation}
\end{widetext}
The helicity dependent field can come from four sources: $\mathcal{A}^+_c(x^-, \mathbf{x})$, $\mathcal{A}^+_a(y^-, \mathbf{y})$, or the Wilson lines $U_{\mathbf{x}}[+\infty, x^-]$ and $U_{\mathbf{y}}[+\infty, y^-]$. In all cases, non-vanishing ensemble averaging always yields the factor $\delta(x^--y^-)$. From the definition in eq.~\eqref{eq:DeltaGWW}, the expression in eq.~\eqref{eq:WW_helicity_dep} is symmetric under the exchange $\mathbf{x}\leftrightarrow \mathbf{y}$. When $x^-=y^-$,  the terms within the double brackets are symmetric under $i\leftrightarrow j$.  Consequently, their produce with $\epsilon^{ij}$ vanishes.  As a result, the WW gluon helicity distribution receives no contribution from the gluon saturation induced helicity effects. 

We have demonstrated that the gluon saturation induced helicity effects do not contribute to single-particle helicity distribution. Single-particle helicity distribution only involves two different transverse coordinates but the ensemble averaging involving gluon saturation induced field strongly favors three or more different transverse coordinates eq.~\eqref{eq:Gamma_func_vanish}.  In the following, we consider two-particle correlated helicity distribution that involves four different transverse coordinates.

\section{Two-Particle Correlated Helicity Distributions}\label{sec:two_particle_dis}
The discussion in the above section suggests that the gluon saturation induced helicity effects are intrinsically two-particle (or multi-particle) correlation effects. One should consider four-point (or higher point) correlation functions.
One explicit example is
\begin{widetext}
\begin{equation}\label{eq:Qi_def}
\mathcal{Q}^i(\mathbf{x}, \mathbf{y}, \mathbf{u}, \mathbf{v})=\llangle \mathrm{tr}\left[V^{i, G[2]}_{\mathbf{x}} V^{\dagger}_{\mathbf{y}}\right]\mathrm{tr}\left[V_{\mathbf{u}}V^{\dagger}_{\mathbf{v}}\right]\rrangle - \llangle \mathrm{tr}\left[V^{i, G[2]}_{\mathbf{x}} V^{\dagger}_{\mathbf{y}}\right]\rrangle \Big\langle\mathrm{tr}\left[V_{\mathbf{u}}V^{\dagger}_{\mathbf{v}}\right]\Big\rangle.
\end{equation}
\end{widetext}
In this definition, the four-point correlation function is constructed by subtracting two-point correlations, isolating the genuine two-particle correlations. This physical quantity plays a key role in the double-spin asymmetry for dijet production in incoherent diffractive scatterings in longitudinally polarized electron-proton/nucleus collisions. We analytically compute $\mathcal{Q}^i(\mathbf{x}, \mathbf{y}, \mathbf{u}, \mathbf{v})$ in the quasi-classical approximation and then evaluate its contribution to double-spin asymmetries in the next section.

\subsection{Direct Helicity Dependence}
We first focus on the part that directly depends on the $A^i$ field. 
\begin{widetext}
\begin{align}
&\llangle \mathrm{tr}\left[V^{i, G[2]}_{\mathbf{x}} V^{\dagger}_{\mathbf{y}}\right]\mathrm{tr}\left[V_{\mathbf{u}}V^{\dagger}_{\mathbf{v}}\right]\rrangle
=\frac{igP^+}{s} \int_{-\infty}^{+\infty} dx^-\llangle \mathrm{tr}\left[V_{\mathbf{x}}[+\infty, x^- ] A^i(x^-, \mathbf{x}) V_{\mathbf{x}}[x^-, -\infty]V^{\dagger}_{\mathbf{y}}\right]\mathrm{tr}\left[V_{\mathbf{u}}V^{\dagger}_{\mathbf{v}}\right]\rrangle \label{eq:4point_Ai}\\
&\qquad\qquad\qquad +\frac{igP^+}{s} \int_{-\infty}^{+\infty} x^-dx^-\llangle \mathrm{tr}\left[V_{\mathbf{x}}[+\infty, x^- ] \partial^i_{\mathbf{x}}A^+(x^-, \mathbf{x}) V_{\mathbf{x}}[x^-, -\infty]V^{\dagger}_{\mathbf{y}}\right]\mathrm{tr}\left[V_{\mathbf{u}}V^{\dagger}_{\mathbf{v}}\right]\rrangle.  \label{eq:4point_A+}
\end{align}
For the term in eq.~\eqref{eq:4point_Ai}, using the MV and the helicity-extended MV models, one obtains
\begin{equation}\label{eq:relating_to_unpolarized}
\begin{split}
&\frac{igP^+}{s} \int_{-\infty}^{+\infty} dx^-\llangle \mathrm{tr}\left[V_{\mathbf{x}}[+\infty, x^- ] A^i(x^-, \mathbf{x}) V_{\mathbf{x}}[x^-, -\infty]V^{\dagger}_{\mathbf{y}}\right]\mathrm{tr}\left[V_{\mathbf{u}}V^{\dagger}_{\mathbf{v}}\right]\rrangle\\
=&\frac{igP^+}{s} \int_{-\infty}^{+\infty} dx^-\llangle \mathrm{tr}\left[V_{\mathbf{x}}[+\infty, x^- ] \Big(-\epsilon^{il}\partial^l_{\mathbf{x}}\beta (x^-, \mathbf{x}) \Big)V_{\mathbf{x}}[x^-, -\infty]V^{\dagger}_{\mathbf{y}}\right]\mathrm{tr}\left[V_{\mathbf{u}}V^{\dagger}_{\mathbf{v}}\right]\rrangle\\
=&-\frac{ig}{s} \epsilon^{il}\int_{-\infty}^{+\infty} dx^-\Big\langle  \mathrm{tr}\left[V_{\mathbf{x}}[+\infty, x^- ]\partial^l_{\mathbf{x}}\alpha (x^-, \mathbf{x})V_{\mathbf{x}}[x^-, -\infty]V^{\dagger}_{\mathbf{y}}\right]\mathrm{tr}\left[V_{\mathbf{u}}V^{\dagger}_{\mathbf{v}}\right]\Big\rangle\\
=&\frac{1}{s} \epsilon^{il} \partial^l_{\mathbf{x}} \Big\langle  \mathrm{tr}\left[V_{\mathbf{x}}V^{\dagger}_{\mathbf{y}}\right]\mathrm{tr}\left[V_{\mathbf{u}}V^{\dagger}_{\mathbf{v}}\right]\Big\rangle.
\end{split}
\end{equation}
It is interesting to observe that, in the quasi-classical approximation using MV and helicity-extended MV models, the polarized Wilson line correlator involving the $A^i$ field can be expressed in terms of its unpolarized counterpart. 
The four-point Wilson line correlator in the unpolarized case has already been computed within the MV model. 
\begin{equation}\label{eq:qqbar_qqbar_correlator}
\begin{split}
&\Big\langle S_{q\bar{q}}(\mathbf{x}, \mathbf{y}) S_{q\bar{q}}(\mathbf{u}, \mathbf{v}) \Big\rangle - \Big\langle S_{q\bar{q}}(\mathbf{x}, \mathbf{y})\Big\rangle \Big\langle S_{q\bar{q}}(\mathbf{u}, \mathbf{v}) \Big\rangle= \frac{1}{N_c^2} \Big\langle \mathrm{tr}[V_{\mathbf{x}}V^{\dagger}_{\mathbf{y}}] \mathrm{tr}[V_{\mathbf{u}}V^{\dagger}_{\mathbf{v}}]\Big\rangle -\frac{1}{N_c^2} \Big\langle \mathrm{tr}[V_{\mathbf{x}}V^{\dagger}_{\mathbf{y}}]\Big\rangle\Big\langle \mathrm{tr}[V_{\mathbf{u}}V^{\dagger}_{\mathbf{v}}]\Big\rangle\\
 \end{split}
\end{equation}
This unpolarized Wilson line structure typically shows up when one considers incoherent diffractive dijet productions in DIS \cite{Kar:2023jkn}.  Within the MV model, it has the following closed form expression \cite{Fujii:2002vh, Blaizot:2004wv, Dominguez:2008aa},
\begin{equation}\label{eq:qqbar_qqbar_correlator_MV}
\begin{split}
\Big\langle S_{q\bar{q}}(\mathbf{x}, \mathbf{y}) S_{q\bar{q}}(\mathbf{u}, \mathbf{v}) \Big\rangle
=&e^{Q_0^2 (1-\frac{1}{N_c^2})\Gamma_1} \Bigg[\left(\frac{F(\mathbf{x}, \mathbf{u}; \mathbf{y}, \mathbf{v}) + \sqrt{\Delta}}{2\sqrt{\Delta}} - \frac{F(\mathbf{x}, \mathbf{y}; \mathbf{u}, \mathbf{v})}{N_c^2 \sqrt{\Delta}}\right)e^{Q_0^2 \sqrt{\Delta}} \\
&- \left(\frac{F(\mathbf{x}, \mathbf{u}; \mathbf{y}, \mathbf{v}) - \sqrt{\Delta}}{2\sqrt{\Delta}} - \frac{F(\mathbf{x}, \mathbf{y}; \mathbf{u}, \mathbf{v})}{N_c^2 \sqrt{\Delta}}\right)e^{-Q_0^2 \sqrt{\Delta}} \Bigg] e^{-Q_0^2 F(\mathbf{x}, \mathbf{u}; \mathbf{y}, \mathbf{v}) + \frac{2}{N_c^2} Q_{0}^2 F(\mathbf{x}, \mathbf{y};\mathbf{u}, \mathbf{v})}.
\end{split}
\end{equation}
\end{widetext}
Here the notations are
\begin{equation}
\begin{split}
&\Gamma_1\equiv \Gamma(\mathbf{x}-\mathbf{y}) + \Gamma(\mathbf{u}-\mathbf{v}),\\
&\Gamma_2\equiv \Gamma(\mathbf{x}-\mathbf{u}) + \Gamma(\mathbf{y}-\mathbf{v}), \\
&\Gamma_3 \equiv \Gamma(\mathbf{x}-\mathbf{v}) + \Gamma(\mathbf{y}-\mathbf{u}), \\
\end{split}
\end{equation}
\begin{subequations}
\begin{align}
&F(\mathbf{x}, \mathbf{y}, \mathbf{u}, \mathbf{v}) \notag\\
=& \frac{1}{2} \Big[\Gamma(\mathbf{x}-\mathbf{u}) + \Gamma(\mathbf{y}-\mathbf{v}) - \Gamma(\mathbf{x}-\mathbf{v}) - \Gamma(\mathbf{y}-\mathbf{u})\Big],\\
&F(\mathbf{x}, \mathbf{u}; \mathbf{y}, \mathbf{v}) \notag \\
 = &\frac{1}{2} \Big[\Gamma(\mathbf{x}-\mathbf{y}) + \Gamma(\mathbf{u}-\mathbf{v}) - \Gamma(\mathbf{x}-\mathbf{v}) - \Gamma(\mathbf{y}-\mathbf{u})\Big],\\
 &F(\mathbf{x}, \mathbf{v}; \mathbf{u}, \mathbf{y}) \notag \\
 =& \frac{1}{2} \Big[\Gamma(\mathbf{x}-\mathbf{u} ) + \Gamma(\mathbf{y}-\mathbf{v}) - \Gamma(\mathbf{x}-\mathbf{y}) - \Gamma(\mathbf{v}-\mathbf{u})\Big]
\end{align}
\end{subequations}
and 
\begin{equation}
\Delta \equiv F^2(\mathbf{x}, \mathbf{u}, \mathbf{y}, \mathbf{v}) + \frac{4}{N_c^2} F(\mathbf{x}, \mathbf{y};\mathbf{u}, \mathbf{v}) F(\mathbf{x}, \mathbf{v};\mathbf{u}, \mathbf{y}).
\end{equation}
 Recall the expression for the unpolarized dipole correlator in the MV model,  
\begin{equation}
\frac{1}{N_c} \left\langle  \mathrm{tr}\left[V_{\mathbf{x}}V_{\mathbf{y}}^{\dagger}\right]\right\rangle  \\
=\mathrm{exp}\left\{ \frac{1}{2} C_F g^2 (L^-\mu_0^2) \Gamma(\mathbf{x}-\mathbf{y})\right\}\\
\end{equation}
with $\Gamma(\mathbf{x}-\mathbf{y}) = 2L(\mathbf{x}-\mathbf{y}) - 2L(\mathbf{0})$ and $\Gamma(\mathbf{r}) 
\simeq  - \frac{1}{4\pi} r^2 \ln (1/\Lambda r)$.
The gluon saturation scale is conventionally introduced in the dipole correlator as 
\begin{equation}
S_{q\bar{q}}(\mathbf{r}) = \mathrm{exp}\left\{-\frac{1}{4}Q_s^2 r^2 \ln \frac{1}{\Lambda r}\right\}.
\end{equation}
One can thus identify the relation between $Q_s^2$ and $\mu_0^2$, 
\begin{equation}
Q_s^2 = \frac{1}{2\pi} C_F g^2 (L^-\mu_0^2)  
\end{equation}
from which one obtains $Q_0^2 = \pi Q_s^2$ in the large-$N_c$ limit.

We are interested in eq.~\eqref{eq:qqbar_qqbar_correlator} in the large-$N_c$ limit and expansions up to the order $\sim (Q_0^2)^3$. Carrying out the expansions, one gets
\begin{equation}\label{eq:qqbar_qqbar_correlator_MV}
\begin{split}
&\Big\langle S_{q\bar{q}}(\mathbf{x}, \mathbf{y}) S_{q\bar{q}}(\mathbf{u}, \mathbf{v}) \Big\rangle - \Big\langle S_{q\bar{q}}(\mathbf{x}, \mathbf{y})\Big\rangle\Big\langle S_{q\bar{q}}(\mathbf{u}, \mathbf{v}) \Big\rangle\\
\simeq & \frac{2}{N_c^2} Q_0^4 F^2(\mathbf{x}, \mathbf{y}; \mathbf{u}, \mathbf{v})+ \frac{2}{N_c^2} Q_0^6 \Gamma_1 F^2(\mathbf{x}, \mathbf{y}; \mathbf{u}, \mathbf{v})\\
&-\frac{4}{3N_c^2} Q_0^6 F(\mathbf{x}, \mathbf{u}; \mathbf{y}, \mathbf{v}) F^2(\mathbf{x}, \mathbf{y}; \mathbf{u}, \mathbf{v}).
\end{split}
\end{equation}
These leading order $\mathcal{O}(Q_0^4)$ and sub-leading order $\mathcal{O}(Q_0^6)$ terms are then substituted into eq.~\eqref{eq:relating_to_unpolarized} to get contributions from direct helicity effect.
\subsection{Gluon Saturation Induced Helicity Dependence}
We proceed to compute the term in eq.~\eqref{eq:4point_A+},
At order $Q_s^4$, three $A^+$ fields are needed from expanding the (partial) Wilson lines and the contribution at this order vanishes due to vanishing color factor. As a result, gluon saturation induced helicity effects start from the sub-leading order $Q_s^6$. To evaluate contribution at the order $Q_s^6$, terms containing $(A^+)^5$ are required. The calculations can be systematically organized based on the number of different transverse coordinates involved.

\subsubsection{Four Transverse Coordinates}
We first consider the case that all four transverse coordinates are involved.  In this case, two of the five $A^+$ fields have the same transverse coordinates. We have the following combinations
\begin{widetext}
\begin{subequations}\label{eq:I4_exp}
\begin{align}
\mathcal{I}_4\equiv &\frac{ig^5P^+}{s} \int_{-\infty}^{+\infty} x^-dx^- \int_{x^-}^{+\infty} dx_1^-\int_{-\infty}^{+\infty} dy^- \int_{-\infty}^{+\infty} du^- \int_{-\infty}^{+\infty} dv^- \mathrm{tr}\left[t^{c_1} t^c t^b\right]\mathrm{tr}\left[t^at^h\right]\notag\\
&\qquad \times \llangle A^+_{c_1}(x_1^-, \mathbf{x}) \partial^i_{\mathbf{x}}A^+_c(x^-, \mathbf{x}) A^+_b(y^-, \mathbf{y}) A^+_a(u^-, \mathbf{u}) A^+_h(v^-, \mathbf{v}) \rrangle \label{eq:I4_a_AP}\\
&+\frac{ig^5P^+}{s} \int_{-\infty}^{+\infty} x^-dx^- \int_{-\infty}^{x^-} dx_1^-\int_{-\infty}^{+\infty} dy^- \int_{-\infty}^{+\infty} du^- \int_{-\infty}^{+\infty} dv^- \mathrm{tr}\left[t^{c} t^{c_1} t^b\right]\mathrm{tr}\left[t^at^h\right]\notag\\
&\qquad \times \llangle A^+_{c_1}(x_1^-, \mathbf{x}) \partial^i_{\mathbf{x}}A^+_c(x^-, \mathbf{x}) A^+_b(y^-, \mathbf{y}) A^+_a(u^-, \mathbf{u}) A^+_h(v^-, \mathbf{v}) \rrangle \label{eq:I4_b_AP}\\
&-\frac{ig^5P^+}{s} \int_{-\infty}^{+\infty} x^- dx^- \int_{-\infty}^{+\infty} dy^- \int_{-\infty}^{y^-} dy_1^- \int_{-\infty}^{+\infty} du^-  \int_{-\infty}^{+\infty} dv^- \mathrm{tr}\left[t^ct^{b_1}t^b\right] \mathrm{tr}\left[t^at^h\right] \notag\\
&\qquad \times \llangle \partial^i_{\mathbf{x}}A^+_c(x^-, \mathbf{x})A^+_{b_1}(y_1^-, \mathbf{y}) A^+_b(y^-, \mathbf{y}) A^+_a(u^-, \mathbf{u}) A^+_h(v^-, \mathbf{v}) \rrangle \label{eq:I4_c_AP}\\
&+\frac{ig^5P^+}{s} \int_{-\infty}^{+\infty} x^- dx^- \int_{-\infty}^{+\infty} dy^-  \int_{-\infty}^{+\infty} du^-  \int_{-\infty}^{u^-} du_1^-\int_{-\infty}^{+\infty} dv^- \mathrm{tr}\left[t^ct^b\right] \mathrm{tr}\left[t^at^{a_1}t^h\right] \notag\\
&\qquad \times \llangle \partial^i_{\mathbf{x}}A^+_c(x^-, \mathbf{x})A^+_b(y^-, \mathbf{y}) A^+_a(u^-, \mathbf{u}) A^+_{a_1}(u^-_1, \mathbf{u}) A^+_h(v^-, \mathbf{v}) \rrangle \label{eq:I4_d_AP}\\
&-\frac{ig^5P^+}{s} \int_{-\infty}^{+\infty} x^- dx^- \int_{-\infty}^{+\infty} dy^-  \int_{-\infty}^{+\infty} du^-  \int_{-\infty}^{+\infty} dv^- \int_{-\infty}^{v^-} dv_1^-\mathrm{tr}\left[t^ct^b\right] \mathrm{tr}\left[t^at^{h_1}t^h\right] \notag\\
&\qquad \times \llangle \partial^i_{\mathbf{x}}A^+_c(x^-, \mathbf{x})A^+_b(y^-, \mathbf{y}) A^+_a(u^-, \mathbf{u}) A^+_{h_1}(v_1^-, \mathbf{v}) A^+_h(v^-, \mathbf{v}) \rrangle\label{eq:I4_e_AP}.
\end{align}
\end{subequations}
Each term contains five $A^+$ fields and any one of the five fields could be the sub-eikonal order gluon saturation induced helicity-dependent field. We have explicitly demonstrated how to carry out ensemble averaging in Sec.~\ref{sec:dipole_TMD}. Calculation details are given in appendix ~\ref{sec:appendix}. We only quote the  final result 
\begin{equation}\label{eq:final_result_xyuv}
\begin{split}
\mathcal{I}_4
= & \frac{-g^6N_c^3}{12s} (L^-\mu_0^2)^3 \Big[ \left[\partial^i_{\mathbf{x}}L(\mathbf{x}-\mathbf{u}) + L(\mathbf{y}-\mathbf{u}) \partial^i_{\mathbf{x}} - L(\mathbf{x}-\mathbf{u}) \partial^i_{\mathbf{x}}\right] \Gamma(\mathbf{x}, \mathbf{y}, \mathbf{v}) \\
&+ \left[ \partial^i_{\mathbf{x}}L(\mathbf{x}-\mathbf{v}) + L(\mathbf{y}-\mathbf{v}) \partial^i_{\mathbf{x}} - L(\mathbf{x}-\mathbf{v}) \partial^i_{\mathbf{x}}\right] \Gamma(\mathbf{x}, \mathbf{y}, \mathbf{u})\\
&+\left[L(\mathbf{y}-\mathbf{v}) \partial^i_{\mathbf{x}} -L(\mathbf{y}-\mathbf{u}) \partial^i_{\mathbf{x}}\right]\Gamma(\mathbf{x}, \mathbf{u}, \mathbf{v}) + \left[\partial^i_{\mathbf{x}}L(\mathbf{x}-\mathbf{u}) - \partial^i_{\mathbf{x}}L(\mathbf{x}-\mathbf{v})\right]\Gamma(\mathbf{y}, \mathbf{u}, \mathbf{v}) \Big].
\end{split}
\end{equation}
It is apparent that eq.~\eqref{eq:final_result_xyuv} is symmetric with the exchange $\mathbf{u}\leftrightarrow \mathbf{v}$.

\subsubsection{Three Transverse Coordinates}
We now consider the cases that only three different transverse coordinates are involved. The three coordinates can be $\{\mathbf{x}, \mathbf{y}, \mathbf{u}\}$, $\{\mathbf{x}, \mathbf{y}, \mathbf{v}\}$ and $\{\mathbf{x}, \mathbf{u}, \mathbf{v}\}$. 

To having five $A^+$ fields distributed among three different transverse coordinates $\{\mathbf{x}, \mathbf{y}, \mathbf{u}\}$(note that, when expanding the Wilson lines, at least two $A^+$ fields should come from $V_{\mathbf{u}}$), the possible combinations are $(1, 2, 2)$, $(2,1, 2)$ and $(1, 1, 3)$. Here each tuple $(n_{\mathbf{x}}, n_{\mathbf{y}}, n_{\mathbf{u}})$ represnts the number of fields assigned to the respective coordinates $\mathbf{x}, \mathbf{y}, $ and $\mathbf{u})$. 
\begin{subequations}\label{eq:I3_xyu_exp}
\begin{align}
\mathcal{I}_{\mathbf{x}\mathbf{y}\mathbf{u}}=&\frac{ig^5 P^+}{s} \int_{-\infty}^{+\infty} x^- dx^- \int_{-\infty}^{+\infty} dy^- \int_{-\infty}^{y^-} dy_1^- \int_{-\infty}^{+\infty} du^- \int_{-\infty}^{u^-} du_1^- \mathrm{tr}\Big[t^ct^{b_1}t^b\Big]\mathrm{tr}\Big[ t^at^{a_1}\Big]\notag\\
&\qquad \times \llangle \partial^i_{\mathbf{x}} A^+_c(x^-, \mathbf{x}) A^+_{b_1} (y_1^-, \mathbf{y}) A^+_b(y^-, \mathbf{y}) A^+_a(u^-, \mathbf{u}) A^+_{a_1}(u_1^-, \mathbf{u}) \rrangle \label{eq:I_xyu_a}\\
& - \frac{ig^5P^+}{s} \int_{-\infty}^{+\infty} x^- dx^- \int_{-\infty}^{+\infty} dy^- \int_{-\infty}^{+\infty} du^- \int_{-\infty}^{u^-} du_1^- \int_{-\infty}^{u^-_1} du_2^- \mathrm{tr}\Big[t^ct^b\Big]\mathrm{tr}\Big[ t^at^{a_1}t^{a_2}\Big]\notag\\
&\qquad \times \llangle \partial^i_{\mathbf{x}} A^+_c(x^-, \mathbf{x}) A^+_b(y^-, \mathbf{y}) A^+_a(u^-, \mathbf{u}) A^+_{a_1}(u_1^-, \mathbf{u}) A^+_{a_2}(u_2^-, \mathbf{u})\rrangle \label{eq:I_xyu_b}\\
&-\frac{ig^5 P^+}{s} \int_{-\infty}^{+\infty} x^- dx^- \int_{x^-}^{+\infty} dx_1^-\int_{-\infty}^{+\infty} dy^- \int_{-\infty}^{+\infty} du^- \int_{-\infty}^{u^-} du_1^- \mathrm{tr}\Big[t^{c_1}t^{c}t^b\Big]\mathrm{tr}\Big[ t^at^{a_1}\Big]\notag\\
&\qquad \times \llangle A^+_{c_1}(x_1^-, \mathbf{x}) \partial^i_{\mathbf{x}} A^+_c(x^-, \mathbf{x})A^+_b(y^-, \mathbf{y}) A^+_a(u^-, \mathbf{u}) A^+_{a_1}(u_1^-, \mathbf{u})  \rrangle \label{eq:I_xyu_c}\\
&-\frac{ig^5 P^+}{s} \int_{-\infty}^{+\infty} x^- dx^- \int^{x^-}_{-\infty} dx_1^-\int_{-\infty}^{+\infty} dy^- \int_{-\infty}^{+\infty} du^- \int_{-\infty}^{u^-} du_1^- \mathrm{tr}\Big[t^{c}t^{c_1}t^b\Big]\mathrm{tr}\Big[ t^at^{a_1}\Big]\notag\\
&\qquad \times \llangle \partial^i_{\mathbf{x}} A^+_c(x^-, \mathbf{x})A^+_{c_1}(x_1^-, \mathbf{x}) A^+_b(y^-, \mathbf{y}) A^+_a(u^-, \mathbf{u}) A^+_{a_1}(u_1^-, \mathbf{u})  \rrangle \label{eq:I_xyu_d}. 
\end{align}
\end{subequations}
The term in eq.~\eqref{eq:I_xyu_b} yields vanishing contribution.
The terms in eq.~\eqref{eq:I_xyu_c} and eq.~\eqref{eq:I_xyu_d} have the same final expression and they add up.  The resulting expression is 
\begin{equation}\label{eq:result_case_xyu}
\mathcal{I}_{\mathbf{x}\mathbf{y}\mathbf{u}} = \frac{g^6 N_c^3}{12s} (L^-\mu_0^2)^3 \Big[ L(\mathbf{y}-\mathbf{u}) \partial^i_{\mathbf{x}}\Gamma(\mathbf{x}, \mathbf{y}, \mathbf{u}) + \left[\partial_{\mathbf{x}}^i L(\mathbf{x}-\mathbf{u}) - L(\mathbf{x}-\mathbf{u}) \partial^i_{\mathbf{x}}\right] \Gamma(\mathbf{x}, \mathbf{y}, \mathbf{u})\Big].
\end{equation}
The result for the case $\{\mathbf{x}, \mathbf{y}, \mathbf{v}\}$ can be obtained from eq.~\eqref{eq:result_case_xyu} by the exchange $\mathbf{u}\leftrightarrow \mathbf{v}$. 
\begin{equation}\label{eq:result_case_xyv}
\mathcal{I}_{\mathbf{x}\mathbf{y}\mathbf{v}} = \frac{g^6 N_c^3}{12s} (L^-\mu_0^2)^3 \Big[ L(\mathbf{y}-\mathbf{v}) \partial^i_{\mathbf{x}}\Gamma(\mathbf{x}, \mathbf{y}, \mathbf{v}) + \left[\partial_{\mathbf{x}}^i L(\mathbf{x}-\mathbf{v}) - L(\mathbf{x}-\mathbf{v}) \partial^i_{\mathbf{x}}\right] \Gamma(\mathbf{x}, \mathbf{y}, \mathbf{v})\Big].
\end{equation}
To have 5 fields at three different transverse coordinates, $\{\mathbf{x}, \mathbf{u}, \mathbf{v}\}$, at least two fields must be at the transverse position $\mathbf{x}$.  We have the possible combinations $(2, 2, 1)$, $(2, 1, 2)$ and $(3, 1, 1)$.  We proceed by computing the corresponding terms for these combinations 
\begin{subequations}\label{eq:I_xuv_full}
\begin{align}
\mathcal{I}_{\mathbf{x}\mathbf{u}\mathbf{v}}
=&-\frac{ig^5P^+}{s} \int_{-\infty}^{+\infty} x^- dx^- \int_{-\infty}^{+\infty} dx_1^- \int_{-\infty}^{+\infty} du^- \int_{-\infty}^{u^-} du_1^- \int_{-\infty}^{+\infty} dv^- \frac{1}{2} \mathrm{tr} \Big[ t^a t^{a_1} t^h\Big] \notag\\
&\qquad \times \llangle A^+_c(x_1^-, \mathbf{x}) \partial^i_{\mathbf{x}} A^+_c(x^-, \mathbf{x}) A^+_a(u^-, \mathbf{u}) A^+_{a_1}(u_1^-, \mathbf{u}) A^+_h(v^-, \mathbf{v})\rrangle      \label{eq:I_xuv_a}\\
&+\frac{ig^5P^+}{s} \int_{-\infty}^{+\infty} x^- dx^- \int_{-\infty}^{+\infty} dx_1^- \int_{-\infty}^{+\infty} du^-  \int_{-\infty}^{+\infty} dv^- \int_{-\infty}^{v^-} dv_1^-\frac{1}{2} \mathrm{tr} \Big[ t^a t^{h_1} t^h\Big] \notag\\
&\qquad \times \llangle A^+_c(x_1^-, \mathbf{x}) \partial^i_{\mathbf{x}} A^+_c(x^-, \mathbf{x}) A^+_a(u^-, \mathbf{u}) A^+_{h_1}(v_1^-, \mathbf{v}) A^+_h(v^-, \mathbf{v})\rrangle.    \label{eq:I_xuv_b}\\
&-\frac{ig^5P^+}{s} \int_{-\infty}^{+\infty} x^- dx^- \int_{-\infty}^{x^-} dx_1^- \int_{-\infty}^{x^-} dx_2^- \theta(x_1^-- x_2^-) \int_{-\infty}^{+\infty} du^- \int_{-\infty}^{+\infty} dv^-  \mathrm{tr} \Big[t^ct^{c_1}t^{c_2}\Big] \mathrm{tr}\Big[ t^a t^h\Big]\notag\\
&\qquad \times \llangle \partial^i_{\mathbf{x}}A^+_c(x^-, \mathbf{x}) A^+_{c_1} (x_1^-, \mathbf{x}) A^+_{c_2}(x_2^-, \mathbf{x}) A^+_a(u^-, \mathbf{u}) A^+_h(v^-, \mathbf{v})\rrangle  \label{eq:I_xuv_c}\\
&-\frac{ig^5P^+}{s} \int_{-\infty}^{+\infty} x^- dx^- \int^{+\infty}_{x^-} dx_1^- \int^{+\infty}_{x^-} dx_2^- \theta(x_1^-- x_2^-) \int_{-\infty}^{+\infty} du^- \int_{-\infty}^{+\infty} dv^-  \mathrm{tr} \Big[t^{c_1}t^{c_2}t^c\Big] \mathrm{tr}\Big[ t^a t^h\Big]\notag\\
&\qquad \times \llangle A^+_{c_1} (x_1^-, \mathbf{x}) A^+_{c_2}(x_2^-, \mathbf{x})\partial^i_{\mathbf{x}}A^+_c(x^-, \mathbf{x})  A^+_a(u^-, \mathbf{u}) A^+_h(v^-, \mathbf{v})\rrangle \label{eq:I_xuv_d}\\
&-\frac{ig^5P^+}{s} \int_{-\infty}^{+\infty} x^- dx^- \int^{+\infty}_{x^-} dx_1^- \int_{-\infty}^{x^-} dx_2^-  \int_{-\infty}^{+\infty} du^- \int_{-\infty}^{+\infty} dv^-  \mathrm{tr} \Big[t^{c_1}t^{c}t^{c_2}\Big] \mathrm{tr}\Big[ t^a t^h\Big]\notag\\
&\qquad \times \llangle A^+_{c_1} (x_1^-, \mathbf{x}) \partial^i_{\mathbf{x}}A^+_c(x^-, \mathbf{x})A^+_{c_2}(x_2^-, \mathbf{x})  A^+_a(u^-, \mathbf{u}) A^+_h(v^-, \mathbf{v})\rrangle \label{eq:I_xuv_e}.
\end{align}
\end{subequations}
Eq.~\eqref{eq:I_xuv_a} and eq.~\eqref{eq:I_xuv_b} are related by the exchange $\mathbf{u}\leftrightarrow \mathbf{v}$. 
The final expressions for eqs.~\eqref{eq:I_xuv_c} and ~\eqref{eq:I_xuv_d} together cancel that for eq.~\eqref{eq:I_xuv_e}. The details of calculations are presented in the appendix. The final result is
\begin{equation}\label{eq:result_case_xuv}
\begin{split}
\mathcal{I}_{\mathbf{x}\mathbf{u}\mathbf{v}}=& \frac{g^6N_c^3}{12s} (L^-\mu_0^2)^3  \Big[ \left[\partial^i_{\mathbf{x}}L(\mathbf{x}-\mathbf{u}) - L(\mathbf{x}-\mathbf{u})\partial^i_{\mathbf{x}}\right]\Gamma(\mathbf{x}, \mathbf{u}, \mathbf{v}) +  \left[\partial^i_{\mathbf{x}}L(\mathbf{x}-\mathbf{v}) - L(\mathbf{x}-\mathbf{v})\partial^i_{\mathbf{x}}\right]\Gamma(\mathbf{x}, \mathbf{v}, \mathbf{u}) \Big]\\
\end{split}
\end{equation}
Combining eqs.~\eqref{eq:result_case_xyu}, ~\eqref{eq:result_case_xyv}, ~\eqref{eq:result_case_xuv}, one obtains the final expression for terms involving three different coordinates
\begin{align}\label{eq:I3_final_res}
\mathcal{I}_3
 =&-\frac{g^6 N_c^3}{12s} (L^-\mu_0^2)^3 \Big[  \left[\partial_{\mathbf{x}}^i L(\mathbf{x}-\mathbf{u})+L(\mathbf{y}-\mathbf{u}) \partial^i_{\mathbf{x}} - L(\mathbf{x}-\mathbf{u}) \partial^i_{\mathbf{x}}\right] \Gamma(\mathbf{x}, \mathbf{y}, \mathbf{u}) \notag \\
&+ \left[\partial_{\mathbf{x}}^i L(\mathbf{x}-\mathbf{v})+ L(\mathbf{y}-\mathbf{v}) \partial^i_{\mathbf{x}} - L(\mathbf{x}-\mathbf{v}) \partial^i_{\mathbf{x}}\right] \Gamma(\mathbf{x}, \mathbf{y}, \mathbf{v}) \notag\\
&+ \left[\partial^i_{\mathbf{x}}L(\mathbf{x}-\mathbf{u}) - L(\mathbf{x}-\mathbf{u})\partial^i_{\mathbf{x}}\right]\Gamma(\mathbf{x}, \mathbf{u}, \mathbf{v})  +  \left[\partial^i_{\mathbf{x}}L(\mathbf{x}-\mathbf{v}) - L(\mathbf{x}-\mathbf{v})\partial^i_{\mathbf{x}}\right]\Gamma(\mathbf{x}, \mathbf{v}, \mathbf{u}) \Big].
\end{align}
Combining eq.~\eqref{eq:I3_final_res}
 and  eq.~\eqref{eq:final_result_xyuv}, 
one gets 
\begin{align}\label{eq:I3+I4}
\mathcal{I}_3 + \mathcal{I}_4 
=&\frac{g^6N_c^3}{12s}(L^-\mu_0^2)^3 \Big[F(\mathbf{x}, \mathbf{y},\mathbf{u}, \mathbf{v}) \partial^i_{\mathbf{x}}G(\mathbf{x}, \mathbf{y}, \mathbf{u}, \mathbf{v}) -\partial^i_{\mathbf{x}}F(\mathbf{x}, \mathbf{y},\mathbf{u}, \mathbf{v}) G(\mathbf{x}, \mathbf{y}, \mathbf{u}, \mathbf{v})  \Big].
\end{align}
Recall the definition
\begin{align}
F(\mathbf{x}, \mathbf{y}, \mathbf{u}, \mathbf{v})
 = L(\mathbf{x}-\mathbf{u}) +L(\mathbf{y}-\mathbf{v})-L(\mathbf{x}-\mathbf{v}) -L(\mathbf{y}-\mathbf{u})
\end{align}
and we have additionally introduced 
\begin{align}
G(\mathbf{x}, \mathbf{y}, \mathbf{u}, \mathbf{v})= \Gamma(\mathbf{x},\mathbf{y},\mathbf{u}) - \Gamma(\mathbf{x}, \mathbf{y}, \mathbf{v}) - \Gamma(\mathbf{y}, \mathbf{u}, \mathbf{v}) +\Gamma(\mathbf{x}, \mathbf{u}, \mathbf{v}). 
\end{align}
A simple sanity check of eq.~\eqref{eq:I3+I4} is that it should vanish when $\mathbf{x}=\mathbf{y}$ or $\mathbf{u}=\mathbf{v}$. 

Summarizing the results in eq.~\eqref{eq:qqbar_qqbar_correlator_MV} and  eq.~\eqref{eq:I3+I4}, one gets
\begin{subequations}\label{eq:final_Qi_exp}
\begin{align}
\mathcal{Q}^i(\mathbf{x}, \mathbf{y}, \mathbf{u}, \mathbf{v}) 
 &\simeq \frac{1}{s} \, 2Q_0^4\, \epsilon^{il}\partial^l_{\mathbf{x}} \Big[ F^2(\mathbf{x}, \mathbf{y}, \mathbf{u}, \mathbf{v})\Big] \label{eq:ave_Q04_direct}\\
& + \frac{1}{s} 2Q_0^6\, \epsilon^{il}\partial^l_{\mathbf{x}} \left[ F^2(\mathbf{x}, \mathbf{y}, \mathbf{u}, \mathbf{v})\left(\Gamma_1 - \frac{2}{3} F(\mathbf{x}, \mathbf{u}, \mathbf{y}, \mathbf{v})\right)\right]     \label{eq:ave_Q06_direct}\\
& + \frac{1}{s} \frac{16}{3} Q_0^6  \Big[ F(\mathbf{x}, \mathbf{y},\mathbf{u}, \mathbf{v}) \partial^i_{\mathbf{x}}G(\mathbf{x}, \mathbf{y}, \mathbf{u}, \mathbf{v}) -\partial^i_{\mathbf{x}}F(\mathbf{x}, \mathbf{y},\mathbf{u}, \mathbf{v}) G(\mathbf{x}, \mathbf{y}, \mathbf{u}, \mathbf{v})  \Big].\label{eq:ave_Q06_induced}
\end{align}
\end{subequations}
We only kept the leading terms in the large-$N_c$ limit, and the power series expansions in $Q_0$ are truncated at the order $\mathcal{O}(Q_0^6)$.  The resulting expression is strictly real-valued. The term in eq.~\eqref{eq:ave_Q04_direct} represents the leading order contribution originating from the direct helicity effect. At the next-to-leading order, there are two contributions. The term in eq.~\eqref{eq:ave_Q06_direct} comes from the direct helicity effect and the term in eq.~\eqref{eq:ave_Q06_induced} is attributed to the gluon saturation induced helicity effect.

\section{Double-Spin Asymmetries for Incoherent Diffractive Dijet Production}\label{sec:ALL_dijet}
We consider double-spin asymmetry for incoherent diffractive dijet production in longitudinally polarized electron-proton/nucleus collisions at high energies, as illustrated in Fig.~\ref{fig:dijet_DIS}. The corresponding cross section can be easily obtained from the calculations on double-spin asymmetry for elastic and inclusive dijet productions presented in \cite{Kovchegov:2024wjs, Kovchegov:2025}. Related calculations can also be found in \cite{Altinoluk:2022jkk, Altinoluk:2024zom, Bhattacharya:2022vvo,Bhattacharya:2024sck}.  Specifically, we focus on events where the longitudinal momentum fractions of the quark and antiquark, $z_1$ and $z_2$ are approximately equal, $z_1 \sim z_2 \simeq \frac{1}{2}$ . In this kinematic region, the contributions from the background field $F^{ij}$ vanishes, leaving only those contributions from the background field $F^{i+}$. Additionally, we assume a purely gluonic background, neglecting contributions from  background quark fields. 
Up to an unimportant overall constant factor, we investigate the following correlation function 
\begin{equation}\label{eq:Cp1p2_def}
\begin{split}
C(\mathbf{p}_1, \mathbf{p}_2) = &\int_{\mathbf{k}_1, \mathbf{k}'_1} \frac{ - i\mathbf{k}_1\times \mathbf{k}'_1}{(\mathbf{k}_1^2+a_f^2)(\mathbf{k}^{\prime 2}_1 + a_f^2)} \int_{\mathbf{x}, \mathbf{y}, \mathbf{u}, \mathbf{v}} e^{-i(\mathbf{p}_1-\mathbf{k}_1)\cdot\mathbf{x} - i(\mathbf{p}_2+\mathbf{k}_1)\cdot\mathbf{y} +i(\mathbf{p}_1-\mathbf{k}'_1)\cdot\mathbf{v}+i(\mathbf{p}_2+\mathbf{k}'_1)\cdot\mathbf{u}}\\
&\times \Big[(\mathbf{p}_1 + \mathbf{k}_1)^i \mathcal{Q}^i(\mathbf{x}, \mathbf{y}, \mathbf{u}, \mathbf{v})  - (-\mathbf{p}_2+\mathbf{k}_1)^i \mathcal{Q}^{i \dagger}(\mathbf{y}, \mathbf{x}, \mathbf{v}, \mathbf{u})\Big] +c.c.
\end{split}
\end{equation}
Here $\mathbf{p}_1$ and $\mathbf{p}_2$ are the final transverse momenta of the quark jet and the antiquark jet, respectively. The $a_f^2 = z_1z_2 Q^2$ with the photon virtuality being $Q^2$. One might also recognize that $\Phi^j(\mathbf{k}_1) = \mathbf{k}_1^j/(\mathbf{k}_1^2+a_f^2)$ is the momentum space  splitting wavefunction for transversely-polarized photon.   The definition of $\mathcal{Q}^i(\mathbf{x}, \mathbf{y}, \mathbf{u}, \mathbf{v})$ was given in eq.~\eqref{eq:Qi_def}. Its expression in the quasi-classical approximation is given in eq.~\eqref{eq:final_Qi_exp}. In the following, we substitute eq.~\eqref{eq:final_Qi_exp} into eq.~\eqref{eq:Cp1p2_def} and evaluate the gluon saturation induced helicity effect as compared to the direct helicity effect.  
\begin{figure*}
    \includegraphics[width=.7\textwidth]{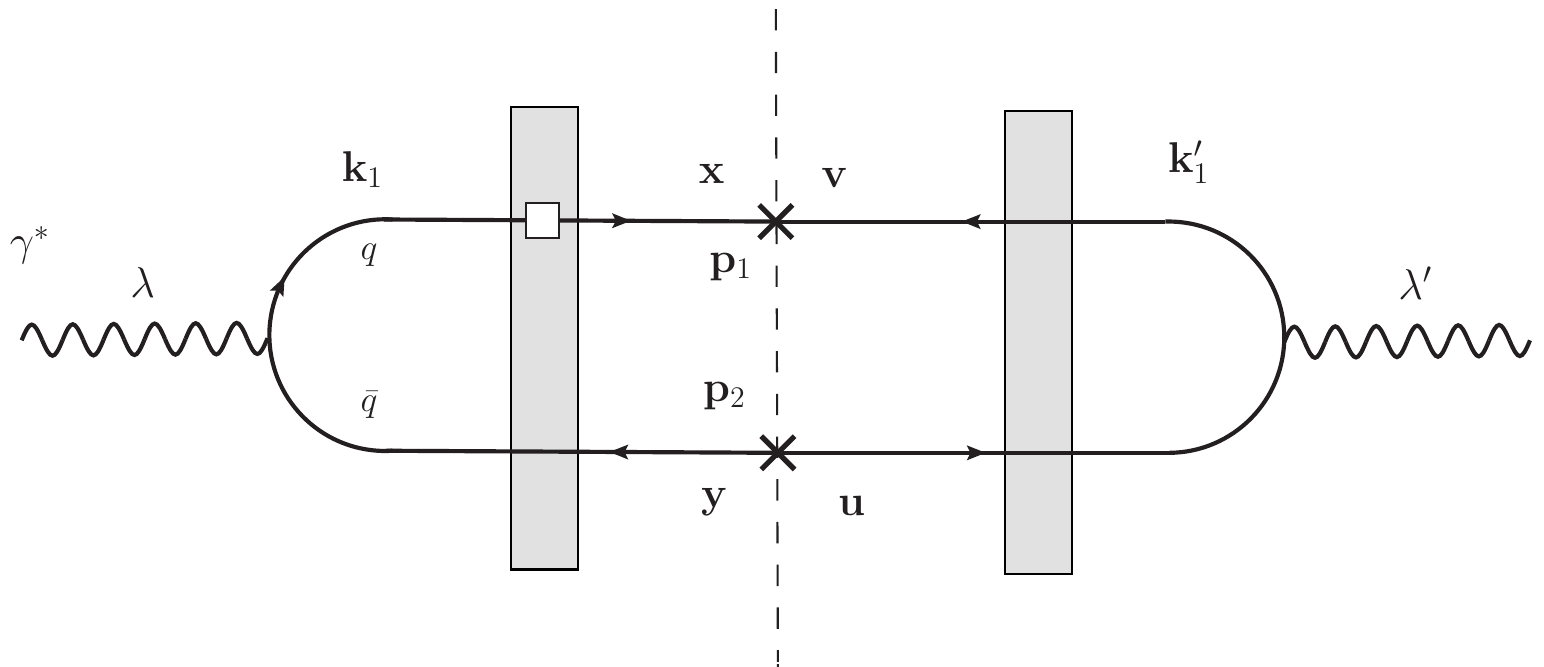}
    \caption{Schematic diagram for quark-antiquark dijet production in longitudinally polarized electron-proton/nucleus collisions. The white box represents sub-eikonal interaction with the nuclear shockwave.}
\label{fig:dijet_DIS}		
\end{figure*} 

\subsection{Direct Helicity Effect in $C(\mathbf{p}_1, \mathbf{p}_2)$}
Direct helicity effect comes from the two terms in eq.~\eqref{eq:ave_Q04_direct} and eq.~\eqref{eq:ave_Q06_direct}. Substituting eqs.~\eqref{eq:ave_Q04_direct} and ~\eqref{eq:ave_Q06_direct} into eq.~\eqref{eq:Cp1p2_def}, performing integration by parts, one obtains
\begin{equation}\label{eq:Cdir_p1p2_exp}
\begin{split}
C_{\mathrm{dir}}(\mathbf{p}_1, \mathbf{p}_2) 
=&\frac{1}{s} \int_{\mathbf{k}_1, \mathbf{k}'_1} \frac{ \mathbf{k}_1\times \mathbf{k}'_1}{(\mathbf{k}_1^2+a_f^2)(\mathbf{k}^{\prime 2}_1 + a_f^2)} \int_{\mathbf{x}, \mathbf{y}, \mathbf{u}, \mathbf{v}} e^{-i(\mathbf{p}_1-\mathbf{k}_1)\cdot\mathbf{x} - i(\mathbf{p}_2+\mathbf{k}_1)\cdot\mathbf{y} +i(\mathbf{p}_1-\mathbf{k}'_1)\cdot\mathbf{v}+i(\mathbf{p}_2+\mathbf{k}'_1)\cdot\mathbf{u}}\\
&\times \Big[2(\mathbf{p}_1-\mathbf{p}_2)\times \mathbf{k}_1\Big]\mathcal{Q}(\mathbf{x}, \mathbf{y}, \mathbf{u}, \mathbf{v}) +c.c.\\
\end{split}
\end{equation}
with the unpolarized quadrupole in the large-$N_c$ limit 
\begin{equation}\label{eq:Qxyuv_up_Q6}
\mathcal{Q}(\mathbf{x}, \mathbf{y}, \mathbf{u},\mathbf{v}) \simeq 2 Q_0^4 F^2(\mathbf{x}, \mathbf{y}, \mathbf{u}, \mathbf{v}) + 2 Q_0^6 F^2(\mathbf{x}, \mathbf{y}, \mathbf{u}, \mathbf{v}) \left( \Gamma_1 - \frac{2}{3} F(\mathbf{x}, \mathbf{u}, \mathbf{y}, \mathbf{v})\right). 
\end{equation}
For the terms at leading order $\mathcal{O}(Q_0^4)$, we need to compute Fourier transformations of $F^2(\mathbf{x}, \mathbf{y}, \mathbf{u}, \mathbf{v}) $. It is a straightforward calculation to obtain
 \begin{equation}\label{eq:Cdir0_final_2}
\begin{split}
C_{\mathrm{dir}}^{(0)}(\mathbf{p}_1, \mathbf{p}_2) 
=& \frac{8}{ s} Q_0^4S_{\perp}\int_{\mathbf{p}} \frac{1}{\mathbf{p}^4 (\mathbf{p}_1+\mathbf{p}_2+\mathbf{p})^4}\epsilon^{ij}\Big[h^i(\mathbf{p}_1) - h^i(\mathbf{p}_1+\mathbf{p}) + h^i(\mathbf{p}_2) - h^i(\mathbf{p}_2+\mathbf{p})\Big]\\
&\qquad \times \Big[\Phi^j(\mathbf{p}_1)- \Phi^j(\mathbf{p}_1+\mathbf{p})- \Phi^j(\mathbf{p}_2) + \Phi^j(\mathbf{p}_2+\mathbf{p})\Big]\\
=& \frac{8}{ s} Q_0^4S_{\perp}\int_{\mathbf{p}} \frac{1}{\mathbf{p}^4 (\mathbf{p}_1+\mathbf{p}_2+\mathbf{p})^4} \,\epsilon^{ij} \, \Psi^i_I(\mathbf{p}_1, \mathbf{p}_2, \mathbf{p}) \Psi^j_F(\mathbf{p}_1, \mathbf{p}_2, \mathbf{p})
\end{split}
\end{equation}
The area is defined as $S_{\perp} = (2\pi)^2 \delta (\mathbf{0})$.
To express the correlation function more compactly and also for the purpose of numerically evaluating the momentum integrals, we introduce the auxiliary function
 \begin{equation}
 h^i(\mathbf{k}_1) = \Phi^i(\mathbf{k}_1) [(\mathbf{p}_1-\mathbf{p}_2)\times \mathbf{k}_1] .
 \end{equation}
In the second equality in eq.~\eqref{eq:Cdir0_final_2},
the following  auxiliary functions are introduced
\begin{equation}
\Psi^j_F(\mathbf{k}_1, \mathbf{k}_2, \mathbf{p}) = \Big[\Phi^j(\mathbf{k}_1)- \Phi^j(\mathbf{k}_1+\mathbf{p})- \Phi^j(\mathbf{k}_2) + \Phi^j(\mathbf{k}_2+\mathbf{p})\Big]
\end{equation}
and 
\begin{equation}
\Psi^i_I(\mathbf{k}_1, \mathbf{k}_2, \mathbf{p}) =  \Big[h^i(\mathbf{k}_1) - h^i(\mathbf{k}_1+\mathbf{p}) + h^i(\mathbf{k}_2) - h^i(\mathbf{k}_2+\mathbf{p})\Big].
\end{equation}
Eq.~\eqref{eq:Cdir0_final_2} is symmetric with respect to $\mathbf{p}_1\leftrightarrow \mathbf{p}_2$. One could further analyze the infrared poles at $\mathbf{p}=0$ and $\mathbf{p}=-\mathbf{p}_1-\mathbf{p}_2$ for the integral over $\mathbf{p}$ in eq.~\eqref{eq:Cdir0_final_2}. It turns out that $C^{(0)}_{\mathrm{dir}}(\mathbf{p}_1, \mathbf{p}_2) \sim 1/(\mathbf{p}_1+\mathbf{p}_2)^2$ around these two poles, indicating that the infrared dependence is quadratic. This observation is not new, similar quadratic infrared sensitivity has also been observed in two gluon correlations in unpolarized collisions \cite{Kovchegov:2012nd}. 

Next, we consider the terms at the sub-leading order $\mathcal{O}(Q_0^6)$ in eq.~\eqref{eq:Qxyuv_up_Q6}.  The final result is
\begin{equation}\label{eq:Cdir(1)_full}
\begin{split}
 C_{\mathrm{dir}}^{(1)}(\mathbf{p}_1, \mathbf{p}_2) 
=&\frac{32}{3 s} Q_0^6S_{\perp}\int_{\mathbf{p}, \mathbf{k}} \frac{1}{\mathbf{p}^4(\mathbf{p}_1+\mathbf{p}_2+\mathbf{p})^4 \mathbf{k}^4}\epsilon^{ij}\Big[ \Psi_I^i(\mathbf{p}_1-\mathbf{k}, \mathbf{p}_2+\mathbf{k}, \mathbf{p}) \Psi_F^j(\mathbf{p}_1, \mathbf{p}_2, \mathbf{p})  \\
&\quad + \Psi_I^i(\mathbf{p}_1, \mathbf{p}_2, \mathbf{p}) \Psi_F^j(\mathbf{p}_1+\mathbf{k}, \mathbf{p}_2-\mathbf{k}, \mathbf{p}) - 3\Psi_I^i(\mathbf{p}_1, \mathbf{p}_2, \mathbf{p}) \Psi_F^j(\mathbf{p}_1, \mathbf{p}_2, \mathbf{p})\Big]\\
&+\frac{16}{3 s} Q_0^6S_{\perp}\int_{\mathbf{p}, \mathbf{k}} \frac{1}{\mathbf{p}^4(\mathbf{p}_1+\mathbf{p}_2+\mathbf{p}+\mathbf{k})^4 \mathbf{k}^4} \epsilon^{ij}\Big[ \, \Psi_I^i(\mathbf{p}_1+\mathbf{k}, \mathbf{p}_2, \mathbf{p}) \Psi_F^j(\mathbf{p}_1+\mathbf{k}, \mathbf{p}_2, \mathbf{p})  \\
&\quad  +  \Psi_I^i(\mathbf{p}_1, \mathbf{p}_2+\mathbf{k}, \mathbf{p}) \Psi_F^j(\mathbf{p}_1, \mathbf{p}_2+\mathbf{k}, \mathbf{p})\Big].
\end{split}
\end{equation}
The pole structures of the integrals over $\mathbf{p}, \mathbf{k}$ could be analyzed analytically, though the calculations are quite tedious.  Instead, we will compute these integral numerically  and check their dependence on the IR cut-off. Eqs.~\eqref{eq:Cdir0_final_2} and ~\eqref{eq:Cdir(1)_full} represent the primary results of this subsection.

\subsection{Gluon Saturation Induced Helicity Effect in $C(\mathbf{p}_1, \mathbf{p}_2)$ }
In the quasi-classical approximation, using the MV and helicity-extended MV models, the averaging over polarized Wilson line correlators yields a real-valued function. Notably, the second term in eq.~\eqref{eq:Cdir_p1p2_exp} can be related to the first term in eq.~\eqref{eq:Cdir_p1p2_exp} by exchanging $\mathbf{p}_1$ and $ \mathbf{p}_2$. For simplicity, we focus on the first term in eq.~\eqref{eq:Cdir_p1p2_exp}. 
Substituting eq.~\eqref{eq:ave_Q06_induced} into the first term in eq.~\eqref{eq:Cdir_p1p2_exp}, one obtains 
\begin{equation}\label{eq:c_p1p2_exp}
\begin{split}
C^{(1)}_{\mathrm{ind}}(\mathbf{p}_1, \mathbf{p}_2)
 =&\frac{1}{s} \frac{16}{3} Q_0^6  \int_{\mathbf{k}_1, \mathbf{k}'_1} \frac{ - i\mathbf{k}_1\times \mathbf{k}'_1}{(\mathbf{k}_1^2+a_f^2)(\mathbf{k}^{\prime 2}_1 + a_f^2)} (\mathbf{p}_1+\mathbf{k}_1)^i  \int_{\mathbf{x}, \mathbf{y}, \mathbf{u}, \mathbf{v}} e^{-i(\mathbf{p}_1-\mathbf{k}_1)\cdot\mathbf{x} - i(\mathbf{p}_2+\mathbf{k}_1)\cdot\mathbf{y}+i(\mathbf{p}_1-\mathbf{k}'_1)\cdot\mathbf{v}+i(\mathbf{p}_2+\mathbf{k}'_1)\cdot\mathbf{u}} \\
 &\qquad \times \Big[(-i(\mathbf{p}_1-\mathbf{k}_1)^i)F(\mathbf{x}, \mathbf{y},\mathbf{u}, \mathbf{v}) G(\mathbf{x}, \mathbf{y}, \mathbf{u}, \mathbf{v}) -2\partial^i_{\mathbf{x}}F(\mathbf{x}, \mathbf{y},\mathbf{u}, \mathbf{v}) G(\mathbf{x}, \mathbf{y}, \mathbf{u}, \mathbf{v})   \Big] \\
  \end{split}
\end{equation}
Here, we demonstrate that the first term within the square bracket in eq.~\eqref{eq:c_p1p2_exp} vanishes.
The explicit expression for this term is given by
\begin{equation}\label{eq:I_functions_simplify}
\begin{split}
& \frac{16}{3s} Q_0^6\int_{\mathbf{p}, \mathbf{q}}\frac{\mathbf{p}\times \mathbf{q}}{\mathbf{p}^4\mathbf{q}^4 (\mathbf{p}+\mathbf{q})^4}\Bigg\{\Big[f^j(\mathbf{p}_1+\mathbf{p})-f^j(\mathbf{p}_2+\mathbf{p})+f^j(\mathbf{p}_2) -f^j(\mathbf{p}_1) \Big]\Psi_F^j(\mathbf{p}_1-\mathbf{q}, \mathbf{p}_2-\mathbf{q}, \mathbf{p}+2\mathbf{q})\, h(\mathbf{p}_1+\mathbf{p}_2+\mathbf{p})\\
&+\Big[f^j(\mathbf{p}_1+\mathbf{p}+\mathbf{q}) - f^j(\mathbf{p}_1-\mathbf{q}) + f^j(\mathbf{p}_2-\mathbf{q}) - f^j(\mathbf{p}+\mathbf{q}+\mathbf{p}_2)\Big]\Psi^j_F(\mathbf{p}_1, \mathbf{p}_2, \mathbf{p})\, h(\mathbf{p}_1+\mathbf{p}_2+\mathbf{p})\\
&+\Big[f^j(\mathbf{p}+\mathbf{q}+\mathbf{p}_2) + f^j(\mathbf{p}_1)-f^j(\mathbf{p}_2) - f^j(\mathbf{p}_1+\mathbf{p}+\mathbf{q})\Big] \Big[\Phi^j(\mathbf{p}+\mathbf{p}_2) + \Phi^j(\mathbf{p}_1+\mathbf{q})\Big]\, h(\mathbf{p}_1+\mathbf{p}_2+\mathbf{p}+\mathbf{q})\\
&-\Big[f^j(\mathbf{p}_1+\mathbf{p}) + f^j(\mathbf{p}_2+\mathbf{q})\Big]\Psi_F^j(\mathbf{p}_1, \mathbf{p}_2, \mathbf{p}+\mathbf{q})\, h(\mathbf{p}_1+\mathbf{p}_2+\mathbf{p}+\mathbf{q})\Bigg\}
\end{split}
\end{equation}
We  introduced another two auxiliary functions 
\begin{equation}
f^{j}(\mathbf{k}_1) = \epsilon^{ij} \Phi^i(\mathbf{k}_1) (\mathbf{p}_1^2-\mathbf{k}_1^2), \qquad  h(\mathbf{k}) = \frac{1}{\mathbf{k}^4}.
\end{equation}
The $f^j(\mathbf{k}_1)$ can be rewritten as
\begin{equation}
f^{j}(\mathbf{k}_1)  = (\mathbf{p}_1^2+a_f^2) \epsilon^{ij} \Phi^i(\mathbf{k}_1) - \epsilon^{ij} \mathbf{k}_1^i.
\end{equation}
Using this expression, the various terms involving the $f^j(\mathbf{k}_1)$ function in eq.~\eqref{eq:I_functions_simplify} can be simplified as
\begin{align}\label{eq:fs_cancel1}
\Big[f^j(\mathbf{p}_1+\mathbf{p})-f^j(\mathbf{p}_2+\mathbf{p})+f^j(\mathbf{p}_2) -f^j(\mathbf{p}_1) \Big] =-(\mathbf{p}_1^2+a_f^2) \epsilon^{ij} \Psi_F^i(\mathbf{p}_1, \mathbf{p}_2, \mathbf{p}),
\end{align} 
\begin{align}\label{eq:fs_cancel2}
\Big[f^j(\mathbf{p}_1+\mathbf{p}+\mathbf{q}) - f^j(\mathbf{p}_1-\mathbf{q}) + f^j(\mathbf{p}_2-\mathbf{q}) - f^j(\mathbf{p}+\mathbf{q}+\mathbf{p}_2)\Big] =-(\mathbf{p}_1^2+a_f^2)\epsilon^{ij} \Psi_F^i(\mathbf{p}_1-\mathbf{q}, \mathbf{p}_2-\mathbf{q}, \mathbf{p}+2\mathbf{q})
\end{align}
and 
\begin{align}
\Big[f^j(\mathbf{p}+\mathbf{q}+\mathbf{p}_2) + f^j(\mathbf{p}_1)-f^j(\mathbf{p}_2) - f^j(\mathbf{p}_1+\mathbf{p}+\mathbf{q})\Big] =(\mathbf{p}_1^2+a_f^2)\epsilon^{ij}\Psi_F^i(\mathbf{p}_1, \mathbf{p}_2, \mathbf{p}+\mathbf{q}).
\end{align}
Substituting eqs.~\eqref{eq:fs_cancel1} and ~\eqref{eq:fs_cancel2} into eq.~\eqref{eq:I_functions_simplify}, the first two terms in eq.~\eqref{eq:I_functions_simplify} cancel each other. 
The last two terms in the curly bracket in eq.~\eqref{eq:I_functions_simplify} can be written as
\begin{equation}
\begin{split}
&\Big((\mathbf{p}_1^2+a_f^2)\epsilon^{ij} \Psi_F^i(\mathbf{p}_1, \mathbf{p}_2, \mathbf{p}+\mathbf{q})\Big[g^j(\mathbf{p}+\mathbf{p}_2) + g^j(\mathbf{p}_1+\mathbf{q})\Big]\\
&-((\mathbf{p}_1^2+a_f^2)\epsilon^{ij}\Big[g^i(\mathbf{p}_1+\mathbf{p}) + g^i(\mathbf{p}_2+\mathbf{q})\Big] -\epsilon^{ij} \left[\mathbf{p}_1+\mathbf{p}_2+\mathbf{p}+\mathbf{q}\right]^i)\Psi_F^j(\mathbf{p}_1, \mathbf{p}_2, \mathbf{p}+\mathbf{q})\Big)\\
&\qquad \times h(\mathbf{p}_1+\mathbf{p}_2+\mathbf{p}+\mathbf{q})\\
= &\Big(-(\mathbf{p}_1^2+a_f^2)\epsilon^{ij} \Big[g^i(\mathbf{p}_1+\mathbf{p}) + g^i(\mathbf{p}_2+\mathbf{q})+g^i(\mathbf{p}+\mathbf{p}_2) + g^i(\mathbf{p}_1+\mathbf{q})\Big] +\epsilon^{ij}[\mathbf{p}_1+\mathbf{p}_2+\mathbf{p}+\mathbf{q}]^i\Big)\\
&\qquad \times \Psi_F^j(\mathbf{p}_1, \mathbf{p}_2, \mathbf{p}+\mathbf{q})\, h(\mathbf{p}_1+\mathbf{p}_2+\mathbf{p}+\mathbf{q})\\
\end{split}
\end{equation}
The above expression is symmetric with respect to $\mathbf{p}\leftrightarrow \mathbf{q}$. However, in the integrand in eq.~\eqref{eq:I_functions_simplify}
, we also have a factor $\mathbf{p}\times \mathbf{q}$, which is antisymmetric wrt $\mathbf{p}\leftrightarrow \mathbf{q}$. As a result, the integral in eq.~\eqref{eq:I_functions_simplify}
 vanishes.

Computing the second term in eq.~\eqref{eq:c_p1p2_exp}, the final result for the correlation function from gluon saturation induced helicity effects is
\begin{equation}\label{eq:Cind_final}
\begin{split}
C^{(1)}_{\mathrm{ind}}(\mathbf{p}_1, \mathbf{p}_2) =& \frac{64}{3s} Q_0^6 S_{\perp}\int_{\mathbf{p}, \mathbf{q}} \frac{\mathbf{p}\times \mathbf{q}}{\mathbf{p}^4 \mathbf{q}^4(\mathbf{p}+\mathbf{q})^2}\left[\frac{\mathcal{K}_1(\mathbf{p}, \mathbf{q})} {(\mathbf{p}_1+\mathbf{p}_2+\mathbf{p})^4} + \frac{\mathcal{K}_2(\mathbf{p}, \mathbf{q})}{ (\mathbf{p}_1+\mathbf{p}_2+\mathbf{p}+\mathbf{q})^4} \right] \\
\end{split}
\end{equation}
with
\begin{equation}
\begin{split}
\mathcal{K}_1(\mathbf{p}, \mathbf{q})
  = &\epsilon^{ij}\Big[-g^i(\mathbf{p}_2-\mathbf{q}, \mathbf{p}, \mathbf{p}_1, \mathbf{p}_2) + g^i(\mathbf{p}_2+\mathbf{p}+\mathbf{q}, \mathbf{p}, \mathbf{p}_1, \mathbf{p}_2)\Big] \Psi^j_F(\mathbf{p}_1, \mathbf{p}_2, \mathbf{p})\\
&+\epsilon^{ij}\Big[ g^i(\mathbf{p}+\mathbf{p}_2, \mathbf{p}, \mathbf{p}_1, \mathbf{p}_2)- g^i(\mathbf{p}_2, \mathbf{p}, \mathbf{p}_1, \mathbf{p}_2)\Big] \Psi^j_F(\mathbf{p}_1-\mathbf{q}, \mathbf{p}_2-\mathbf{q}, \mathbf{p}+2\mathbf{q})\\
&+\{\mathbf{p}_1\leftrightarrow \mathbf{p}_2\},  
\end{split}
\end{equation}
and 
\begin{equation}
\begin{split}
\mathcal{K}_2(\mathbf{p}, \mathbf{q})
 = &\epsilon^{ij} \, g^i(\mathbf{p}_2+\mathbf{q}, \mathbf{p}+\mathbf{q}, \mathbf{p}_1, \mathbf{p}_2)\, \Psi_F^j(\mathbf{p}_1, \mathbf{p}_2, \mathbf{p}+\mathbf{q}) \\
&+\frac{1}{2}\epsilon^{ij}\Big[ g^i(\mathbf{p}_2, \mathbf{p}+\mathbf{q}, \mathbf{p}_1, \mathbf{p}_2) - g^i(\mathbf{p}_2+\mathbf{p}+\mathbf{q}, \mathbf{p}+\mathbf{q}, \mathbf{p}_1, \mathbf{p}_2)\Big]\Psi_F^j(\mathbf{p}_1+\mathbf{q}, \mathbf{p}_2+\mathbf{q}, \mathbf{p}-\mathbf{q})\\
&+\{\mathbf{p}_1\leftrightarrow \mathbf{p}_2\}. 
\end{split}
\end{equation}
Here $\{\mathbf{p}_1\leftrightarrow \mathbf{p}_2\}$ represents the corresponding terms with the exchange $\mathbf{p}_1 \leftrightarrow \mathbf{p}_2$. 
We have introduced another auxiliary function 
\begin{equation}
g^i(\mathbf{k}, \mathbf{p}, \mathbf{p}_1, \mathbf{p}_2) =  \Big[(\mathbf{p}_1-\mathbf{k})\cdot(\mathbf{p}_1+\mathbf{p}_2+\mathbf{p})\Big]\, \Phi^i(\mathbf{k}).
\end{equation}
Eq.~\eqref{eq:Cind_final} is the main result of this subsection. It can be computed numerically using Monte Carlo numerical integrations. 
\end{widetext}

\subsection{Numerical Calculations}
To evaluate the effect of gluon saturation induced helicity-dependent field, we numerically compute the integrals in eq.~\eqref{eq:Cdir0_final_2}, eq.~\eqref{eq:Cdir(1)_full}
 and  eq.~\eqref{eq:Cind_final}. We use the CUBA library for multidimensional numerical integration \cite{Hahn:2004fe}. 

Before numerically evaluating these integrals, it is essential to address their sensitivity to the infrared (IR) cut-off. Fixed-order calculations using the MV model, where $\mu_0^2$ is treated as constant, are well-known to have power-law sensitivity to the IR cut-off \cite{Gelis:2009wh, Kovner:2018azs}. It is expected that all order resummation is needed to reduce the power-law divergence to logarithmical sensitivity. An explicit demonstration of how saturation effects can regularize the power-law IR divergence was given in \cite{Kovchegov:2013ewa}. In \cite{Dumitru:2008wn, Dusling:2009ar, Gelis:2009wh}, it was argued that the IR cut-off should be of the same order as the gluon saturation scale $Q_s$,  as this scale corresponds to the typical transverse momentum of saturated gluons. The IR cut-off is thus parameterized as $\Lambda^2_{IR} = \kappa Q_s^2$, where $\kappa <1 $ is a numerical constant. Additionally, quantum evolutions renders $\mu_0^2$  momentum-dependent, further regularizing IR behavior \cite{Iancu:2002aq}. In this paper, we adopt a simple regularization prescription, which is effectively to make the replacement in eq.~\eqref{eq:L(x)_exp},  as proposed in \cite{McLerran:2015sva}, 
\begin{equation}
L(\mathbf{q}) = \frac{1}{\mathbf{q}^4} \rightarrow \frac{1}{\mathbf{q}^2(\mathbf{q}^2+\Lambda_{IR}^2)}.
\end{equation}
It affects the ensemble averagings given in eq.~\eqref{eq:MV} and eq.~\eqref{eq:hMV}. 
We regularize the IR behavior using the physical gluon saturation scale. This prescription also guarantees color neutrality for the MV model. 

For the three integrals in  eq.~\eqref{eq:Cdir0_final_2}, eq.~\eqref{eq:Cdir(1)_full}
 and  eq.~\eqref{eq:Cind_final}, we factorize out the common factor $8 Q_0^4 S_{\perp}/s$. In the following, we plot the correlation function $C(\mathbf{p}_1, \mathbf{p}_2)$ as a function of the azimuthal angle $\Delta \phi$ between the two transverse momenta $\mathbf{p}_1, \mathbf{p}_2$. The magnitudes $|\mathbf{p}_1|, |\mathbf{p}_2|$ are treated  as input parameters. Additionally, we will also need to input the gluon saturation scale $Q_s$ and the photon virtuality $Q^2$. 
 
 \begin{figure}[!t]
    \includegraphics[width=.45\textwidth]{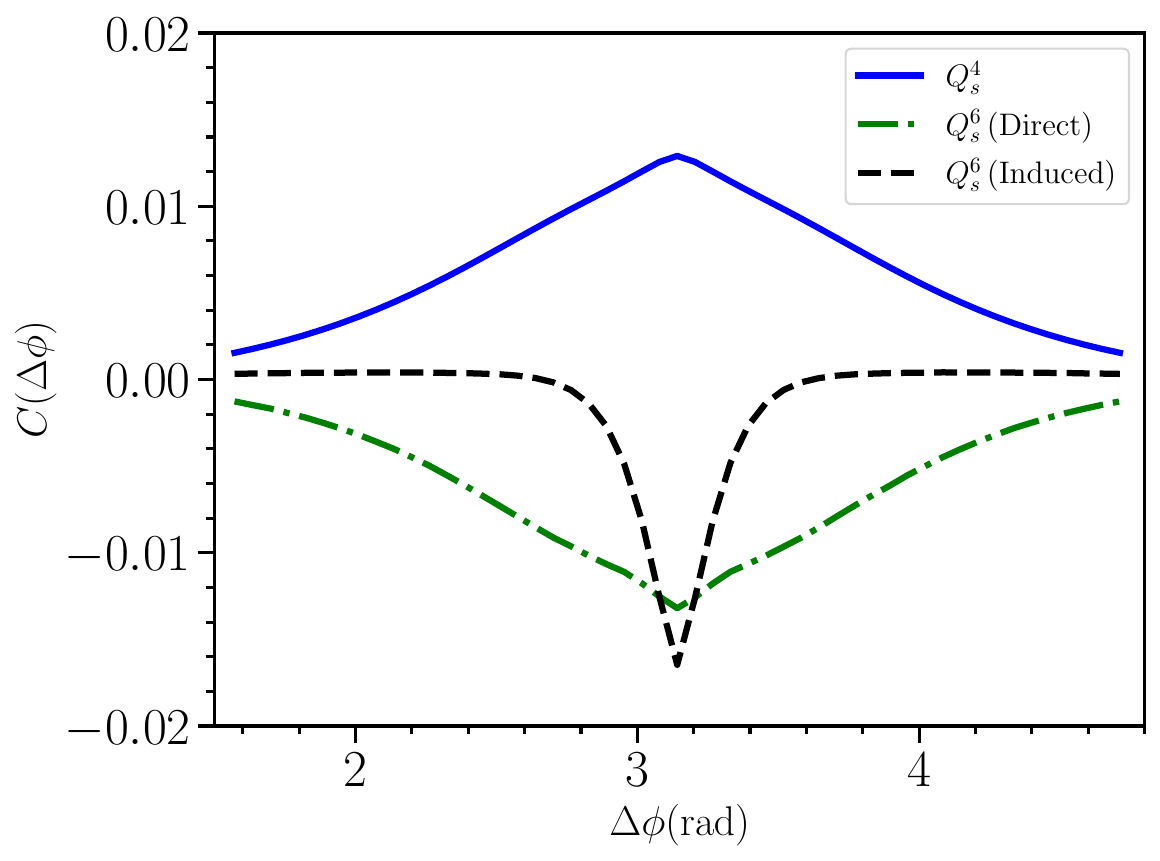}
    \caption{Comparison of the leading order $\mathcal{O}(Q_s^4)$ and sub-leading $\mathcal{O}(Q_s^6)$ contributions to the dijet momentum correlation as a function of the azimuthal angle $\Delta \phi$. Both the direct helicity effect and the gluon saturation induced helicity effect at the order $\mathcal{O}(Q_s^6)$ are plotted. }
\label{fig:dijet_Qs4vsQs6}		
\end{figure}

To illustrate the main properties of these three integrals, we choose the following representative values as inputs: $Q_s = 1.0 \, \mathrm{GeV}$, $Q^2 = 4.0\,\mathrm{GeV}^2$,  $|\mathbf{p}_1|=|\mathbf{p}_2| = 2.0 \,\mathrm{GeV}$ and the IR parameter $\kappa = 0.6$.  One plots the leading order $\mathcal{O}(Q_s^4)$ and the sub-leading order $\mathcal{O}(Q_s^6)$ contributions to the correlation function in the range $[\pi/2, 3\pi/2]$ in Fig.~\ref{fig:dijet_Qs4vsQs6}.  The leading order correlation function is positive, peaking at the back-to-back region, $\Delta \phi = \pi$, as expected. At the sub-leading order, both the direct helicity effect and the gluon saturation induced helicity effect are negative, with their peaks also located at $\Delta \phi = \pi$. Notably, their peak values are comparable (note that the y-axis is measured in units of $\mathrm{GeV}^{-4}$). Higher order terms (beyond the $\mathcal{O}(Q_s^4)$ order) are anticipated to smear the leading order back-to-back peak. However, quantitatively assessing the amount of smearing at the order $\mathcal{O}(Q_s^6)$ is challenging as it requires computing contributions from the order $\mathcal{O}(Q_s^8)$, which is expected to be positive. These series in powers of $Q_s$ alternate in signs.  What one can learn from Fig.~\ref{fig:dijet_Qs4vsQs6} is that, like the direct helicity effect, gluon saturation induced helicity effect suppresses the back-to-back peak. The suppression region for the gluon saturation induced helicity effect is much narrower around $\Delta \phi =\pi $ as compared to that of the direct helicity effect. 

\begin{figure*}
\centering
\begin{subfigure}{0.49\textwidth}
  \centering
  \includegraphics[width=0.9\textwidth]{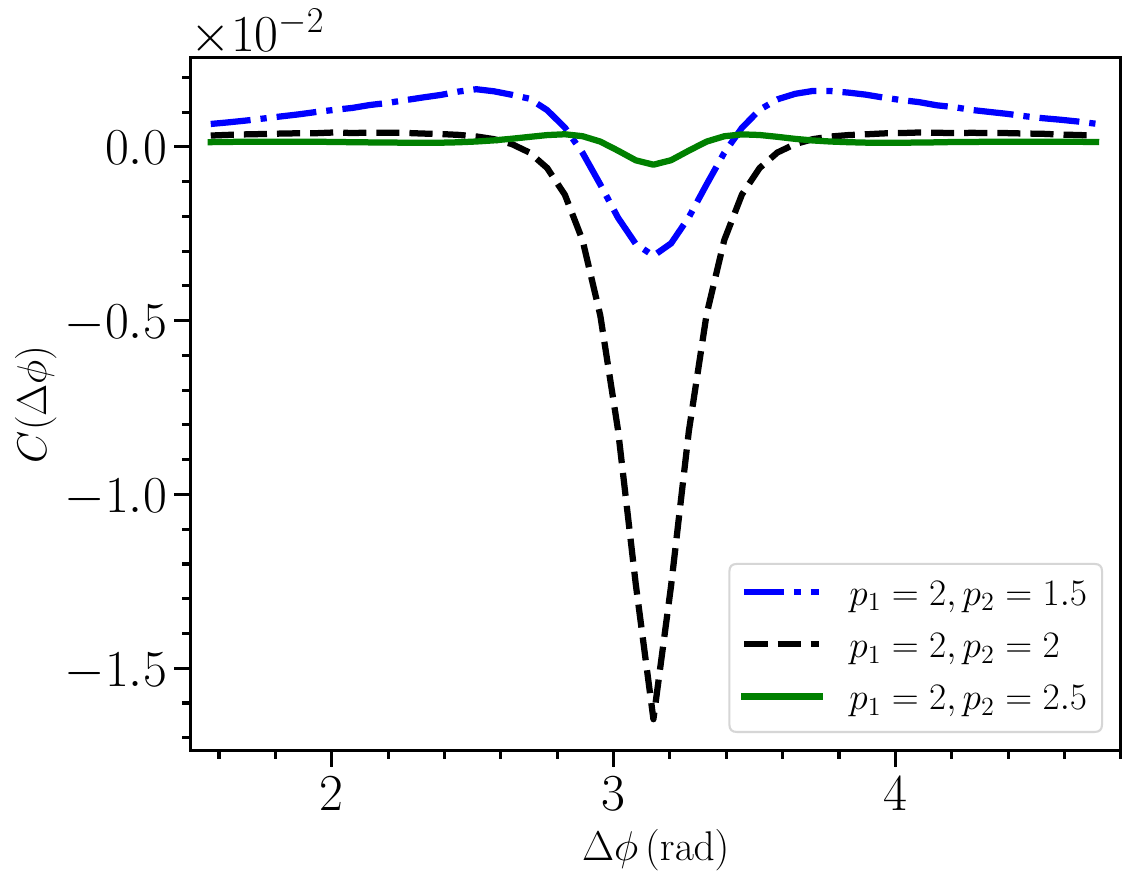}
  \caption{}
  \label{fig:induced_diff_p1p2}
\end{subfigure}
\begin{subfigure}{0.49\textwidth}
  \centering
  \includegraphics[width=0.9\textwidth]{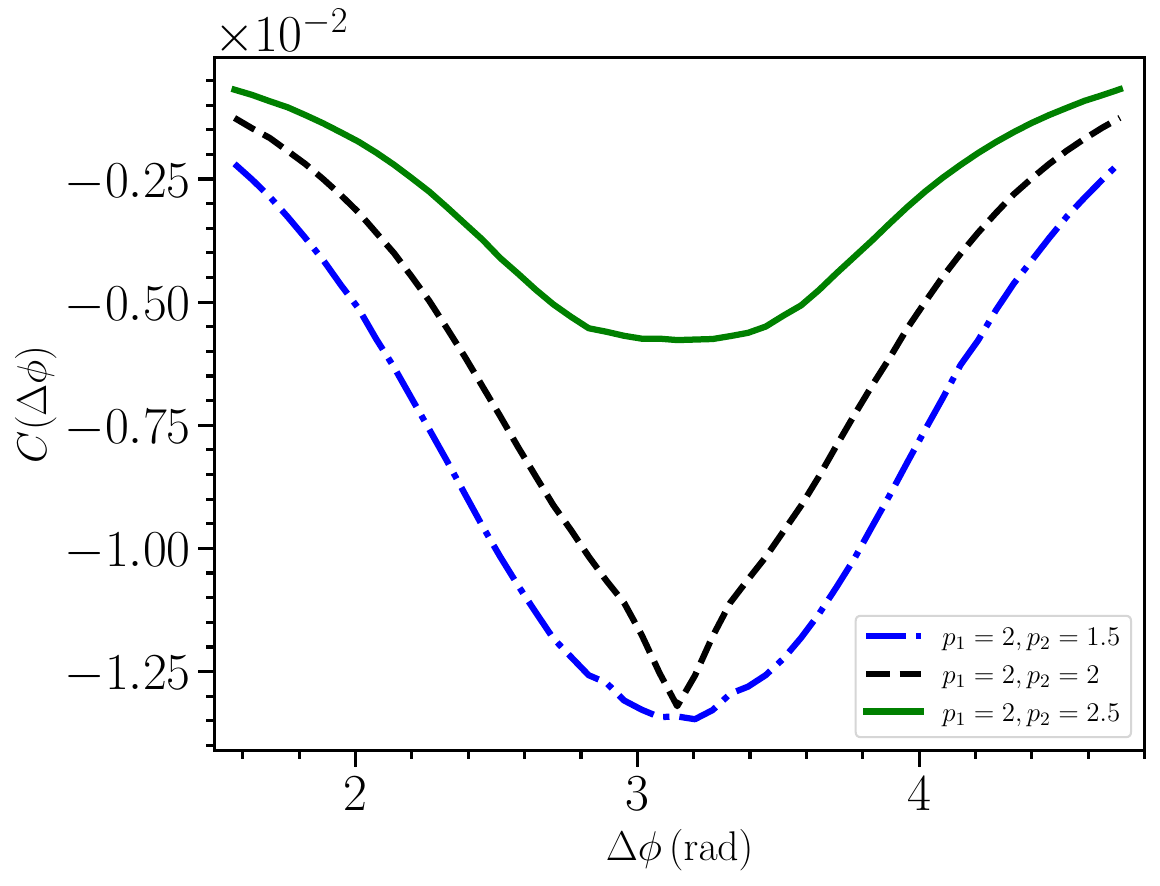}
  \caption{}
  \label{fig:direct_diff_p1p2}
\end{subfigure}
\caption{Correlation functions from (a) gluon saturation induced helicity effect and (b) direct helicity effect for different values of $\mathbf{p}_1, \mathbf{p}_2$. For fixed $|\mathbf{p}_1|=2\, \mathrm{GeV}$, $|\mathbf{p}_2| = 1.5, \, 2, \, 2.5\, \mathrm{GeV}$.}
\label{fig:corr_diff_p1p2}
\end{figure*}
\begin{figure*}
\centering
\begin{subfigure}{0.5\textwidth}
  \centering
  \includegraphics[width=0.9\textwidth]{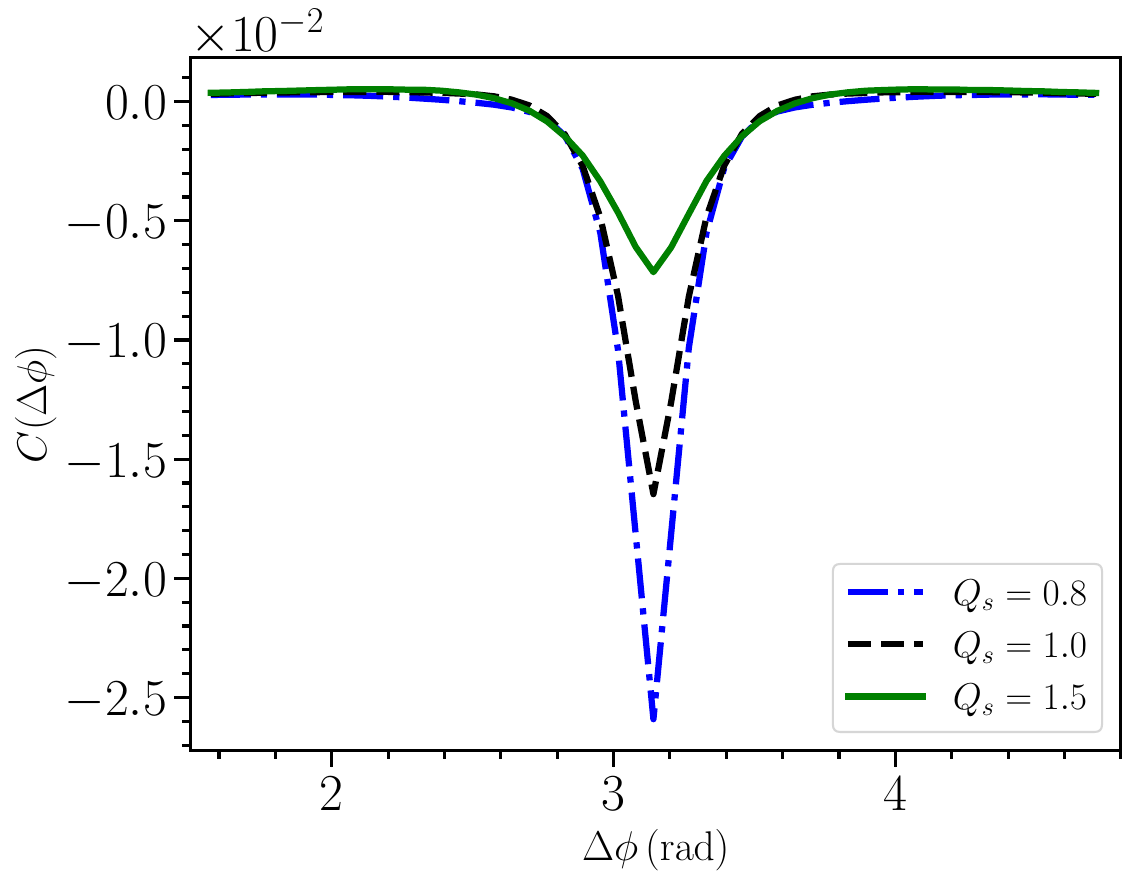}
  \caption{}
  \label{fig:induced_diff_Qs}
\end{subfigure}%
\begin{subfigure}{0.5\textwidth}
  \centering
  \includegraphics[width=0.9\textwidth]{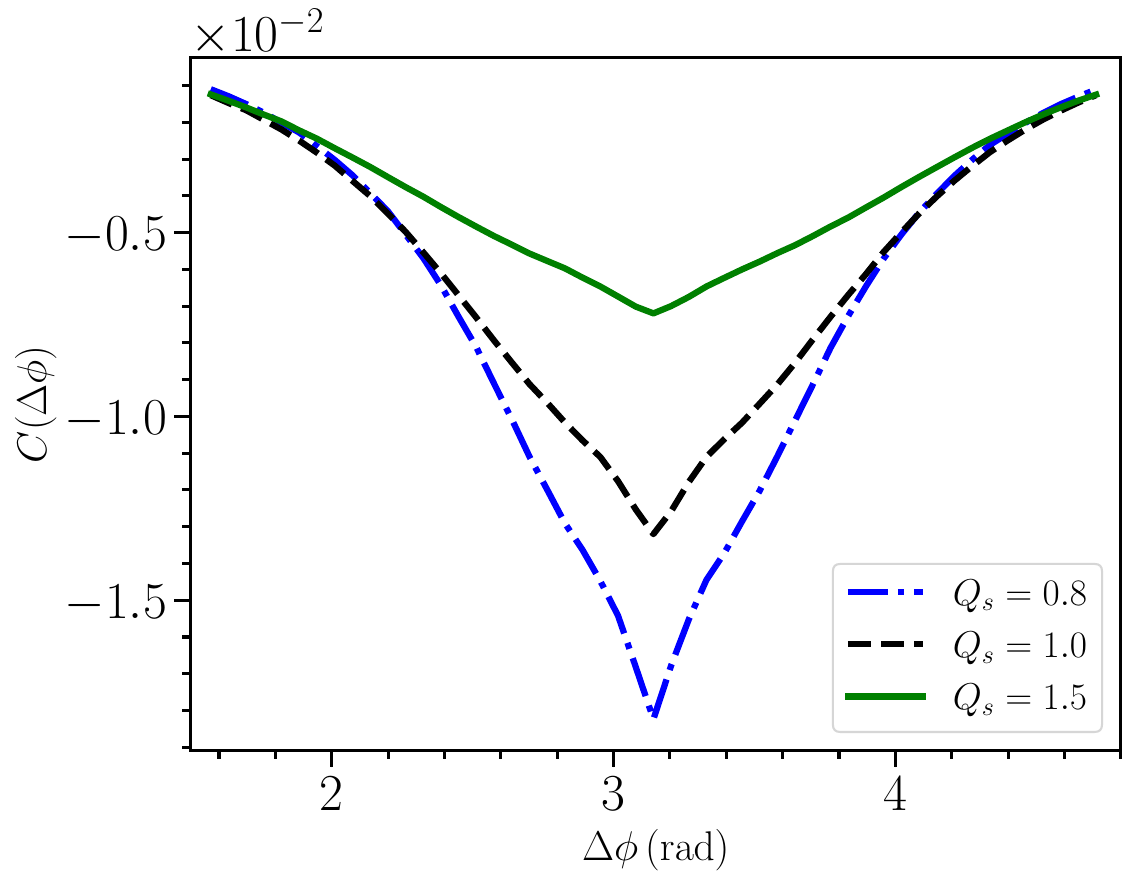}
  \caption{}
  \label{fig:direct_diff_Qs}
\end{subfigure}
\caption{Correlation functions from (a) gluon saturation induced helicity effect and (b) direct helicity effect for three different values of $Q_s = 0.8, \, 1.0,\,  1.5\, \mathrm{GeV}$.}
\label{fig:corr_diff_Qs}
\end{figure*}

Fig.~\ref{fig:corr_diff_p1p2} shows the sub-leading order correlation functions for different values of $|\mathbf{p}_1|$ and $ |\mathbf{p}_2|$. As shown in Fig.~\ref{fig:induced_diff_p1p2}, the azimuthal angle correlation from gluon saturation induced helicity effect is very sensitive to the relative magnitudes between $|\mathbf{p}_1|$ and $|\mathbf{p}_2|$. In contrast, azimuthal angle correlation from the direct helicity effect exhibits only mild sensitivity to changes of $|\mathbf{p}_1|, |\mathbf{p}_2|$ as illustrated in Fig.~\ref{fig:direct_diff_p1p2}.

In Fig.~\ref{fig:corr_diff_Qs}, the correlation functions are plotted for different values of the saturation scale $Q_s$. Fig.~\ref{fig:induced_diff_Qs} illustrates the gluon saturation induced helicity effect. The peak values roughly decreases quadratically as one increase the value of $Q_s$.  A similar trend is observed for the direct helicity effect, as shown in Fig.~\ref{fig:direct_diff_Qs}. As discussed at the beginning of this subsection, the saturation scale regularizes the IR behavior. Both the leading order and sub-leading order contribution to the correlation function are expected to be quadratically dependent on $Q_s$.

\section{Conclusion and Outlook}\label{sec:summary_outlook}
In this paper, we investigated a novel helicity-dependent effect induced by gluon saturation at sub-eikonal order in the high energy limit. At the sub-eikonal order, the transverse gluon field,  $A_{\mathrm{sub}}^i(x^-, \mathbf{x})$, exhibits explicit helicity dependence. Furthermore, the sub-eikonal order longitudinal gluon field, $A_{\mathrm{sub}}^+(x^-, \mathbf{x})$, acquires helicity dependence induced by gluon saturation. This effect emerges through the nonlinear interaction between the eikonal-order gluon field, $A^+_{\mathrm{eik}}(x^-, \mathbf{x})$, and sub-eikonal order gluon field, $A^i_{\mathrm{sub}}(x^-, \mathbf{x})$. The corresponding analytic expression is given in eq.~\eqref{eq:induced_eik+_subi}.  We derived the helicity-dependent field induced by gluon saturation using two complementary approaches: solving the classical Yang-Mills equations and performing explicit diagrammatic calculations. A distinguishing feature of this field is its dependence on two distinct color sources. While these sources share the same longitudinal coordinate, they must differ in their transverse coordinates and their colors.

We evaluated the gluon saturation induced helicity effects in several physical quantities using simple helicity-extended MV model in the quasi-classical approximation. We found no contributions to the single-particle helicity distributions, specifically the dipole gluon helicity TMD and the WW gluon helicity TMD. Instead, the gluon saturation induced helicity effects manifest as genuine two-particle (or multi-particle) correlation effects, rather than as single-particle distribution effects. 

We analyzed the four-point polarized Wilson line correlation function, which involves a transverse chromo-electrically polarized Wilson line dipole and an unpolarized Wilson line dipole. At leading order $\mathcal{O}( Q_s^4)$, only direct helicity effects contribute. At sub-leading order $\mathcal{O}(Q_s^6)$, we derived an additional contribution arising from gluon saturation-induced helicity effects, specifically in the large-$N_c$ limit. The analytic expression for this result is provided in eq.~\eqref{eq:final_Qi_exp}. 

The four-point polarized Wilson line correlation function can be investigated through the double-spin asymmetry in incoherent diffractive quark-antiquark dijet production in longitudinally polarized electron-proton/nucleus collisions at high energies. The quark-antiquark correlated production, as a function of the azimuthal angle between the two jet momenta, has been numerically computed. At leading order, the correlation function exhibits a prominent positive peak in the back-to-back region around $\Delta \phi = \pi$. At sub-leading order, both the helicity-induced gluon saturation effect and the direct helicity effect contribute negatively, also peaking in the back-to-back region. These two helicity effects are of comparable magnitudes and are expected to suppress the leading-order back-to-back peak. Notably, the azimuthal correlation function from the gluon saturation-induced helicity effect is more narrowly distributed around the back-to-back region compared to that from the direct helicity effect. To quantitatively determine the suppression of the back-to-back peak caused by the gluon saturation-induced helicity effect, it will be necessary to compute sub-sub-leading terms of order $\mathcal{O}(Q_s^8)$ and potentially need an exponentiation scheme to include all-order contributions. While the elementary procedures for quasi-classical averaging have been developed, completing this calculation poses significant challenges and is deferred to future studies.

In this paper, gluon saturation induced helicity effects are evaluated within the quasi-classical approximation. Incorporating quantum evolution will be essential for future phenomenological applications. While significant progress has been made in deriving small-$x$ helicity evolution equations for polarized Wilson line correlators  \cite{Kovchegov:2015pbl,Kovchegov:2016weo, Kovchegov:2017lsr, Kovchegov:2018znm, Chirilli:2021lif, Cougoulic:2022gbk, Borden:2024bxa}, these efforts primarily focus on the dilute regime without accounting for gluon saturation. Investigating the interplay between gluon saturation and helicity effects in the evolution equations would be an intriguing direction for future research. 

At the sub-eikonal order, background classical quark fields also carry helicity information \cite{Cougoulic:2020tbc}. The helicity effects induced by gluon saturation are also expected to arise from the nonlinear interactions between $A^+_{\mathrm{eik}}(x^-, \mathbf{x})$ and $\psi_{\mathrm{sub}}(x^-, \mathbf{x})$. To provide a more complete picture, it is necessary to extend the current analysis to include the contributions of background quark fields.

To study the gluon saturation induced helicity effects, we mainly focused on the observable of double-spin asymmetry for incoherent diffractive dijet production in longitudinally polarized electron-proton/nucleus collisions. Additionally, it would be valuable to explore other promising observables \cite{Kovner:2021lty, Kovner:2023yas} and generalize them to polarized collisions . The guiding principle is to look for two-particle (or multi-particle) correlations while subtracting single-particle distributions. It is worth noting that double-spin asymmetry for dijet production has also been measured in longitudinally polarized proton-proton collisions at RHIC \cite{STAR:2018yxi, STAR:2019yqm, STAR:2021mqa}. However, it is probably much more challenging to look for saturation induced helicity effect in polarized proton-proton collisions. Recently, the theoretical framework to study particle production and correlations in longitudinally polarized proton-proton collisions within the small $x$ formalism was introduced in \cite{Kovchegov:2024aus}.

The helicity effect induced by gluon saturation reveals that, at sub-eikonal order, the traditional picture of multiple independent $t$-channel gluon exchanges, with a single sub-eikonal order helicity-dependent insertion, requires modification. Beyond the independent exchange of gluons, helicity-independent gluons can nonlinearly interact with helicity-dependent gluons, merging to form a new $t$-channel gluon before subsequent exchanges. These genuinely nonlinear interactions, driven by gluon saturation, may provide novel insights for probing gluon saturation in polarized collisions.\\

\begin{acknowledgments}
I thank Yuri Kovchegov for stimulating and helpful  discussions.
The work is supported by the U.S. Department of Energy, Office of
Science, Office of Nuclear Physics under Award Number DE-SC0004286.
\end{acknowledgments}

\appendix
\begin{widetext}
\section{Explicit Calculations}\label{sec:appendix}
In this appendix, we provide more details in calculating $\mathcal{I}_4$ in eq.~\eqref{eq:I4_exp} and $\mathcal{I}_3$ in eq.~\eqref{eq:I3_xyu_exp}.
\subsection{Calculating $\mathcal{I}_4$}
We analyze eq.~\eqref{eq:I4_a_AP} in detail. Other terms can be similarly calculated. 
We decompose the five-field averaging into a product of three-field averaging and two-field averaging. The non-vanishing combinations are
\begin{equation}\label{eq:A2_app}
\begin{split}
&\llangle A^+_{c_1}(x_1^-, \mathbf{x}) \partial^i_{\mathbf{x}}A^+_c(x^-, \mathbf{x}) A^+_b(y^-, \mathbf{y}) A^+_a(u^-, \mathbf{u}) A^+_h(v^-, \mathbf{v}) \rrangle\\
=&\Big\langle A^+_{c_1}(x_1^-, \mathbf{x})A^+_a(u^-, \mathbf{u})  \Big\rangle \llangle  \partial^i_{\mathbf{x}}A^+_c(x^-, \mathbf{x}) A^+_b(y^-, \mathbf{y})  A^+_h(v^-, \mathbf{v}) \rrangle\\
&+\Big\langle  \partial^i_{\mathbf{x}}A^+_c(x^-, \mathbf{x})A^+_a(u^-, \mathbf{u}) \Big\rangle \llangle   A^+_{c_1}(x_1^-, \mathbf{x})A^+_b(y^-, \mathbf{y})  A^+_h(v^-, \mathbf{v}) \rrangle\\
&+\Big\langle A^+_{c_1}(x_1^-, \mathbf{x}) A^+_h(v^-, \mathbf{v})  \Big\rangle \llangle  \partial^i_{\mathbf{x}}A^+_c(x^-, \mathbf{x}) A^+_b(y^-, \mathbf{y}) A^+_a(u^-, \mathbf{u}) \rrangle\\
&+\Big\langle  \partial^i_{\mathbf{x}}A^+_c(x^-, \mathbf{x}) A^+_h(v^-, \mathbf{v})\Big\rangle \llangle   A^+_{c_1}(x_1^-, \mathbf{x})A^+_b(y^-, \mathbf{y}) A^+_a(u^-, \mathbf{u})  \rrangle\\
=&\frac{-4g\mu_0^6}{P^+}\, \delta^{c_1a}f^{cbh}\,  L(\mathbf{x}-\mathbf{u}) \partial^i_{\mathbf{x}}\Gamma(\mathbf{x}, \mathbf{y}, \mathbf{v}) \, \delta(x_1^--u^-) \delta(x^--y^-) \delta(x^--v^-)\\
&+\frac{-4g\mu_0^6}{P^+} \delta^{ca} f^{c_1bh} \, \partial^i_{\mathbf{x}}L(\mathbf{x}-\mathbf{u}) \Gamma(\mathbf{x}, \mathbf{y}, \mathbf{v}) \, \delta(x^--u^-)\delta(x_1^--y^-)\delta(x_1^--v^-)\\
&+\frac{-4g\mu_0^6}{P^+} \delta^{c_1h} f^{cba}\, L(\mathbf{x}-\mathbf{v}) \partial^i_{\mathbf{x}} \Gamma(\mathbf{x}, \mathbf{y}, \mathbf{u})\, \delta(x_1^--v^-) \delta(x^--y^-)\delta(x^--u^-)\\
&+\frac{-4g\mu_0^6}{P^+} \delta^{ch}f^{c_1ba} \, \partial^i_{\mathbf{x}}L(\mathbf{x}-\mathbf{v}) \Gamma(\mathbf{x}, \mathbf{y}, \mathbf{u})\, \delta(x^--v^-) \delta(x_1^--y^-)\delta(x_1^--u^-)
\end{split}
\end{equation}
Substituting eq.~\eqref{eq:A2_app} into eq.~\eqref{eq:I4_a_AP}, the integrals over longitudinal coordinates all give the same factor $-\frac{(L^-)^3}{12}$. The color factors can also be easily computed, for example,  
\begin{equation}
\mathrm{tr}\left[t^{c_1} t^c t^b\right]\mathrm{tr}\left[t^at^h\right] \delta^{c_1a} f^{cbh} = \frac{1}{2}f^{cbc_1}\mathrm{tr}\left[t^{c_1} t^c t^b\right] = \frac{1}{2} \frac{i}{4} N_c(N_c^2-1) = \frac{i}{4}N_c^2 C_F.
\end{equation} 
The other color factors can be easily obtained. The final result for eq.~\eqref{eq:I4_a_AP} is 
\begin{equation}
\begin{split}
-\frac{1}{12s} g^6(L^-\mu_0^2)^3N_c^2 C_F&\Big[ \left(L(\mathbf{x}-\mathbf{u})\partial^i_{\mathbf{x}} - \partial^i_{\mathbf{x}}L(\mathbf{x}-\mathbf{u})\right) \Gamma(\mathbf{x}, \mathbf{y}, \mathbf{v} +  \left(L(\mathbf{x}-\mathbf{v})\partial^i_{\mathbf{x}} - \partial^i_{\mathbf{x}}L(\mathbf{x}-\mathbf{v})\right) \Gamma(\mathbf{x}, \mathbf{y}, \mathbf{u})\Big]
\end{split}
\end{equation}
The expression is symmetric with $\mathbf{u}\leftrightarrow \mathbf{v}$. The term in eq.~\eqref{eq:I4_b_AP} can be computed to have exactly the same result as the term in eq.~\eqref{eq:I4_a_AP}.

For the term in eq.~\eqref{eq:I4_c_AP}, one gets the non-vanishing combinations
\begin{equation}\label{eq:A5_app}
\begin{split}
&\llangle \partial^i_{\mathbf{x}}A^+_c(x^-, \mathbf{x})A^+_{b_1}(y_1^-, \mathbf{y}) A^+_b(y^-, \mathbf{y}) A^+_a(u^-, \mathbf{u}) A^+_h(v^-, \mathbf{v}) \rrangle\\
=&\Big\langle A^+_{b_1}(y_1^-, \mathbf{y})A^+_a(u^-, \mathbf{u}) \Big\rangle \llangle \partial^i_{\mathbf{x}}A^+_c(x^-, \mathbf{x}) A^+_b(y^-, \mathbf{y})  A^+_h(v^-, \mathbf{v}) \rrangle\\
&+\Big\langle A^+_b(y^-, \mathbf{y})A^+_a(u^-, \mathbf{u}) \Big\rangle \llangle \partial^i_{\mathbf{x}}A^+_c(x^-, \mathbf{x})  A^+_{b_1}(y_1^-, \mathbf{y}) A^+_h(v^-, \mathbf{v}) \rrangle\\
&+\Big\langle A^+_{b_1}(y_1^-, \mathbf{y}) A^+_h(v^-, \mathbf{v})\Big\rangle \llangle \partial^i_{\mathbf{x}}A^+_c(x^-, \mathbf{x}) A^+_b(y^-, \mathbf{y}) A^+_a(u^-, \mathbf{u})   \rrangle\\
&+\Big\langle A^+_b(y^-, \mathbf{y})A^+_h(v^-, \mathbf{v})  \Big\rangle \llangle \partial^i_{\mathbf{x}}A^+_c(x^-, \mathbf{x}) A^+_{b_1}(y_1^-, \mathbf{y}) A^+_a(u^-, \mathbf{u}) \rrangle\\
=& -\frac{4g\mu_0^6}{P^+} \delta^{b_1a} f^{cbh} L(\mathbf{y}-\mathbf{u}) \partial^i_{\mathbf{x}}\Gamma(\mathbf{x}, \mathbf{y}, \mathbf{v}) \delta(y_1^--u^-) \delta(x^--y^-) \delta(x^--v^-) \\
&  -\frac{4g\mu_0^6}{P^+} \delta^{ba} f^{cb_1h} L(\mathbf{y}-\mathbf{u}) \partial^i_{\mathbf{x}} \Gamma(\mathbf{x}, \mathbf{y}, \mathbf{v})\delta(y^--u^-) \delta(x^--y_1^-) \delta(x^--v^-)\\
& -\frac{4g\mu_0^6}{P^+} \delta^{b_1h} f^{cba} L(\mathbf{y}-\mathbf{v}) \partial^i_{\mathbf{x}}\Gamma(\mathbf{x}, \mathbf{y}, \mathbf{u}) \delta(y_1^--v^-) \delta(x^--y^-) \delta(x^--u^-)\\
& -\frac{4g\mu_0^6}{P^+}\delta^{bh} f^{cb_1a} L(\mathbf{y}-\mathbf{v}) \partial^i_{\mathbf{x}}\Gamma(\mathbf{x}, \mathbf{y}, \mathbf{u}) \delta(y^--v^-) \delta(x^--y_1^-)\delta(x^--u^-). \\
\end{split}
\end{equation}
Substituting eq.~\eqref{eq:A5_app} into eq.~\eqref{eq:I4_c_AP}, working out the integrals over longitudinal coordinates and the color factors, one obtains
\begin{equation}
+\frac{1}{12s} g^6 (L^-\mu_0^2)^3 N_c^2 C_F \Big[2L(\mathbf{y}-\mathbf{u}) \partial^i_{\mathbf{x}}\Gamma(\mathbf{x}, \mathbf{y}, \mathbf{v}) + 2L(\mathbf{y}-\mathbf{v}) \partial^i_{\mathbf{x}}\Gamma(\mathbf{x}, \mathbf{y}, \mathbf{u}) \Big].
\end{equation}
Again, it is symmetric with respect to $\mathbf{u}\leftrightarrow \mathbf{v}$.

For the term in eq.~\eqref{eq:I4_d_AP}, the non-vanishing combinations are 
\begin{equation}\label{eq:A7_app}
\begin{split}
&\llangle \partial^i_{\mathbf{x}}A^+_c(x^-, \mathbf{x})A^+_b(y^-, \mathbf{y}) A^+_a(u^-, \mathbf{u}) A^+_{a_1}(u^-_1, \mathbf{u}) A^+_h(v^-, \mathbf{v}) \rrangle \\
=&\Big\langle \partial^i_{\mathbf{x}}A^+_c(x^-, \mathbf{x}) A^+_a(u^-, \mathbf{u}) \Big\rangle \llangle A^+_b(y^-, \mathbf{y}) A^+_{a_1}(u^-_1, \mathbf{u}) A^+_h(v^-, \mathbf{v}) \rrangle\\
&+\Big\langle \partial^i_{\mathbf{x}}A^+_c(x^-, \mathbf{x})A^+_{a_1}(u^-_1, \mathbf{u})  \Big\rangle \llangle A^+_b(y^-, \mathbf{y})A^+_a(u^-, \mathbf{u})  A^+_h(v^-, \mathbf{v}) \rrangle\\
&+\Big\langle A^+_b(y^-, \mathbf{y}) A^+_a(u^-, \mathbf{u}) \Big\rangle \llangle \partial^i_{\mathbf{x}}A^+_c(x^-, \mathbf{x})  A^+_{a_1}(u^-_1, \mathbf{u}) A^+_h(v^-, \mathbf{v}) \rrangle\\
&+\Big\langle A^+_b(y^-, \mathbf{y})A^+_{a_1}(u^-_1, \mathbf{u})  \Big\rangle \llangle \partial^i_{\mathbf{x}}A^+_c(x^-, \mathbf{x}) A^+_a(u^-, \mathbf{u})  A^+_h(v^-, \mathbf{v}) \rrangle\\
=&-\frac{4g\mu_0^6}{P^+} \delta^{ca} f^{ba_1h} \partial^i_{\mathbf{x}}L(\mathbf{x}-\mathbf{u}) \Gamma(\mathbf{y}, \mathbf{u}, \mathbf{v}) \delta(x^--u^-) \delta(y^--u_1^-) \delta(y^--v^-) \\
&-\frac{4g\mu_0^6}{P^+} \delta^{ca_1} f^{bah} \partial^i_{\mathbf{x}} L(\mathbf{x}-\mathbf{u}) \Gamma(\mathbf{y}, \mathbf{u}, \mathbf{v})\delta(x^--u_1^-) \delta(y^--u^-) \delta(y^--v^-)\\
&-\frac{4g\mu_0^6}{P^+} \delta^{ba} f^{ca_1h} L(\mathbf{y}-\mathbf{u}) \partial^i \Gamma(\mathbf{x}, \mathbf{u}, \mathbf{v}) \delta(y^--u^-) \delta(x^--u_1^-) \delta(x^--v^-) \\
&-\frac{4g\mu_0^6}{P^+} \delta^{ba_1} f^{cah} L(\mathbf{y}-\mathbf{u}) \partial^i_{\mathbf{x}} \Gamma(\mathbf{x}, \mathbf{u}, \mathbf{v})  \delta(y^--u_1^-) \delta(x^--u^-) \delta(x^--v^-)
 \end{split}
\end{equation}
Substituting eq.~\eqref{eq:A7_app} into eq.~\eqref{eq:I4_d_AP}, computing the integrals over longitudinal coordinates and the color factors, one obtains 
\begin{equation}
+\frac{1}{12s} g^6 (L^-\mu_0^2)^3 N_c^2 C_F \Big[ 2\partial^i_{\mathbf{x}}L(\mathbf{x}-\mathbf{u}) \Gamma(\mathbf{y}, \mathbf{u}, \mathbf{v}) - 2L(\mathbf{y}-\mathbf{u}) \partial^i_{\mathbf{x}}\Gamma(\mathbf{x}, \mathbf{u}, \mathbf{v})\Big].
\end{equation}

For the term in eq.~\eqref{eq:I4_e_AP}, the non-vanishing combinations are 
\begin{equation}\label{eq:A9_app}
\begin{split}
& \llangle \partial^i_{\mathbf{x}}A^+_c(x^-, \mathbf{x})A^+_b(y^-, \mathbf{y}) A^+_a(u^-, \mathbf{u}) A^+_{h_1}(v_1^-, \mathbf{v}) A^+_h(v^-, \mathbf{v}) \rrangle\\
=& \Big\langle \partial^i_{\mathbf{x}}A^+_c(x^-, \mathbf{x})A^+_{h_1}(v_1^-, \mathbf{v})\Big\rangle \llangle A^+_b(y^-, \mathbf{y}) A^+_a(u^-, \mathbf{u}) A^+_h(v^-, \mathbf{v}) \rrangle\\
+& \Big\langle \partial^i_{\mathbf{x}}A^+_c(x^-, \mathbf{x})A^+_h(v^-, \mathbf{v}) \Big\rangle \llangle A^+_b(y^-, \mathbf{y}) A^+_a(u^-, \mathbf{u})A^+_{h_1}(v_1^-, \mathbf{v})  \rrangle\\ 
+&\Big\langle A^+_b(y^-, \mathbf{y})  A^+_{h_1}(v_1^-, \mathbf{v})\Big\rangle \llangle\partial^i_{\mathbf{x}}A^+_c(x^-, \mathbf{x}) A^+_a(u^-, \mathbf{u}) A^+_h(v^-, \mathbf{v}) \rrangle\\
+& \Big\langle A^+_b(y^-, \mathbf{y})A^+_h(v^-, \mathbf{v}) \Big\rangle \llangle \partial^i_{\mathbf{x}}A^+_c(x^-, \mathbf{x})  A^+_a(u^-, \mathbf{u})A^+_{h_1}(v_1^-, \mathbf{v})  \rrangle\\
=&-\frac{4g\mu_0^6}{P^+} \delta^{ch_1} f^{bah} \partial^i_{\mathbf{x}}L(\mathbf{x}-\mathbf{v}) \Gamma(\mathbf{y}, \mathbf{u}, \mathbf{v}) \delta(x^--v_1^-) \delta(y^--u^-) \delta(y^--v^-)\\
& -\frac{4g\mu_0^6}{P^+} \delta^{ch} f^{bah_1} \partial^i_{\mathbf{x}}L(\mathbf{x}-\mathbf{v}) \Gamma(\mathbf{y}, \mathbf{u}, \mathbf{v}) \delta(x^--v^-) \delta(y^--u^-) \delta(y^--v_1^-)\\
& -\frac{4g\mu_0^6}{P^+} \delta^{bh_1} f^{cah} L(\mathbf{y}-\mathbf{v}) \partial^i_{\mathbf{x}}\Gamma(\mathbf{x}, \mathbf{u}, \mathbf{v}) \delta(y^--v_1^-) \delta(x^--u^-) \delta(x^--v^-)\\
& -\frac{4g\mu_0^6}{P^+} \delta^{bh} f^{cah_1} L(\mathbf{y}-\mathbf{v}) \partial^i_{\mathbf{x}}\Gamma(\mathbf{x}, \mathbf{u}, \mathbf{v}) \delta(y^--v^-) \delta(x^--u^-)\delta(x^--v_1^-).
\end{split}
\end{equation}
Substituting eq.~\eqref{eq:A9_app} into eq.~\eqref{eq:I4_e_AP}, after computing the integrals involving longitudinal coordinates and the associated color factors, one gets
\begin{equation}
-\frac{1}{12s} g^6(L^-\mu_0^2)^3 N_c^2C_F\Big[2\partial^i_{\mathbf{x}}L(\mathbf{x}-\mathbf{v}) \Gamma(\mathbf{y}, \mathbf{u}, \mathbf{v}) - 2L(\mathbf{y}-\mathbf{v}) \partial^i_{\mathbf{x}}\Gamma(\mathbf{x}, \mathbf{u}, \mathbf{v}) \Big]. 
\end{equation}
From the above term by term calculation, one get the final expression for eq.~\eqref{eq:I4_exp}
\begin{equation}
\begin{split}
\mathcal{I}_4 = -\frac{1}{6s} g^6(L^-\mu_0^2)^3N_c^2 C_F&\Big[ \left(L(\mathbf{x}-\mathbf{u})\partial^i_{\mathbf{x}} - \partial^i_{\mathbf{x}}L(\mathbf{x}-\mathbf{u})\right) \Gamma(\mathbf{x}, \mathbf{y}, \mathbf{v}) +  \left(L(\mathbf{x}-\mathbf{v})\partial^i_{\mathbf{x}} - \partial^i_{\mathbf{x}}L(\mathbf{x}-\mathbf{v})\right) \Gamma(\mathbf{x}, \mathbf{y}, \mathbf{u})\\
&-L(\mathbf{y}-\mathbf{u}) \partial^i_{\mathbf{x}}\Gamma(\mathbf{x}, \mathbf{y}, \mathbf{v})- L(\mathbf{y}-\mathbf{v}) \partial^i_{\mathbf{x}}\Gamma(\mathbf{x}, \mathbf{y}, \mathbf{u})-\partial^i_{\mathbf{x}}L(\mathbf{x}-\mathbf{u}) \Gamma(\mathbf{y}, \mathbf{u}, \mathbf{v}) \\
&+L(\mathbf{y}-\mathbf{u}) \partial^i_{\mathbf{x}}\Gamma(\mathbf{x}, \mathbf{u}, \mathbf{v})+\partial^i_{\mathbf{x}}L(\mathbf{x}-\mathbf{v}) \Gamma(\mathbf{y}, \mathbf{u}, \mathbf{v}) - L(\mathbf{y}-\mathbf{v}) \partial^i_{\mathbf{x}}\Gamma(\mathbf{x}, \mathbf{u}, \mathbf{v})\Big].\\
\end{split}
\end{equation}

\subsection{Calculating $\mathcal{I}_3$}
For the term in eq.~\eqref{eq:I_xyu_a}, the non-vanishing combinations in the averaging are
\begin{equation}\label{eq:A13_app}
\begin{split}
&\llangle \partial^i_{\mathbf{x}} A^+_c(x^-, \mathbf{x}) A^+_{b_1} (y_1^-, \mathbf{y}) A^+_b(y^-, \mathbf{y}) A^+_a(u^-, \mathbf{u}) A^+_{a_1}(u_1^-, \mathbf{u}) \rrangle \\
=&\Big\langle A^+_{b_1} (y_1^-, \mathbf{y})A^+_{a_1}(u_1^-, \mathbf{u}) \Big\rangle  \llangle \partial^i_{\mathbf{x}} A^+_c(x^-, \mathbf{x})  A^+_b(y^-, \mathbf{y}) A^+_a(u^-, \mathbf{u})  \rrangle \\
&+\Big\langle A^+_b(y^-, \mathbf{y}) A^+_a(u^-, \mathbf{u}) \Big\rangle \llangle \partial^i_{\mathbf{x}} A^+_c(x^-, \mathbf{x}) A^+_{b_1} (y_1^-, \mathbf{y})  A^+_{a_1}(u_1^-, \mathbf{u}) \rrangle\\
=&-\frac{4g\mu_0^6}{P^+} \delta^{b_1a_1} f^{cba} L(\mathbf{y}-\mathbf{u}) \partial^i_{\mathbf{x}}\Gamma(\mathbf{x}, \mathbf{y}, \mathbf{u}) \delta(y_1^--u_1^-) \delta(x^--u^-) \delta(x^--y^-)\\
&-\frac{4g\mu_0^6}{P^+} \delta^{ba} f^{cb_1a_1} L(\mathbf{y}-\mathbf{u}) \partial^i_{\mathbf{x}}\Gamma(\mathbf{x}, \mathbf{y}, \mathbf{u}) \delta(y^--u^-) \delta(x^--y_1^-)\delta(x^--u_1^-)
\end{split}
\end{equation}
Substituting eq.~\eqref{eq:A13_app} into eq.~\eqref{eq:I_xyu_a}, one gets the result
\begin{equation}
-\frac{1}{12s} g^6(L^-\mu_0^2)^3 N_c^2 C_F \Big[ 2L(\mathbf{y}-\mathbf{u}) \partial^i_{\mathbf{x}}\Gamma(\mathbf{x}, \mathbf{y}, \mathbf{u})\Big].
\end{equation}
The term in eq.~\eqref{eq:I_xyu_b} has vanishing contribution. For the term in eq.~\eqref{eq:I_xyu_c}, the non-vanishing combinations in evaluating the five-field averaging are 
\begin{equation}
\begin{split}
&\llangle A^+_{c_1}(x_1^-, \mathbf{x}) \partial^i_{\mathbf{x}} A^+_c(x^-, \mathbf{x})A^+_b(y^-, \mathbf{y}) A^+_a(u^-, \mathbf{u}) A^+_{a_1}(u_1^-, \mathbf{u})  \rrangle \\
=&\Big\langle A^+_{c_1}(x_1^-, \mathbf{x}) A^+_a(u^-, \mathbf{u}) \Big\rangle \llangle \partial^i_{\mathbf{x}} A^+_c(x^-, \mathbf{x})A^+_b(y^-, \mathbf{y}) A^+_{a_1}(u_1^-, \mathbf{u})  \rrangle\\
&+\Big\langle \partial^i_{\mathbf{x}} A^+_c(x^-, \mathbf{x})A^+_{a_1}(u_1^-, \mathbf{u}) \Big\rangle \llangle A^+_{c_1}(x_1^-, \mathbf{x}) A^+_b(y^-, \mathbf{y}) A^+_a(u^-, \mathbf{u})   \rrangle\\
=&-\frac{4g\mu_0^6}{P^+} \delta^{c_1a} f^{cba_1} L(\mathbf{x}-\mathbf{u}) \partial^i_{\mathbf{x}}\Gamma(\mathbf{x}, \mathbf{y}, \mathbf{u}) \delta(x_1^--u^-) \delta(x^--y^-) \delta(x^--u_1^-)\\
&-\frac{4g\mu_0^6}{P^+}  \delta^{ca_1}f^{c_1ba} \partial^i_{\mathbf{x}}L(\mathbf{x}-\mathbf{u}) \Gamma(\mathbf{x}, \mathbf{y}, \mathbf{u}) \delta(x^--u_1^-) \delta(x_1^--y^-)\delta(x_1^--u^-).
 \end{split}
\end{equation}
Substituting the above expression into eq.~\eqref{eq:I_xyu_c}, one gets the result 
\begin{equation}
\frac{1}{12s} g^6(L^-\mu_0^2)^3 N_c^2 C_F \Big[L(\mathbf{x}-\mathbf{u}) \partial^i_{\mathbf{x}}\Gamma(\mathbf{x}, \mathbf{y}, \mathbf{u}) - \partial^i_{\mathbf{x}}L(\mathbf{x}-\mathbf{u}) \Gamma(\mathbf{x}, \mathbf{y}, \mathbf{u}) \Big].
\end{equation}
For the term in eq.~\eqref{eq:I_xyu_d}, the non-vanishing combinations are 
\begin{equation}
\begin{split}
& \llangle \partial^i_{\mathbf{x}} A^+_c(x^-, \mathbf{x})A^+_{c_1}(x_1^-, \mathbf{x}) A^+_b(y^-, \mathbf{y}) A^+_a(u^-, \mathbf{u}) A^+_{a_1}(u_1^-, \mathbf{u})  \rrangle \\
=&\Big\langle A^+_{c_1}(x_1^-, \mathbf{x})A^+_{a_1}(u_1^-, \mathbf{u})  \Big\rangle  \llangle \partial^i_{\mathbf{x}} A^+_c(x^-, \mathbf{x}) A^+_b(y^-, \mathbf{y}) A^+_a(u^-, \mathbf{u}) \rrangle\\
&+\Big\langle \partial^i_{\mathbf{x}} A^+_c(x^-, \mathbf{x})   A^+_a(u^-, \mathbf{u}) \Big\rangle \llangle A^+_{c_1}(x_1^-, \mathbf{x}) A^+_b(y^-, \mathbf{y})  A^+_{a_1}(u_1^-, \mathbf{u})  \rrangle\\
=&-\frac{4g\mu_0^6}{P^+} \delta^{c_1a_1} f^{cba} L(\mathbf{x}-\mathbf{u}) \partial^i_{\mathbf{x}}\Gamma(\mathbf{x}, \mathbf{y}, \mathbf{u}) \delta(x_1^--u_1^-) \delta(x^--y^-) \delta(x^--u^-)\\
&- \frac{4g\mu_0^6}{P^+} \delta^{ca} f^{c_1ba_1} \partial^i_{\mathbf{x}}L(\mathbf{x}-\mathbf{u})\Gamma(\mathbf{x},\mathbf{y}, \mathbf{u})\delta(x^--u^-) \delta(x_1^--y^-)\delta(x_1^--u_1^-).
 \end{split}
\end{equation}
Substituting the above expression into eq.~\eqref{eq:I_xyu_d}, one gets the result 
\begin{equation}
\frac{1}{12s} g^6(L^-\mu_0^2)^3 N_c^2 C_F \Big[L(\mathbf{x}-\mathbf{u}) \partial^i_{\mathbf{x}}\Gamma(\mathbf{x}, \mathbf{y}, \mathbf{u}) - \partial^i_{\mathbf{x}}L(\mathbf{x}-\mathbf{u}) \Gamma(\mathbf{x}, \mathbf{y}, \mathbf{u}) \Big].
\end{equation}
Collecting the term by term calculations, one gets
\begin{equation}\label{eq:I_xyu_final}
\mathcal{I}_{\mathbf{x}\mathbf{y}\mathbf{u}} = -\frac{1}{6s} g^6(L^-\mu_0^2)^3 N_c^2 C_F \Big[ L(\mathbf{y}-\mathbf{u}) \partial^i_{\mathbf{x}}\Gamma(\mathbf{x}, \mathbf{y}, \mathbf{u}) +\left(\partial^i_{\mathbf{x}}L(\mathbf{x}-\mathbf{u})-L(\mathbf{x}-\mathbf{u}) \partial^i_{\mathbf{x}}\right)\Gamma(\mathbf{x}, \mathbf{y}, \mathbf{u})  \Big].
\end{equation}
The result for the case $\{\mathbf{x}, \mathbf{y}, \mathbf{v}\}$ can be obtained from eq.~\eqref{eq:I_xyu_final} by the exchange $\mathbf{u}\rightarrow \mathbf{v}$. 
\begin{equation}
\mathcal{I}_{\mathbf{x}\mathbf{y}\mathbf{v}} = -\frac{1}{6s} g^6(L^-\mu_0^2)^3 N_c^2 C_F \Big[ L(\mathbf{y}-\mathbf{v}) \partial^i_{\mathbf{x}}\Gamma(\mathbf{x}, \mathbf{y}, \mathbf{v})+\left(\partial^i_{\mathbf{x}}L(\mathbf{x}-\mathbf{v})-L(\mathbf{x}-\mathbf{v}) \partial^i_{\mathbf{x}}\right)\Gamma(\mathbf{x}, \mathbf{y}, \mathbf{v}) \Big].
\end{equation}

For the case $\{\mathbf{x}, \mathbf{u}, \mathbf{v}\}$, one needs to calculate eq.~\eqref{eq:I_xuv_full}. 
For the term in eq.~\eqref{eq:I_xuv_a}, the non-vanishing combinations when evaluating the five-field averaging are 
\begin{equation}
\begin{split}
&\llangle A^+_c(x_1^-, \mathbf{x}) \partial^i_{\mathbf{x}} A^+_c(x^-, \mathbf{x}) A^+_a(u^-, \mathbf{u}) A^+_{a_1}(u_1^-, \mathbf{u}) A^+_h(v^-, \mathbf{v})\rrangle\\
=&\Big\langle A^+_c(x_1^-, \mathbf{x}) A^+_a(u^-, \mathbf{u}) \Big\rangle \llangle  \partial^i_{\mathbf{x}} A^+_c(x^-, \mathbf{x})  A^+_{a_1}(u_1^-, \mathbf{u}) A^+_h(v^-, \mathbf{v})\rrangle\\
&+ \Big\langle  \partial^i_{\mathbf{x}} A^+_c(x^-, \mathbf{x}) A^+_a(u^-, \mathbf{u})\Big\rangle \llangle A^+_c(x_1^-, \mathbf{x})  A^+_{a_1}(u_1^-, \mathbf{u}) A^+_h(v^-, \mathbf{v})\rrangle\\
&+\Big\langle A^+_c(x_1^-, \mathbf{x}) A^+_{a_1}(u_1^-, \mathbf{u}) \Big\rangle \llangle  \partial^i_{\mathbf{x}} A^+_c(x^-, \mathbf{x})   A^+_a(u^-, \mathbf{u})A^+_h(v^-, \mathbf{v})\rrangle\\
&+ \Big\langle  \partial^i_{\mathbf{x}} A^+_c(x^-, \mathbf{x})A^+_{a_1}(u_1^-, \mathbf{u})  \Big\rangle \llangle A^+_c(x_1^-, \mathbf{x}) A^+_a(u^-, \mathbf{u}) A^+_h(v^-, \mathbf{v})\rrangle\\
=&-\frac{4g\mu_0^6}{P^+} \delta^{ca} f^{ca_1h} L(\mathbf{x}-\mathbf{u}) \partial^i_{\mathbf{x}}\Gamma(\mathbf{x}, \mathbf{u}, \mathbf{v})\delta(x_1^--u^-) \delta(x^--u_1^-) \delta(x^--v^-) \\
 &-\frac{4g\mu_0^6}{P^+} \delta^{ca} f^{ca_1h} \partial^i_{\mathbf{x}}L(\mathbf{x}-\mathbf{u}) \Gamma(\mathbf{x}, \mathbf{u}, \mathbf{v}) \delta(x^--u^-) \delta(x_1^--u_1^-) \delta(x_1^--v^-) \\
 &-\frac{4g\mu_0^6}{P^+} \delta^{ca_1} f^{cah} L(\mathbf{x}-\mathbf{u}) \partial^i_{\mathbf{x}} \Gamma(\mathbf{x}, \mathbf{u}, \mathbf{v}) \delta(x_1^--u_1^-) \delta(x^--u^-) \delta(x^--v^-) \\
 &-\frac{4g\mu_0^6}{P^+} \delta^{ca_1} f^{cah} \partial^i_{\mathbf{x}}L(\mathbf{x}-\mathbf{u}) \Gamma(\mathbf{x}, \mathbf{u}, \mathbf{v}) \delta(x^--u_1^-) \delta(x_1^--u^-) \delta(x_1^--v^-)
   \end{split}
\end{equation}
Substituting the above expression into eq.~\eqref{eq:I_xuv_a}, working out the integral over longitudinal coordinates and the color factors, one obtains 
\begin{equation}\label{eq:res_Ixuv_a}
+\frac{1}{6s} g^6(L^-\mu_0^2)^3 N_c^2 C_F \left[L(\mathbf{x}-\mathbf{u}) \partial^i_{\mathbf{x}} - \partial^i_{\mathbf{x}}L(\mathbf{x}-\mathbf{u})\right] \Gamma(\mathbf{x}, \mathbf{u}, \mathbf{v}).
\end{equation}
The result for eq.~\eqref{eq:I_xuv_b} can be obtained from eq.~\eqref{eq:res_Ixuv_a} by $\mathbf{u}\leftrightarrow \mathbf{v}$. 
\begin{equation}
+\frac{1}{6s} g^6(L^-\mu_0^2)^3 N_c^2 C_F \left[L(\mathbf{x}-\mathbf{v}) \partial^i_{\mathbf{x}} - \partial^i_{\mathbf{x}}L(\mathbf{x}-\mathbf{v})\right] \Gamma(\mathbf{x}, \mathbf{v}, \mathbf{u}).
\end{equation}
For the terms in eq.~\eqref{eq:I_xuv_c}, eq.~\eqref{eq:I_xuv_d} and eq.~\eqref{eq:I_xuv_e}, the non-vanishing combinations in evaluating the five-field averaging are the same, they are 
\begin{equation}
\begin{split}
&\llangle \partial^i_{\mathbf{x}}A^+_c(x^-, \mathbf{x}) A^+_{c_1} (x_1^-, \mathbf{x}) A^+_{c_2}(x_2^-, \mathbf{x}) A^+_a(u^-, \mathbf{u}) A^+_h(v^-, \mathbf{v})\rrangle\\
=&\Big\langle A^+_{c_1} (x_1^-, \mathbf{x})A^+_a(u^-, \mathbf{u})  \Big\rangle \llangle \partial^i_{\mathbf{x}}A^+_c(x^-, \mathbf{x}) A^+_{c_2}(x_2^-, \mathbf{x}) A^+_h(v^-, \mathbf{v})\rrangle\\
&+\Big\langle A^+_{c_2} (x_2^-, \mathbf{x})A^+_a(u^-, \mathbf{u})  \Big\rangle \llangle \partial^i_{\mathbf{x}}A^+_c(x^-, \mathbf{x}) A^+_{c_1}(x_1^-, \mathbf{x}) A^+_h(v^-, \mathbf{v})\rrangle\\
&+\Big\langle A^+_{c_1} (x_1^-, \mathbf{x}) A^+_h(v^-, \mathbf{v}) \Big\rangle \llangle \partial^i_{\mathbf{x}}A^+_c(x^-, \mathbf{x}) A^+_{c_2}(x_2^-, \mathbf{x})A^+_a(u^-, \mathbf{u})\rrangle\\
&+\Big\langle A^+_{c_2} (x_2^-, \mathbf{x})A^+_h(v^-, \mathbf{v}) \Big\rangle \llangle \partial^i_{\mathbf{x}}A^+_c(x^-, \mathbf{x}) A^+_{c_1}(x_1^-, \mathbf{x})A^+_a(u^-, \mathbf{u})  \rrangle\\
=&-\frac{4g\mu_0^6}{P^+} L(\mathbf{x}-\mathbf{u}) \Big[\partial_{\mathbf{x}}^i \Gamma(\mathbf{x}, \mathbf{y}, \mathbf{v})\Big]_{\mathbf{y}\rightarrow \mathbf{x}} \Big[ \delta^{c_1a} f^{cc_2h}\delta(x_1^--u^-) \delta(x^--x_2^-)\delta(x^--v^-)\\
&\qquad +\delta^{c_2a} f^{cc_1h} \delta(x_2^--u^-)\delta(x^--x_1^-)\delta(x^--v^-)\Big]\\
&-\frac{4g\mu_0^6}{P^+} L(\mathbf{x}-\mathbf{v}) \Big[\partial_{\mathbf{x}}^i \Gamma(\mathbf{x}, \mathbf{y}, \mathbf{u})\Big]_{\mathbf{y}\rightarrow \mathbf{x}} \Big[ \delta^{c_1h} f^{cc_2a}\delta(x_1^--v^-) \delta(x^--x_2^-)\delta(x^--u^-)\\
&\qquad +\delta^{c_2h} f^{cc_1a} \delta(x_2^--v^-)\delta(x^--x_1^-)\delta(x^--u^-)\Big]
\end{split}
\end{equation}
It is not difficult to check that the expressions for eq.~\eqref{eq:I_xuv_c} and eq.~\eqref{eq:I_xuv_d} together cancel the expression for eq.~\eqref{eq:I_xuv_e}. 
From the above term by term computations, one gets
\begin{equation}
\begin{split}
\mathcal{I}_{\mathbf{x}\mathbf{u}\mathbf{v}} =& +\frac{1}{6s} g^6(L^-\mu_0^2)^3 N_c^2 C_F\Big(\left[L(\mathbf{x}-\mathbf{u}) \partial^i_{\mathbf{x}} - \partial^i_{\mathbf{x}}L(\mathbf{x}-\mathbf{u})\right] \Gamma(\mathbf{x}, \mathbf{u}, \mathbf{v})\\
&+\left[L(\mathbf{x}-\mathbf{v}) \partial^i_{\mathbf{x}} - \partial^i_{\mathbf{x}}L(\mathbf{x}-\mathbf{v})\right] \Gamma(\mathbf{x}, \mathbf{v}, \mathbf{u})\Big)
\end{split}
\end{equation}
The final result from terms involving three different transverse coordinates is
\begin{equation}
\begin{split}
\mathcal{I}_3 = &\mathcal{I}_{\mathbf{x}\mathbf{y}\mathbf{u}} + \mathcal{I}_{\mathbf{x}\mathbf{y}\mathbf{v}} + \mathcal{I}_{\mathbf{x}\mathbf{u}\mathbf{v}}\\\
=& -\frac{1}{6s} g^6(L^-\mu_0^2)^3 N_c^2 C_F \Big[ L(\mathbf{y}-\mathbf{u}) \partial^i_{\mathbf{x}}\Gamma(\mathbf{x}, \mathbf{y}, \mathbf{u}) +\left[\partial^i_{\mathbf{x}}L(\mathbf{x}-\mathbf{u})-L(\mathbf{x}-\mathbf{u}) \partial^i_{\mathbf{x}}\right]\Gamma(\mathbf{x}, \mathbf{y}, \mathbf{u}) \\
&+L(\mathbf{y}-\mathbf{v}) \partial^i_{\mathbf{x}}\Gamma(\mathbf{x}, \mathbf{y}, \mathbf{v})+\left[\partial^i_{\mathbf{x}}L(\mathbf{x}-\mathbf{v})-L(\mathbf{x}-\mathbf{v}) \partial^i_{\mathbf{x}}\right]\Gamma(\mathbf{x}, \mathbf{y}, \mathbf{v}) \\
&+\left[\partial^i_{\mathbf{x}}L(\mathbf{x}-\mathbf{u})-L(\mathbf{x}-\mathbf{u}) \partial^i_{\mathbf{x}} \right] \Gamma(\mathbf{x}, \mathbf{u}, \mathbf{v})+\left[\partial^i_{\mathbf{x}}L(\mathbf{x}-\mathbf{v})-L(\mathbf{x}-\mathbf{v}) \partial^i_{\mathbf{x}} \right] \Gamma(\mathbf{x}, \mathbf{v}, \mathbf{u})\Big].
\end{split}
\end{equation}

\end{widetext}


\newpage
 \bibliography{trijet, softgluon}
\end{document}